%% file: class_S_chiral.tex
\RequirePackage{pdf14}
\documentclass[letterpaper]{article}
\pdfoutput=1

\usepackage[table]{xcolor}
\usepackage{array,setspace,mathrsfs,amsfonts,amsmath,yfonts,dsfont,bbm,colonequals,amscd}
\usepackage{subfig}
\usepackage{./auxiliary/tikz-cd}
\usepackage{./auxiliary/jheppub}

\input{./auxiliary/topmatter}
\input{./sections/titlepage}

\begin{document}
\setcounter{tocdepth}{2}
\maketitle

\input{./sections/S1}
\input{./sections/Section_2/S2}
\input{./sections/Section_3/S3}
\input{./sections/Section_4/S4}
\input{./sections/Section_5/S5}
\input{./sections/acknowledge}
\appendix
\input{./appendices/app_A}
\input{./appendices/app_B}
\input{./appendices/app_C}

\bibliographystyle{./auxiliary/JHEP}
\bibliography{class_S_chiral}

\end{document}

%% file: auxiliary/topmatter.tex
\newcommand{\blue}[1]{\textcolor{blue}{#1}}

\newcommand{\includegraphicsif}[2]{\IfFileExists{#2}{\includegraphics[#1]{#2}}{\includegraphics[scale=.5]{./figures/dummy.pdf}}} 
\newtheorem{conj}{Conjecture}
\newcommand{\vev}[1]{\langle#1\rangle}
\newcommand{\high}[1]{\raisebox{.5pt}{$\scriptstyle #1$}}
\newcommand{\bigslant}[2]{{\raisebox{.2em}{$#1$}\left/\raisebox{-.2em}{$#2$}\right.}}

\newcommand\qq{\mathbbmtt{Q}}
\newcommand\eea{\end{eqnarray}}
\newcommand\ee{\end{equation}}

\newcommand\ceq{\colonequals}
\newcommand\eqc{\equalscolon}

\newcommand{\Bo}{{\rm Bo}}
\newcommand{\PE}{{\rm PE}}
\newcommand{\goodchi}{\protect\raisebox{1pt}{$\chi$}}

\newcommand\mf{\mathfrak}

\newcommand\nn{\nonumber}
\newcommand\ph{\phantom}
\newcommand\wt{\widetilde}
\newcommand\wh{\widehat}
\newcommand\eg{{\it e.g.}}
\newcommand\ie{{\it i.e.}}
\newcommand\cf{{\it cf.}}

\newcommand\e{\epsilon}

\newcommand\bOn{\mathbf{1}}

\newcommand\bTh{\mathbf{3}}

\newcommand\Ab{\mathbb{A}}

\newcommand\Cb{\mathbb{C}}

\newcommand\Pb{\mathbb{P}}

\newcommand\Rb{\mathbb{R}}
\newcommand\Sb{\mathbb{S}}
\newcommand\Zb{\mathbb{Z}}

\renewcommand\SS{\mathcal{S}}

\newcommand\BB{\mathcal{B}}
\newcommand\CC{\mathcal{C}}
\newcommand\DD{\mathcal{D}}

\newcommand\HH{\mathcal{H}}
\newcommand\II{\mathcal{I}}
\newcommand\JJ{\mathcal{J}}

\newcommand\NN{\mathcal{N}}
\newcommand\OO{\mathcal{O}}

\newcommand\QQ{\mathcal{Q}}
\newcommand\RR{\mathcal{R}}
\newcommand\TT{\mathcal{T}}

\newcommand\WW{\mathcal{W}}

\newcommand\ZZ{\mathcal{Z}}

\newcommand\zb{{\bar z}}

\def\QQ{\mathcal{Q}}

\def\Tr{\mbox{Tr}}

\def\a{\alpha}
\def\b{\beta}
\def\g{\gamma}
\def\dt{\delta}

\def\ab{\textbf{a}}

\def\ph{\phantom}

\def\QQ{\mathcal{Q}}

\def\hf{\frac{1}{2}}

\def\d{\partial}
\def\del{\partial}

\def \mf{\mathfrak}
\def \gf{\mf{g}}
\def \uf{\mf{u}}
\def \ff{\mf{f}}
\def \slf{\mf{sl}}
\def \suf{\mf{su}}
\def \sof{\mf{so}}

\def \uspf{\mf{usp}}

\def \of{\mf{o}}
\def \ef{\mf{e}}
\def \hhf{\mf{h}}
\def \Rf{\mf{R}}

\newcommand\restr[2]{{
  \left.\kern-\nulldelimiterspace
  #1
  \vphantom{\big|}
  \right|_{#2}
  }}
\newcommand\fverb{\setbox\fverbbox=\hbox\bgroup\verb}
\newcommand\fverbdo{\egroup\medskip\noindent%
			\fbox{\unhbox\fverbbox}\ }
\newcommand\fverbit{\egroup\item[\fbox{\unhbox\fverbbox}]}
\newbox\fverbbox

\newcommand{\rep}[1]{\ensuremath{\mathbf{#1}}}

\numberwithin{equation}{section}

\def\repa{\raise4pt\hbox{$\square$}\mkern-14mu\raise-4pt\hbox{$\square$}}
\def\repab{\overline{\raise4pt\hbox{$\square$}\mkern-14mu\raise-4pt\hbox{$\square$}\mkern-1mu}}

\def\smileface{\ensuremath{\hbox{\large$\bigcirc$}\mkern-15mu\raise-1pt\hbox{\scriptsize$\smallsmile$}%
\mkern-10mu\raise4pt\hbox{..}\mkern4mu}}
\def\frownface{\ensuremath{\hbox{\large$\bigcirc$}\mkern-15mu\raise-1pt\hbox{\scriptsize$\smallfrown$}%
\mkern-10mu\raise4pt\hbox{..}\mkern4mu}}

\usepackage{bm}

%% file: sections/titlepage.tex
\title{Chiral algebras of class \texorpdfstring{$\SS$}{S}}

\author[\!1]{Christopher Beem,}
\author[\!2]{Wolfger Peelaers,}
\author[\!2]{Leonardo Rastelli,}
\author[3]{and Balt C. van Rees}

\affiliation[1]{Institute for Advanced Study, Einstein Dr., Princeton, NJ 08540, USA}
\affiliation[2]{C.~N.~Yang Institute for Theoretical Physics, SUNY, Stony Brook, NY 11794-3840, USA}
\affiliation[3]{Theory Group, Physics Department, CERN, CH-1211 Geneva 23, Switzerland}

\preprint{YITP-SB-14-30,  CERN-PH-TH-2014-165}
\bigskip
\abstract{Four-dimensional $\NN=2$ superconformal field theories have families of protected correlation functions that possess the structure of two-dimensional chiral algebras. In this paper, we explore the chiral algebras that arise in this manner in the context of theories of class $\SS$. The class $\SS$ duality web implies nontrivial associativity properties for the corresponding chiral algebras, the structure of which is best summarized in the language of generalized topological quantum field theory. We make a number of conjectures regarding the chiral algebras associated to various strongly coupled fixed points.}

\keywords{Supersymmetry, conformal field theory, class \texorpdfstring{$\SS$}{S}, chiral algebra}

%% file: sections/S1.tex

\section{Introduction}
\label{sec:intro}

A large and interesting class of interacting quantum field theories are the \emph{theories of class $\SS$} \cite{Gaiotto:2009we,Gaiotto:2009hg}. These are superconformal field theories (SCFTs) with half-maximal (\ie, $\NN=2$) supersymmetry in four dimensions. The most striking feature of this class of theories is that they assemble into vast duality webs that are neatly describable in the language of two-dimensional conformal geometry. This structure follows from the defining property of theories of class $\SS$: they can be realized as the low energy limits of (partially twisted) compactifications of six-dimensional CFTs with $(2,0)$ supersymmetry on punctured Riemann surfaces.

Generic theories of class $\SS$ are strongly interacting. (In many cases they possess generalized weak-coupling limits wherein the neighborhood of a certain limit point on their conformal manifold can be described by a collection of \emph{isolated} strongly coupled SCFTs with weakly gauged flavor symmetries.) It is remarkable, then, that one can say much of anything about these theories in the general case. One classic and successful approach has been to restrict attention to the weakly coupled phases of these theories by, for example, studying the physics of Coulomb branch vacua at the level of the low energy effective Lagrangian and the spectrum of BPS states. Relatedly, one may utilize brane constructions of these theories to extract some features of the Coulomb branch physics \cite{Witten:1997sc,Benini:2009gi}.

An alternative -- and perhaps more modern -- tactic is to try to constrain or solve for various aspects of these theories using consistency conditions that follow from duality. This approach was successfully carried out in \cite{Gaiotto:2012xa} (building on the work of \cite{Gadde:2009kb, Gadde:2010te, Gadde:2011ik, Gadde:2011uv}) to compute the superconformal index of a very general set of class $\SS$ fixed points (see also \cite{Lemos:2012ph,Mekareeya:2012tn} for extensions to even more general cases). Subsequently, the framework for implementing this approach to study the (maximal) Higgs branch was established in \cite{Moore:2011ee}. The general aspiration in this sort of program is that the consistency conditions imposed by generalized $S$-duality and the (known) behavior of these theories under certain partial Higgsing and weak gauging operations may be sufficient to completely determine certain nice observables. In this sense the approach might be thought of as a sort of ``theory space bootstrap''. One expects that this approach has the greatest probability of success when applied to observables of class $\SS$ theories that are protected against corrections when changing exactly marginal couplings, thus leading to objects that are labelled by topological data and have no dependence on continuous parameters.%
\footnote{Observables with a manageable dependence on the marginal couplings, such as $\Rb^4_{\epsilon_1\epsilon_2}$ and $\Sb^4$ partition functions, also provide natural settings for this type of argument.}

A new class of protected observables for four-dimensional $\NN=2$ SCFTs was introduced in \cite{Beem:2013sza}. There it was shown that certain carefully selected local operators, restricted to be coplanar and treated at the level of cohomology with respect to a particular nilpotent supercharge, form a closed subalgebra of the operator algebra. Moreover their operator product expansions and correlation functions are meromorphic functions of the operator insertion points on the plane. This subalgebra consequently acquires the structure of a two-dimensional chiral algebra. The spectrum and structure constants of this chiral algebra are subject to a non-renormalization theorem that renders them independent of marginal couplings. The existence of this sector can formally be summarized by defining a map that associates to any $\NN=2$ SCFT in four dimensions the chiral algebra that computes the appropriate protected correlation functions,
\begin{equation*}
\goodchi~:~\bigg\{\bigslant{\text{$\NN=2$ SCFTs}}{\text{Marginal deformations}}\bigg\}~\longrightarrow~\bigg\{\text{Chiral algebras}\bigg\}~.
\end{equation*}
Chiral algebras with the potential to appear on the right hand side of this map are not generic -- they must possess a number of interesting properties that reflect the physics of their four-dimensional ancestors.

In this paper we initiate the investigation of chiral algebras that are associated in this manner with four-dimensional theories of class $\SS$. For lack of imagination, we refer to the chiral algebras appearing in this fashion as \emph{chiral algebras of class $\SS$}. For a general strongly interacting SCFT, there is at present no straightforward method for identifying the associated chiral algebra. Success in this task would implicitly fix an infinite amount of protected CFT data (spectral data and three-point couplings) that is generally difficult to determine. However, given the rigid nature of chiral algebras, one may be optimistic that chiral algebras of class $\SS$ can be understood in some detail by leveraging the constraints of generalized $S$-duality and the wealth of information already available about the protected spectrum of these theories. In the present work, we set up the abstract framework of this bootstrap problem in the language of generalized topological quantum field theory, and put into place as many ingredients as possible to define the problem concretely. We perform some explicit calculations in the case of theories of rank one and rank two, and formulate a number of conjectures for the higher rank case. One of our main results is a general prescription to obtain the chiral algebra of a theory with sub-maximal punctures given that of the related theory with all maximal punctures. We demonstrate that the reduction in the rank of a puncture is accomplished in the chiral algebra by quantum Drinfeld-Sokolov reduction, with the chiral algebra procedure mirroring the corresponding four-dimensional procedure involving a certain Higgsing of flavor symmetries.

Ultimately we believe that the bootstrap problem for chiral algebras of class $\SS$ may prove solvable, and we hope that the existence of this remarkable structure will pique the interest of readers with a passion for vertex operator algebras. Characterizing these algebras should prove to be both mathematically and physically rewarding.

The organization of this paper is as follows. Section \ref{sec:review} is a two-part review: first of the protected chiral algebra of $\NN=2$ SCFTs, and then of $\NN=2$ SCFTs of class $\SS$. In Section \ref{sec:TQFT}, we outline the structure of the chiral algebras of class $\SS$, using the $A_1$ and $A_2$ cases as examples. We also take some steps to formalize the TQFT structure of the chiral algebras of class $\SS$ so as to emphasize that the structures outlined here are susceptible to rigorous mathematical analysis. In Section \ref{sec:reducing}, we describe the generalization of our story to the case of theories with sub-maximal punctures. In the process, we are led to consider the problem of quantum Drinfeld-Sokolov reduction for modules of affine Lie algebras. In Section \ref{sec:cyl_and_cap}, we offer some comments on unphysical chiral algebras that are expected to exist at a formal level in order to complete the TQFT structure. A number of technical details having to do with rank two theories are included in Appendix \ref{app:level_by_level}. Details having to do with unphysical cylinder and cap chiral algebras appear in Appendix \ref{app:cylinders_and_caps}. Finally, in Appendix \ref{app:spectral_sequences} we review the methods for computing the cohomology of a double complex using spectral sequences. These methods are instrumental to the analysis of Section \ref{sec:reducing}.

%% file: sections/Section_2/S2.tex

\section{Background}
\label{sec:review}

We begin with a review of the two main topics being synthesized in this paper: the protected chiral algebras of $\NN=2$ SCFTs and superconformal theories of class $\SS$. Readers who have studied our first paper on protected chiral algebras \cite{Beem:2013sza} should be fine skipping Section \ref{subsec:chiral_review}, while those familiar with the class $\SS$ literature (for example, \cite{Gaiotto:2009we,Gadde:2009kb,Gaiotto:2012xa,Chacaltana:2010ks}) may safely skip Section \ref{subsec:class_S_review}.

\medskip
\input{./sections/Section_2/S2_1}
\medskip
\input{./sections/Section_2/S2_2}
\bigskip

%% file: sections/Section_2/S2_1.tex

\subsection{Review of protected chiral algebras}
\label{subsec:chiral_review}

The observables we aim to study for class $\SS$ fixed points are those described by the protected chiral algebras introduced in \cite{Beem:2013sza} (see also \cite{Beem:2014kka} for the extension to six dimensions). The purpose of this section is to provide a short overview of how those chiral algebras come about and the properties that were deduced for them in the original papers. We simply state the facts in this section; the interested reader is encouraged to consult the original work for explanations.

The starting point is the $\NN=2$ superconformal algebra $\mf{su}(2,2|2)$. The fermionic generators of the algebra are Poincar\'e supercharges $\{\QQ^{\II}_{\alpha},\tilde\QQ_{\dot\alpha \JJ}\}$ and special conformal supercharges $\{\SS^{\alpha}_{\II},\tilde\SS^{\dot\alpha\JJ}\}$. From these, one can form two interesting nilpotent supercharges that are mixtures of Poincar\'e and special conformal supercharges,
\begin{equation}
\label{eq:square_supercharges}
\qq_{\,1}\ceq \QQ^1_{-}+\tilde\SS^{\dot{-}2}~,\qquad\qq_{\,2}\ceq \tilde\QQ_{\dot{-}2}+\SS_1^{-}~.
\end{equation}
These supercharges have the following interesting property. Let us define the subalgebra of the four-dimensional conformal symmetry algebra that acts on a plane $\Rb^2\subset\Rb^4$ as $\slf(2)\times\overline{\slf(2)}$. Let us further denote the complexification of the $\suf(2)_R$ $R$-symmetry as $\slf(2)_R$. These subalgebras have the following nice relationship to the supercharges $\qq_{\,i}$,
\begin{equation}
\label{eq:q_exact_commutators}
[\qq_{\,i},\slf(2)]=0~,\qquad \{\qq_{\,i},\cdot\}=\mathrm{diag}\left[\overline{\slf(2)}\times\slf(2)_R\right]~.
\end{equation}
It follows from these relations that operators that are $\qq\,$-closed must behave as meromorphic operators in the plane. They have meromorphic operator product expansions (modulo $\qq\,$-exact terms) and their correlation functions are meromorphic functions of the positions. Restricting from the full $\NN=2$ SCFT to $\qq\,$-cohomology therefore defines a two-dimensional chiral algebra. For a pedagogical discussion of chiral algebras, see \cite{Bouwknegt:1992wg}.

The conditions for a local operator to define a nontrivial $\qq\,$-cohomology element were worked out in \cite{Beem:2013sza}. It turns out that such operators are restricted to lie in the \emph{chiral algebra plane}: $\{x_3=x_4=0\}$. When inserted at the origin, an operator belongs to a well-defined cohomology class if
and only if it obeys the conditions 
\begin{equation}
\label{schurconditions}
\hat h \ceq \frac{E-(j_1+j_2)}{2}-R = 0~,\qquad \ZZ \ceq j_1-j_2+r = 0~.
\end{equation}
Unitarity of the superconformal representation requires $\hat h \geqslant \frac{ | \ZZ | }{2}$, so the first condition actually implies the second. We refer to operators obeying $\hat h = 0$ as \emph{Schur operators}. All Schur operators are necessarily $\suf(2)_R$ highest weight states. Indeed, if the $\suf(2)_R$ raising generator did \emph{not} annihilate a Schur operator, it would generate an operator with $\hat h < 0$, which would violate unitarity.

As $\overline {\slf(2)}$ does not commute with $\qq$, ordinary translations of Schur operators in the chiral algebra plane fail to be $\qq\,$-closed away from the origin. Rather, we translate operators using the twisted translation generator $\widehat{L}_{-1}\ceq \overline{L}_{-1}+\RR_{-}$, where $\RR_{-}$ is the lowering operator of $\suf(2)_R$. As shown in Eqn.~\eqref{eq:q_exact_commutators}, this is a $\qq\,$-exact operation. We find that local operators defining nontrivial $\qq\,$-cohomology classes can be written in the form
\begin{equation}
\label{eq:twisted_translated}
\OO(z,\zb)\ceq u_{\II_1}(\zb) \cdots u_{\II_k}(\zb)\OO^{\{\II_1\cdots\II_k\}}(z,\zb)~,\qquad \text{where}\qquad u_{\II}(\zb)\ceq\binom{\,1\,}{\zb}~.
\end{equation}
Here $\OO^{1\cdots1}(0)$ is a Schur operator, and we are suppressing Lorentz indices. It is these \emph{twisted-translated} Schur operators, taken at the level of cohomology, that behave as meromorphic operators in two dimensions,
\begin{equation}
\label{eq:twisted_translated_to_meromorphic}
\OO(z) \ceq [\OO(z,\zb)]_{\qq_{\,i}}~.
\end{equation}
We now turn to a recap of the various types of four-dimensional operators that may satisfy the Schur condition, and thus participate in the protected chiral algebra.

\renewcommand{\arraystretch}{1.55}
\begin{table}
\centering
\begin{tabular}{|l|l|l|l|l|}
\hline \hline
Multiplet  			   & $\OO_{\rm Schur}$  		    						 & $h\ceq\frac{E+j_1+j_2}{2}$ & $\ph{-}r$  		  	& Lagrangian ``letters''\\ 
\hline 
$\hat\BB_R$  		   & $\Psi^{11\dots 1}$   				   					 & $R$ 		 				  & $\ph{-}0$ 			& $Q$, $\tilde Q$ \\ 
\hline
$\bar\DD_{R(j_1,0)}$   & ${\QQ}^1_+\Psi^{11\dots 1}_{+\dots+}$ 					 & $R+j_1+1$ 				  & $-j_1-\frac12~$ 	& $Q$, $\tilde Q$, $\lambda^1_+$ \\
\hline
$\DD_{R(0,j_2)}$  	   & $\wt{\QQ}^1_{\dot +}\Psi^{11\dots 1}_{\dot+\dots\dot+}$ & $R+j_2+1$				  & $\ph{-}j_2+\frac12$ & $Q$, $\tilde Q$, $\tilde\lambda^1_{\dot+}$ \\
\hline
$\hat\CC_{R(j_1,j_2)}$ & $\QQ^1_{+} \wt{\QQ}^1_{\dot+}\Psi^{1\dots1}_{+\dots+\,\dot+\dots\dot+}$ & $R+j_1+j_2 +2$ & $\ph{-}j_2-j_1$ & $D_{+\dot+}^n Q$, $D_{+\dot+}^n \tilde Q$, $D_{+\dot+}^n \lambda^1_+$, $D_{+\dot+}^n \tilde\lambda^1_{\dot+}$\\
\hline
\end{tabular}
\caption{\label{schurTable} This table summarizes the manner in which Schur operators fit into short multiplets of the $\NN=2$ superconformal algebra. We use the naming conventions for supermultiplets of Dolan and Osborn \cite{Dolan:2002zh}. For each supermultiplet, we denote by $\Psi$ the superconformal primary. There is then a single conformal primary Schur operator ${\OO}_{\rm Schur}$, which in general is obtained by the action of some Poincar\'e supercharges on $\Psi$. The holomorphic dimension ($h$) and $U(1)_r$ charge ($r$) of ${\OO}_{\rm Schur}$ are determined in terms of the quantum numbers $(R,j_1,j_2)$ that label the shortened multiplet. We also indicate the schematic form that ${\OO}_{\rm Schur}$ can take in a Lagrangian theory by enumerating the elementary ``letters'' from which the operator may be built. We denote by $Q$ and $\tilde Q$ the complex scalar fields of a hypermultiplet, by $\lambda_{\alpha}^\II$ and $\tilde \lambda_{\dot \alpha}^\II$ the left- and right-handed fermions of a vector multiplet, and by $D_{\alpha \dot \alpha}$ the gauge-covariant derivatives. Note that while in a Lagrangian theory Schur operators are built from these letters, the converse is false -- not \emph{all} gauge-invariant words of this kind are Schur operators. Only the special combinations with vanishing anomalous dimensions retain this property at finite coupling.} 
\end{table}

\subsubsection{Taxonomy of Schur operators}
\label{subsubsec:schur_taxonomy}

A Schur operator is annihilated by two Poincar\'e supercharges of opposite chiralities ($\QQ_-^1$ and $\widetilde \QQ_{2 \dot -}$ in our conventions). A summary of the different classes of Schur operators, organized according to how they fit in shortened multiplets of the superconformal algebra, is given in Table \ref{schurTable} (reproduced from \cite{Beem:2013sza}). Let us briefly discuss each row in turn.

The first row describes half-BPS operators that are a part of the Higgs branch chiral ring. These have $E = 2R$ and $j_1 = j_2 = 0$. In a Lagrangian theory, operators of this type schematically take the form $QQ\cdots\tilde Q\tilde Q$. A special case is when $R = 1$, in which case a conserved current is amongst the super-descendants of the primary. The half-BPS primary is then the ``moment map'' operator $\mu_A$ which has dimension two and transforms in the adjoint representation of the flavor symmetry. The $\suf(2)_R$ highest weight state of the moment map is a Schur operator.

The operators in the second row are more general $\NN=1$ chiral operators, obeying $E = 2 R+|r|$ and $r= -j_1 -\frac12$. Together with the Higgs branch chiral ring operators (which can be regarded as the special case with $r=0$), they make up the so-called \emph{Hall-Littlewood chiral ring}. These are precisely the operators that are counted by the Hall-Littlewood limit of the superconformal index \cite{Gadde:2011uv}. In a Lagrangian theory, these operators are obtained by constructing gauge-invariant words out of $Q$, $\tilde Q$, and the gaugino field $\lambda^1_+$ (the bottom component of the field strength chiral superfield $W_\alpha$ with $\alpha = +$). In complete analogy, the third line describes $\NN=1$ \emph{anti}-chiral operators obeying $E= 2 R+|r|$, $r= j_2 +\frac12$, which belong to the Hall-Littlewood anti-chiral ring. The second and third lines are CPT conjugate to each other. It is believed that $\DD$ and $\overline\DD$ type operators are absent in any theory arising from a (generalized) quiver description with no loops (\ie, an \emph{acyclic quiver}). These are theories for which the Hall-Littlewood superconformal index matches the ``Hilbert series'' for the Higgs branch \cite{Gadde:2011uv,Benvenuti:2010pq}. Equivalently, these are the theories for which the maximal Higgs branch is an honest Higgs branch, with no low-energy abelian gauge field degrees of freedom surviving.

The fourth line describes the most general type of Schur operators, which belong to supermultiplets that obey less familiar semi-shortening conditions. An important operator in this class is the conserved current for $\suf(2)_R$, which belongs to the $\hat \CC_{0(0, 0)}$ supermultiplet which also contains the stress-energy tensor and is therefore universally present in any $\NN=2$ SCFT. This current has one component with $E= 3$, $R=1$, $j_1 = j_2 = \frac12$ which is a Schur operator.

Finally, let us point out the conspicuous absence of half-BPS operators that belong to the \emph{Coulomb} branch chiral ring (these take the form $\Tr\,\phi^k$ in a Lagrangian theory, where $\phi$ is the complex scalar of the $\NN=2$ vector multiplet). These operators are in many ways more familiar than those appearing above due to their connection with Coulomb branch physics. The protected chiral algebra is thus complementary, rather than overlapping, with a Coulomb branch based analysis of class $\SS$ physics.

\subsubsection{The $4d/2d$ dictionary}
\label{subsubsec:4d_2d_dict}

There is a rich dictionary relating properties of a four-dimensional SCFT with properties of its associated chiral algebra. Let us briefly review some of the universal entries in this dictionary that were worked out in \cite{Beem:2013sza}. Interested readers should consult that reference for more detailed explanations.

\subsubsection*{Virasoro symmetry}

The stress tensor in a four-dimensional $\NN=2$ SCFT lives in the $\hat \CC_{0(0, 0)}$ supermultiplet, which contains as a Schur operator a component of the $\suf(2)_R$ conserved current $\JJ^{(\II\JJ)}_{\alpha\dot\alpha}$. The corresponding twisted-translated operator gives rise in cohomology to a two-dimensional meromorphic operator of dimension two, which acts as a two-dimensional stress tensor, $T(z)\ceq [\JJ_{+\dot+}(z,\bar z)]_{\qq}$. As a result, the global $\slf(2)$ symmetry that is inherited from four dimensions is always enhanced to a local Virasoro symmetry acting on the chiral algebra. From the current-current OPE, which is governed by superconformal Ward identities, one finds a universal expression for the Virasoro central charge,
\begin{equation}
\label{eq:cc_relation}
c_{2d} = -12\,c_{4d}~,
\end{equation}
where $c_{4d}$ is the conformal anomaly coefficient of the four-dimensional theory associated to the square of the Weyl tensor. Note that the chiral algebra is necessarily non-unitary due to the negative sign in Eqn.~\eqref{eq:cc_relation}.

\subsubsection*{Affine symmetry}

Similarly, continuous global symmetries of the four-dimensional SCFT (when present) are enhanced to local affine symmetries at the level of the associated chiral algebra. This comes about because the conserved flavor symmetry current sits in the $\hat\BB_1$ supermultiplet, whose bottom component is the moment-map operator discussed above. The $\suf(2)_R$ highest weight component of the moment map operator then gives rise to an affine current, $J_A(z)\ceq [\mu_A(z,\bar z)]_{\qq}$. The level of the affine current algebra is related to the four-dimensional flavor central charge by another universal relation,
\begin{equation}
\label{eq:kk_relation}
k_{2d} = -\frac12 k_{4d}~.
\end{equation}

\subsubsection*{Hall-Littlewood ring generators as chiral algebra generators}

Identifying chiral algebra generators is of crucial importance if one is to find an intrinsic characterization of any particular chiral algebra without reference to its four-dimensional parent. A very useful fact is that generators of the Hall-Littlewood chiral ring (and in particular those of the Higgs branch chiral ring) necessarily give rise to generators of the protected chiral algebra after passing to $\qq\,$-cohomology. This follows from $\suf(2)_R$ and $\uf(1)_r$ selection rules, which forbid such an operator from appearing in any non-singular OPEs. A special case is the aforementioned affine currents, which arise from Higgs branch moment map operators with $E=2R=2$. With the exception of theories with free hypermultiplets, these are always generators.
 
\subsubsection*{Exactly marginal gauging}

Given an SCFT $\TT$ with a flavor symmetry $G$ that has flavor central charge $k_{4d}=4h^{\vee}$, one may form a new family of SCFTs $\TT_G$ by introducing an $\NN=2$ vector multiplet in the adjoint representation of $G$ and gauging the symmetry. This specific value of the flavor central charge ensures that the gauge coupling beta function vanishes, so the procedure preserves conformal invariance.

There exists a corresponding procedure at the level of chiral algebras that produces the chiral algebra $\goodchi[\TT_G]$ given that of the original theory $\goodchi[\TT]$. In parallel with the introduction of a $G$-valued vector multiplet, one introduces a dimension $(1,0)$ ghost system $(b_A,c^A)$ with $A=1,\ldots,\dim G$. In the tensor product of this ghost system and the chiral algebra $\goodchi[\TT]$, one may form a canonical nilpotent BRST operator given by
\begin{equation}
\label{eq:BRST_def}
Q_{\rm BRST} \ceq \oint\frac{dz}{2\pi i}\,j_{\rm BRST}(z)~,\qquad j_{\rm BRST}(z) \ceq \left(c^A\left[J_A-\frac12 f_{AB}^{\ph{AB}C}\,c^B\,b_C\right]\right)(z)~,
\end{equation}
where the affine currents $J_A(z)$ are those associated with the $G$ symmetry of $\goodchi[\TT]$, and $f_{AB}^{\ph{AB}C}$ are the structure constants for $G$. Nilpotency of this BRST operator depends on the precise value of the affine level $k_{2d}=-2h^{\vee}$, and so the self-consistency of this procedure is intimately connected with the preservation of conformal invariance in four dimensions. The gauged chiral algebra is then obtained as the cohomology of the BRST operator relative to the $b\,$-ghost zero modes,
\begin{equation}
\label{eq:gauged_cohomology}
\goodchi[\TT_G] = H^{\star}_{\rm BRST} \left[\psi\in\goodchi[\TT]\otimes\goodchi_{(b,c)}~|~b^A_{0}\psi=0\right]
\end{equation}

\subsubsection*{Superconformal index}

The superconformal index of a superconformal field theory is the Witten index of the radially quantized theory, refined by a set of fugacities that keep track of the maximal set of charges commuting with each other and with a chosen supercharge. For our purposes, we consider the specialization of the index of an $\NN=2$ SCFT known as the Schur index \cite{Gadde:2011ik,Gadde:2011uv}. The trace formula for the Schur index reads
\begin{equation}
\label{eq:SchurSCI}
\II^{\rm (Schur)}(q; {\bf x}) = \Tr_{\HH[\Sb^3]}(-1)^F q^{E-R}\prod_i {x_i}^{f_i}~,
\end{equation}
where $F$ denotes the fermion number and $\{f_i\}$ the Cartan generators of the flavor group. The Schur index counts (with signs) precisely the operators obeying the condition (\ref{schurconditions}). Moreover, for Schur operators $E-R$ coincides with the left-moving conformal weight $h$ (the eigenvalue of $L_0$),
\begin{equation}
\label{eq:schu_op_dimension}
E-R ~=~ \frac{E+j_1+j_2}{2} ~\eqc~ h~.
\end{equation}
It follows that the graded character of the chiral algebra is identical to the Schur index,
\begin{equation}
\label{charschur}
\II_{\chi}(q; {\bf x}) ~\ceq~ \Tr_{\HH_{\chi}}\,(-1)^F q^{L_0} ~=~ \II^{\rm Schur}(q; {\bf x})~,
\end{equation}
where $\HH_{\chi}$ denotes the state space of the chiral algebra. Note that this object is not interpreted as an index when taken as a partition function of the chiral algebra, because (with the exception of chiral algebras associated to $\NN=4$ theories in four dimensions) the protected chiral algebra itself is not supersymmetric.

%% file: sections/Section_2/S2_2.tex

\subsection{Review of theories of class \texorpdfstring{$\SS$}{S}}
\label{subsec:class_S_review}

Four-dimensional superconformal field theories of class $\SS$ may be realized as the low-energy limit of twisted compactifications of an $\NN=(2,0)$ superconformal field theory in six dimensions on a Riemann surface, possibly in the presence of half-BPS codimension-two defect operators. The resulting four-dimensional theory is specified by the following data:\footnote{We restrict our attention in this note to \emph{regular} theories. A larger class of theories can be obtained by additionally allowing for \emph{irregular} punctures \cite{Witten:2007td}. Still more possibilities appear when the UV curve is decorated with outer automorphisms twist lines \cite{Tachikawa:2010vg,Chacaltana:2012ch}.} 
\begin{itemize}
\item A simply-laced Lie algebra $\gf = \{A_n, D_n, E_6, E_7, E_8\}$. This specifies the choice of six-dimensional $(2,0)$ theory.
\item A (punctured) Riemann surface ${\CC}_{g,s}$ known as the \emph{UV curve}, where $g$ indicates the genus and $s$ the number of punctures. In the low energy limit, only the complex structure of ${\CC}_{g,s}$ plays a role. The complex structure moduli of the curve are identified with exactly marginal couplings in the SCFT.
\item A choice of embedding $\Lambda_i: \suf(2) \to \gf$ (up to conjugacy) for each puncture $i=1,\ldots,s$. These choices reflect the choice of codimension-two defect that is present at each puncture in the six-dimensional construction. The centralizer $\hhf_{\Lambda_i} \subset \gf$ of the embedding is the global symmetry associated to the defect. The theory enjoys a global flavor symmetry algebra given by at least $\oplus_{i=1}^s\hhf_{\Lambda_i}$.\footnote{In some exceptional cases the global symmetry of the theory is enhanced due to the existence of additional symmetry generators that are not naturally associated to an individual puncture.}
\end{itemize}
When necessary, we will label the corresponding four-dimensional SCFT as $\TT[\gf; \CC_{g,s}; \{\Lambda_i\}]$. Because we are ultimately only interested in theories modulo their exactly marginal couplings, we will not keep track of a point in the complex structure moduli space of the UV curve.

For the sake of simplicity, we will restrict our attention to theories where $\gf$ is in the $A$ series. The generalization to $D$ and $E$ series theories (at least in the abstract discussion) should be possible to carry out without a great deal of additional difficulty. In the $A_{n-1}$ case -- \ie, $\gf=\suf(n)$ -- the data at punctures can be reformulated as a partition of $n$: $[n_1^{\ell_1}\,n_2^{\ell_2}\,\dots\,n_k^{\ell_k}]$ with $\sum_i \ell_i n_i = n$ and $n_i > n_{i+1}$. Such a partition indicates how the fundamental representation $\ff$ of $\suf(n)$ decomposes into irreps of $\Lambda(\suf(2))$,
\begin{equation}
\label{eq:fundamental_decomposition_partition}
\ff \rightarrow \bigoplus_{i=1}^{k} \ell_i V_{\frac12(n_i-1)}~,
\end{equation}
where $V_j$ denotes the spin $j$ representation of $\suf(2)$. An equivalent description comes from specifying a nilpotent element $e$ in $\suf(n)$, \ie, an element for which $\left({\rm ad}_e\right)^r=0$ for some positive integer $r$. The Jordan normal form of such a nilpotent element is given by
\begin{equation}
\label{eq:jordan_normal_form}
e = \bigoplus_{i=1}^{k} \overbrace{J_{n_i}\oplus\cdots\oplus J_{n_i}}^{\ell_i~{\rm times}}~,
\end{equation}
where $J_m$ is the elementary Jordan block of size $m$, \ie, a sparse $m\times m$ matrix with only ones along the superdiagonal. Thus every nilpotent element specifies a partition of $n$ and vice versa. The $\suf(2)$ embedding comes from defining $\suf(2)$ generators $t_0, t_\pm$ and demanding that $\Lambda(t_-)=e$.

The trivial embedding is identified with the partition $[1^n]$ and leads to a defect with maximal flavor symmetry $\hhf = \suf(n)$. A puncture labelled by this embedding is called \emph{full} or \emph{maximal}. The opposite extreme is the principal embedding, which has partition $[n^1]$. This embedding leads to $\hhf = \varnothing$, and the puncture is effectively absent. Another important case is the subregular embedding, with partition $[n-1,1]$, which leads to $\hhf = \uf(1)$ (as long as $n>2$). Punctures labelled by the subregular embedding are called \emph{minimal} or \emph{simple}.

The basic entities of class $\SS$ are the theories associated to thrice-punctured spheres, or \emph{trinions}. The designations of these theories are conventionally shortened as 
\begin{equation}
\label{eq:trinion_theory_label}
T_n^{\Lambda_1\Lambda_2\Lambda_3} \ceq \TT[\suf(n); \CC_{0, 3}; \{\Lambda_1\,\Lambda_2\,\Lambda_3\}]~.
\end{equation} 
For the special case of all maximal punctures, the convention is to further define $T_n\ceq T_n^{[1^n][1^n][1^n]}$. All of the trinion theories are isolated SCFTs -- they have no marginal couplings. For most of these theories, no Lagrangian description is known. An important class of exceptions are the theories with two maximal punctures and one minimal puncture: $T_n^{[1^n][1^n][n-1,1]}$. These are theories of $n^2$ free hypermultiplets, which in this context are naturally thought of as transforming in the bifundamental representation of $\suf(n)\times\suf(n)$. In the case $n=2$, the minimal and maximal punctures are the same and the theory of four free hypermultiplets (equivalently, eight free half-hypermultiplets) is the $T_2$ theory. In this case the global symmetry associated to the punctures is $\suf(2)\times\suf(2)\times\suf(2)$ which is a subgroup of the full global symmetry $\uspf(8)$.

At the level of two-dimensional topology, an arbitrary surface $\CC_{g,s}$ can be assembled by taking $2g-2+s$ copies of the three-punctured sphere, or ``pairs of pants'', and gluing legs together pairwise $3g-3+s$ times. Each gluing introduces a complex \emph{plumbing parameter} and for a given construction of this type the plumbing parameters form a set of coordinates for a patch of the Teichmuller space of Riemann surfaces of genus $g$ with $s$ punctures. A parallel procedure is used to construct the class $\SS$ theory associated to an arbitrary UV curve using the basic trinion theories. Starting with $2g-2+s$ copies of the trinion theory $T_n$, one glues along maximal punctures by gauging the diagonal subgroup of the $\suf(n)\times\suf(n)$ flavor symmetry associated to the punctures. This introduces an $\suf(n)$ gauge group in the four-dimensional SCFT, and the marginal gauge coupling is related to the plumbing parameter. If one wants, the remaining maximal punctures can then be reduced to sub-maximal punctures using the Higgsing procedure described below.\footnote{In terms of the low energy SCFT, the operations of Higgsing at external punctures and gauging of internal ones commute, so one may equally well think of gluing together trinions some of whose punctures are not maximal. Our presentation here is not meant to convey the full depth of what is possible in class $\SS$.} To a given pants decomposition of a UV curve, one associates a ``weakly coupled'' frame of the corresponding SCFT in which the flavor symmetries of a collection of trinion theories are being weakly gauged. The equivalence of different pants decompositions amounts to $S$-duality. It is only in very special cases that a weakly coupled duality frame of this type will actually be described by a Lagrangian field theory.

By now, quite a few general facts are known about theories of class $\SS$. Here we simply review some relevant ones while providing pointers to the original literature. The list is not meant to be comprehensive in any sense.

\medskip
\subsubsection*{Central charges}

The $a$ and $c$ conformal anomalies have been determined for all of the regular $A$-type theories in \cite{Chacaltana:2010ks,Chacaltana:2012zy}. The answer takes the following form,
\begin{equation}
\label{eq:4dcentralcharges}
c_{4d}=\frac{2 n_v + n_h}{12}~, \qquad a=\frac{5 n_v + n_h}{24}~,
\end{equation}
where 
\begin{align}\label{eq:nv_nh_defs}
\begin{split}
n_v &~=~ \sum_{i=1}^s n_v(\Lambda_i) + (g-1)\left( \tfrac{4}{3}h^\vee \dim \gf + \text{rank } \gf \right)~, \\
n_h &~=~ \sum_{i=1}^s n_h(\Lambda_i) + (g-1)\left( \tfrac{4}{3}h^\vee \dim \gf \right)~,
\end{split}\end{align}
and
\begin{align}\label{eq:nv_nh_defs_2}
\begin{split}
n_v(\Lambda) &~=~ 8\left(\rho\cdot \rho - \rho \cdot \Lambda(t_0)\right) + \tfrac{1}{2}(\text{rank } \gf - \dim \gf_{0})~,\\
n_h(\Lambda) &~=~ 8\left(\rho\cdot \rho - \rho \cdot \Lambda(t_0)\right) + \tfrac{1}{2} \dim \gf_{\frac{1}{2}}~.
\end{split}\end{align}
In these equations, $\rho$ is the Weyl vector of $\suf(n)$ and $h^\vee$ is the dual coxeter number, which is equal to $n$ for $\gf=\suf(n)$. The Freudenthal-de Vries strange formula states that $|\rho|^2 = \frac{h^\vee}{12}\dim\gf$, which is useful in evaluating these expressions. Additionally, the embedded Cartan generator $\Lambda(t_0)$ has been used to define a grading on the Lie-algebra, 
\begin{equation}
\label{eq:lie_algebra_grading}
\gf = \bigoplus_{m\in\frac{1}{2}\Zb} \gf_m~,\qquad \gf_m\ceq\left\{ t \in \gf\ |\ {\rm ad}_{\Lambda(t_0)}t = mt \right\}~.
\end{equation}
This grading will make another appearance in Sec. \ref{sec:reducing}.

The $\suf(n)$ flavor symmetry associated to a full puncture comes with flavor central charge $k_{\suf(n)}=2n$. This is a specialization of the general formula $k_{\rm ADE}=2h^{\vee}$. For a non-maximal puncture, the flavor central charge for a given simple factor $\hhf_{\rm simp}\subseteq\hhf$ is given by \cite{Chacaltana:2012zy},
\begin{equation}
\label{eq:flavor_central_charge_reduced}
k_{\hhf_{\rm simp}}\delta_{AB}= 2 \sum_j \Tr_{\RR_j^{(\text{adj})}}T_A T_B~,
\end{equation}
where $T_A,T_B$ are generators of $\hhf_{\rm simp}$ satisfying the normalization $\Tr_{\hhf_{\rm simp}} T_A T_B = h^\vee_{\hhf_{\rm simp}}\delta_{AB}$ and we have introduced the decomposition of the adjoint representation of $\suf(n)$ into representations of $\hhf_\Lambda \otimes\Lambda(\suf(2))$,
\begin{equation}
\label{eq:adjdecomposition}
\mathrm{adj}_{\gf} = \bigoplus_j \RR_j^{(\mathrm{adj})} \otimes V_j~.
\end{equation}
In cases where there are global symmetries that extend the symmetries associated to punctures, the central charge can be deduced in terms of the embedding index.

\medskip
\subsubsection*{Higgs branch chiral ring and their relations}

Operators in an $\NN=2$ SCFT whose conformal dimension is equal to twice their $\suf(2)_R$ spin ($E=2R$) form a ring called the \emph{Higgs branch chiral ring}. This ring is generally believed to be the ring of holomorphic functions (in a particular complex structure) on the Higgs branch of the moduli space of vacua of the theory. It is expected to be finitely generated, with the generators generally obeying nontrivial algebraic relations. For theories of class $\SS$ the most general such relations have not been worked out explicitly to the best of our knowledge. However, certain cases of the relations can be understood.

For any puncture there is an associated global symmetry $\hhf$, and the conserved currents for that global symmetry will lie in superconformal representations that include \emph{moment map} operators $\mu^A$, $A=1,\ldots,\dim\hhf$ that belong to the Higgs branch chiral ring. Of primary interest to us are the relations that involve solely these moment map operators. Let us specialize to the case where all punctures are maximal, so $\hhf_i=\gf$ for all $i=1,\ldots,s$. There are then chiral ring relations given by
\begin{equation}
\label{eq:moment_map_relation}
\Tr\mu_1^k=\Tr\mu_2^k=\cdots=\Tr\mu_s^k~,\qquad k=1,2\ldots~.
\end{equation}
There are additional Higgs branch chiral ring generators for a general class $\SS$ theory of the form
\begin{equation}
\label{eq:higgs_branch_extra_generators}
Q_{(k)}^{\II^{(k)}_1\cdots\II^{(k)}_s}~,\qquad k=1,\ldots,n-1~,
\end{equation}
of dimension $E_k=2R_k=\frac12 k(n-k)(2g-2+s)$. The multi-indices $\II^{(k)}$ index the $k$-fold antisymmetric tensor representation of $\suf(n)$. There are generally additional chiral ring relations involving these $Q_{(k)}$ operators, some of which mix them with the moment maps \cite{Gadde:2013fma}. The complete form of these extra relations has not been worked out -- a knowledge of such relations would characterize the Higgs branch of that theory as a complex algebraic variety, and such a characterization is presently lacking for all but a small number of special cases. We will not make explicit use of such additional relations in what follows.

\medskip
\subsubsection*{Higgsing and reduction of punctures: generalities}

Theories with non-maximal punctures can be obtained by starting with a theory with maximal punctures and going to a particular locus on the Higgs branch \cite{Benini:2009gi,Tachikawa:2013kta,Chacaltana:2012zy,Maruyoshi:2013hja}. The flavor symmetry associated to a puncture is reflected in the existence of the above-mentioned half-BPS moment map operators, $\mu_A$, that transform in the adjoint representation of the flavor symmetry with corresponding index $A=1,\ldots,n^2-1$. In reducing the flavor symmetry via Higgsing, one aims to give an expectation value to one of the $\mu_i$'s, say $\mu_1$, while keeping $\vev{\mu_{i \neq 1} }=0$. Consistency with Eqn. \eqref{eq:moment_map_relation} then requires that $\vev{\Tr\mu_1^k}=0$ for any $k$, or put differently, $\vev{\mu_1}$ is a nilpotent $\suf(n)$ matrix. Since any nilpotent element can be realized as the image of $t_-\in\suf(2)$ with respect to some embedding $\Lambda:\suf(2)\hookrightarrow\suf(n)$, the relevant loci on the Higgs branch are characterized by such an embedding, where we have
\begin{equation}
\label{eq:invariant_vev}
\vev{\mu_1} = v\, \Lambda(t_-)~.
\end{equation}
The expectation value breaks the $\suf(n)$ flavor symmetry associated with the puncture to $\hhf_\Lambda$, the centralizer of the embedded $\suf(2)$, as well as the $\suf(2)_R$ symmetry (and also conformal symmetry). It will be important in the following that a linear combination of the flavor and $\suf(2)_R$ Cartan generators remains unbroken,\footnote{We suspect not only this Cartan generator, but the full diagonal subalgebra of $\suf(2)_R$ and the embedded $\suf(2)$ is preserved on the sublocus of the Higgs branch in question. It should be possible to prove such a thing using the hyperkahler structure on nilpotent cones described in \cite{Swann}. We thank D. Gaiotto, A. Neitzke, and Y. Tachikawa for helpful conversations on this point.} namely
\begin{equation}
\label{eq:new_R_cartan}
\tilde R \ceq R + J_0~,\qquad J_0 \ceq \Lambda (t_0)~.
\end{equation}
In such a vacuum, the low energy limit of the theory is described by the interacting class $\SS$ SCFT with the same UV curve as the original theory, but with the first puncture replaced by a puncture of type $\Lambda$. Additionally there will be decoupled free fields arising from the Nambu-Goldstone fields associated to the symmetry breaking \cite{Maruyoshi:2013hja,Tachikawa:2013kta}. We identify $\tilde R$ as the Cartan generator of the $\suf(2)_{\tilde R}$ symmetry of the infrared fixed point.

It will prove useful to introduce notation to describe the breaking of $\suf(n)$ symmetry in greater detail. The generators of $\suf(n)$ can be relabeled according to the decomposition of Eqn. \eqref{eq:adjdecomposition},
\begin{equation}
\label{eq:generator_expansion}
T_A ~~\Longrightarrow~~ T_{j,m;\WW(\RR_j)}~,
\end{equation}
where $m=-j, -j+1,\ldots,+j$ is the eigenvalue of the generator with respect to $\Lambda(t_0)$, and $\WW(\RR_j)$ runs over the various weights of the representation $\RR_j$ of $\hhf_{\Lambda}$. Expanding $\mu_1$ around its expectation value, we have
\begin{equation}
\label{eq:mu_expansion}
\mu_1 = v\, \Lambda(t_-) + \sum_j \sum_{m=-j}^{+j} \sum_{\WW(\RR_j)} (\tilde\mu_1)_{j;m,\WW(\RR_j)}T_{j;m,\WW(\RR_j)}~.
\end{equation}
The operators $(\tilde\mu_1)_{j;m,\WW(\RR_j)}$ with $m<j$ become the field operators of the Nambu-Goldstone modes. Their number is given by $\dim_{\Cb}O_{\Lambda(t_-)}^{\gf}$ -- the complex dimension of the nilpotent orbit of $\Lambda(t_-)$. They are ultimately organized into $\hf\dim_{\Cb}O_{\Lambda(t_-)}^{\gf}$ free hypermultiplets.

\medskip
\subsubsection*{Superconformal index}

The superconformal index of an SCFT is an invariant on its conformal manifold. For theories of class $\SS$, this means that the index does not depend on the complex structure moduli of the UV curve. On general grounds, one then expects the class $\SS$ index to be computed by a topological quantum field theory living on the UV curve \cite{Gadde:2009kb}. This expectation is borne out in detail, with a complete characterization of the requisite TQFT achieved in a series of papers \cite{Gadde:2011ik, Gadde:2011uv, Gaiotto:2012xa}. Our interest is in the Schur specialization of the index, which is identical to the graded character of the protected chiral algebra, see Eqn \eqref{charschur}. In \cite{Gadde:2011ik}, the corresponding TQFT was recognized as a $q$-deformed version of two-dimensional Yang-Mills theory in the zero-area limit. Here we will summarize this result and introduce appropriate notation that will be useful in Sec. \ref{subsec:reduced_index}.

For the class $\SS$ theory $\TT[\gf; \CC_{g,s}; \{\Lambda_i\}]$, the Schur index takes the form\footnote{Not every possible choice of   Riemann surface decorated by a choice
of $\{ {\Lambda}_i \}$ at the punctures
corresponds to a physical SCFT.  
An indication that a choice of decorated surface may be unphysical is if the sum  in (\ref{eq:SchurindexUVcurve}) diverges,
which happens when the  flavor  symmetry is ``too small''. There are subtle borderline
cases where the sum diverges, but the theory is perfectly physical.
These cases have to be treated with more care~\cite{Gaiotto:2012uq}.}

\begin{equation}
\label{eq:SchurindexUVcurve}
\II^{\rm Schur}(q; {\bf x} )= \sum_{\Rf} C_\Rf(q)^{2g-2+s}\prod_{i=1}^s \psi_{\Rf}^{\Lambda_i}({\bf x}_{\Lambda_i} ;q)~.
\end{equation}
The sum runs over the set of finite-dimensional irreducible representations $\Rf$ of the Lie algebra $\gf$. Each puncture contributes a ``wavefunction'' $\psi_{\Rf}^{\Lambda_i}({\bf x}_{\Lambda_i};q)$, while the Euler character of the UV curve determines the power of the ``structure constants'' $C_\Rf(q)$ that appear. Each wavefunction depends on fugacities ${\bf x}_\Lambda$ conjugate to the Cartan generators of the flavor group $\hhf_\Lambda$ associated to the puncture in question. Note that by definition, the structure constants are related to wave functions for the principal embedding, which corresponds to having no puncture at all, \ie,
\begin{equation}
\label{eq:wave_function_to_structure_const}
C_\Rf(q)^{-1} \equiv \psi^{\rho}_\Rf (q)~,
\end{equation}
where $\rho$ denotes the principal embedding.\footnote{The discussion so far applies to a general simply-laced Lie algebra $\gf$. Recall that when $\gf = \suf(n)$, the principal embedding corresponds to the partition $[n^1]$.}

To write down the general wavefunction we need to discuss some group theory preliminaries. Under the embedding $\Lambda:\suf(2)\hookrightarrow\gf$, a generic representation $\Rf$ of $\gf$ decomposes into $\hhf_\Lambda \otimes\Lambda(\suf(2))$ representations,
\begin{equation}
\label{eq:generaldecomposition}
\Rf = \bigoplus_j \RR_j^{(\Rf)} \otimes V_j~,
\end{equation}
where $\RR_j^{(\Rf)}$ is some (generically reducible) representation of $\hhf_\Lambda$. We define the fugacity assignment $\mathrm{fug}_{\Lambda}(\mathbf{x}_{\Lambda}; q)$ as the solution (for ${\bf x}$) of the following character decomposition equation,\footnote{For $\gf = \suf(n)$ the solution is unique up to the action of the Weyl group.}
\begin{equation}
\label{eq:define_fugs}
\chi_{\ff}^{\gf}(\mathbf{x}) = \sum_j \ \chi_{\RR_j^{( \ff )}}^{\hhf_{{\Lambda}}}(\mathbf{\mathbf{x}}_{{\Lambda}}) \, \chi_{V_j}^{\suf(2)}(q^{\frac{1}{2}}\, ,q^{-\frac{1}{2}})\; ,
\end{equation}
where $\chi_{\ff}^{\gf}(\mathbf{x})$ is the character of $\gf$ in the fundamental representation (denoted by $\ff$), and the right hand side is determined by the decomposition of Eqn. \eqref{eq:generaldecomposition} with $\Rf \equiv \ff$. Note that ${\bf x} = \mathrm{fug}_{\Lambda}(\mathbf{x}_{\Lambda};q)$ also solves the more general character equation
\begin{equation}
\label{eq:general_fugacity_relation}
\goodchi_\Rf^{\gf}(\mathbf{x}) = \sum_j\,\goodchi_{\RR_j^{(\Rf)}}^{\hhf_{\Lambda}}(\mathbf{\mathbf{x}}_{\Lambda})\,\goodchi_{V_j}^{\suf(2)}(q^{\frac{1}{2}},q^{-\frac{1}{2}})~,
\end{equation}
for any other representation $\Rf$. A couple of simple examples help to clarify these definitions. Taking $\gf = \suf(2)$ and $\Lambda: \suf(2) \hookrightarrow \suf(2)$ the principal embedding -- in this case just the identity map --
the centralizer is trivial and Eqn. \eqref{eq:define_fugs} becomes
\begin{equation}
\label{eq:su2_fugacity_relation}
a + a^{-1} = q^{\hf} + q^{-\hf}~, 
\end{equation}
which has the two solutions $a=q^{\hf}$ and $a=q^{-\hf}$, which are related to each other by the action of the Weyl group $a \leftrightarrow a^{-1}$. A more complicated example is $\gf = \suf(3)$ and $\Lambda$ the subregular embedding, which corresponds to the partition $[2^1, 1^1]$. The centralizer is $\hhf_\Lambda = \uf(1)$. Given $\suf(3)$ fugacities $(a_1,a_2,a_3)$ with $a_1 a_2 a_3 = 1$, we denote the $\uf(1)$ fugacity by $b$ and then Eqn. \eqref{eq:define_fugs} takes the form
\begin{equation}
\label{eq:su3_fugacity_relation}
a_1 + a_2 + a_3 = b^{-2} + b \, ( q^{\hf} + q^{-\hf} )~.
\end{equation}
Up to the action of the Weyl group, which permutes the $a_i$, the unique solution is given by $(a_1,a_2,a_3)=(q^{\hf} b,q^{-\hf} b,b^{-2})$. 

The wavefunction for a general choice of embedding and representation now takes the following form,
\begin{equation}
\label{eq:schur_wave_functions}
\psi_{\Rf}^{\Lambda}({\bf x}_\Lambda ;q) \ceq K_{\Lambda}({\bf x}_{\Lambda}; q)\,\goodchi_{\Rf}^\gf ( \mathrm{fug}_{\Lambda}(\mathbf{x}_{\Lambda}; q) )~.
\end{equation}
The \emph{$K$-factors} admit a compact expression as a plethystic exponential \cite{Mekareeya:2012tn},
\begin{equation}
\label{eq:K-factor}
K_{\Lambda}(\mathbf{x}_{\Lambda};q) \ceq \PE \left[ \sum_j \frac{q^{j+1}}{1-q} \goodchi^{\hhf_\Lambda} _{\RR_j^{(\mathrm{adj})}}(\mathbf{x}_{\Lambda}) \right]~,
\end{equation}
where the summation is over the terms appearing in the decomposition of Eqn. \eqref{eq:generaldecomposition} applied to the adjoint representation,
\begin{equation}
\label{eq:adjdecomposition2}
\mathrm{adj}_{\gf} = \bigoplus_j \RR_j^{(\mathrm{adj})} \otimes V_j~.
\end{equation}
Note that $\RR_0^{(\mathrm{adj})} = \mathrm{adj}_{\hhf_\Lambda }\oplus \text{singlets}$. For the maximal puncture, corresponding to the trivial embedding $\Lambda_{\rm max} \equiv 0$, the wavefunction reads
\begin{equation} 
\label{eq:psi_max}
\psi_{\Rf}^{ \Lambda_{\rm max} }({\bf x} ;q) =K_{\rm max} ({\bf x}; q) \, \chi_\Rf^\gf ( {\bf x} ) \,, \quad K_{\rm max} ({\bf x}; q) \ceq \PE \left[ \frac{q}{1-q} \chi_{\rm adj} ^\gf ( {\bf x} ) \right]~.
\end{equation}
At the other extreme, for the principal embedding $\Lambda = \rho$, the decomposition of Eqn. \eqref{eq:adjdecomposition2} reads
\begin{equation}
\label{eq:adjoint_decomposition_principal}
\mathrm{adj}_{\gf} = \bigoplus_{i=1}^{{\rm rank} \, \gf} V_{d_i -1}~,
\end{equation}
where $\{ d_i \}$ are the degrees of invariants of $\gf$, so in particular $d_i = i +1$ for $\suf(n)$. We then find
\begin{equation}
\label{eq:psi_rho}
\psi^\rho_\Rf (q) = \PE \left[ \sum_{i}^{{\rm rank} \, \gf} \frac{q^{d_i} }{1-q} \right] \chi^\gf_\Rf ( \mathrm{fug}_{\rho}(q) )~.
\end{equation}
For $\gf = \suf(n)$, the fugacity assignment associated to the principal embedding takes a particularly simple form,
\begin{equation}
\label{eq:principle_fugacities}
\mathrm{fug}_{\rho}(q) = (q^{\frac{n-1}{2} }, q^{\frac{n-3}{2}}, \dots q^{-\frac{n-1}{2} })~.
\end{equation}
Together, Eqns. \eqref{eq:wave_function_to_structure_const}, \eqref{eq:psi_rho}, and \eqref{eq:principle_fugacities} provide an explicit expression for the structure constants $C_\Rf (q)$.

Finally, let us recall the procedure for gluing two theories along maximal punctures at the level of the index. Consider two theories $\TT_1 = \TT[\gf; \CC_{g_1,s_1}; \{\Lambda_i\}]$ and $\TT_2 = \TT[\gf; \CC_{g_2,s_2}; \{\Lambda_i\}]$, each of which are assumed to have at least one maximal puncture. We denote their Schur indices as $\II^{\rm Schur}_{\TT_1}(q;{\bf a},\ldots)$ and $\II^{\rm Schur}_{\TT_2}(q;{\bf b},\ldots)$, where we have singled out the dependence on flavor fugacities ${\bf a}$ and ${\bf b}$ of the two maximal punctures that are going to be glued. As usual, gluing corresponds to gauging the diagonal subgroup of the flavor symmetry $G \times G$ associated to the two maximal punctures. The index of the glued theory is then given by
\begin{equation}
\label{eq:indexgluing}
\oint [d{\bf a}] \Delta({\bf a})\ I_V(q; {\bf a})\ \II^{\rm Schur}_{\TT_1}(q;{\bf a},\ldots)\ \II^{\rm Schur}_{\TT_2}(q;{\bf a}^{-1},\ldots)~,
\end{equation}
where $[d{\bf a}] \ceq \prod_{j=1}^{r} \frac{da_j}{2\pi i a_j}$, $\Delta({\bf a})$ is the Haar measure, and $I_V(q; {\bf a})$ is the index of an $\NN=2$ vector multiplet in the Schur limit,
\begin{align}
\label{eq:vector_multiplet_index}
I_V(q; {\bf a})=\PE\left[\frac{-2q}{1-q}\goodchi_{\text{adj}}({\bf a}) \right] = K_{\text{max}}(\mathbf{a};q)^{-2}~.
\end{align}
If we write the indices of $\TT_1$ and $\TT_2$ in the form dictated by Eqn. \eqref{eq:SchurindexUVcurve}, then the contour integral is rendered trivial because the $K$-factors in the wave functions that are being glued cancel against the index of the vector multiplet and the characters $\goodchi_{\Rf}^\gf$ are orthonormal with respect to the Haar measure. The result is that we obtain an expression that takes the form of Eqn. \eqref{eq:SchurindexUVcurve}, but with $g = g_1 + g_2$ and $s = s_1 + s_2 -2$.

\medskip
\subsubsection*{Higgsing and reduction of punctures: superconformal index}

We will now argue that the expression given in Eqn.~\eqref{eq:schur_wave_functions} for the general wavefunction of type $\Lambda$ is dictated by the Higgsing procedure if one takes for granted the formula given in Eqn.~\eqref{eq:psi_max} for the maximal wavefunction. In fact, the argument we are about to present should be applicable outside of the narrow context under consideration here, so for some parts of the argument we will use a fairly general language.

We are interested in the relationship between the Schur limit of the superconformal index of an $\NN=2$ SCFT and that of the low energy theory at a point on the Higgs branch. It is a familiar feature of supersymmetric indices that in some sense the only difference between the indices of UV and IR fixed points is a possible redefinition of fugacities. In particular, if a renormalization group flow is triggered by a vev that breaks some global symmetry, then the fugacities dual to the broken generators must be set to zero. Furthermore, if the index is to be interpreted as a \emph{superconformal} index of the IR fixed point, then the appropriate $R$-symmetries that appear in the superconformal algebra of that fixed point must be identified and the fugacities redefined appropriately. 

There are two related obstacles to applying this simple reasoning in many cases. One is the appearance of accidental symmetries at the IR fixed point. Fugacities dual to the generators of accidental symmetries cannot be introduced in the UV description of the index, and so in particular if the superconformal $R$-symmetry in the IR mixes with accidental symmetries, then the superconformal index is inaccessible. The second obstacle is the possible presence of decoupled free fields in addition to the degrees of freedom of interest at low energies. These two issues are related because whenever decoupled free fields emerge at low energies, there will necessarily be an accidental global symmetry that acts just on those fields, and this symmetry will generally contribute to the superconformal $R$-symmetry. 

In nice cases it is possible to overcome these obstacles and write the superconformal index of the IR theory in terms of that of the UV fixed point in a fairly simple way. Sufficient conditions for us to be able to do this are:
\begin{enumerate}
\item[$\bullet$] The only accidental symmetries at the IR fixed point are those associated to the decoupled Nambu-Goldstone bosons of spontaneous symmetry breaking.
\item[$\bullet$] The Cartan generator of the $\suf(2)_R$ symmetry of the IR fixed point, when restricted to act on operators in the interacting sector, can be identified and written as a linear combination of UV symmetries.
\item[$\bullet$] The Higgs branch chiral ring operators that become the field operators for Nambu-Goldstone bosons in the infrared are identifiable, and their quantum numbers with respect to UV symmetries known.
\end{enumerate}
When these conditions are met, the prescription for computing the index of the IR fixed point is simple, and amounts to subtracting out the contributions of the decoupled free fields to the index,
\begin{equation}
\label{eq:schematic_index_reduction}
\II_{\rm IR}(q;{\bf x}_{\rm IR})=\lim_{{\bf x}_{\rm UV} \to {\bf x}_{\rm IR}}~\frac{\II_{\rm UV}(q;{\bf x}_{\rm UV})}{\II_{\rm NGB}(q;{\bf x}_{\rm UV})}~.
\end{equation}
Here ${\bf x}_{\rm UV}$ are the fugacities dual to the UV global symmetries, while ${\bf x}_{\rm IR}$ are those dual to the IR global symmetries. The two sets of fugacities are related to one another by a specialization. The denominator on the right hand side is the index of $\frac12N_{\rm NGB}$ free hypermultiplets, where $N_{\rm NGB}$ is the number of complex Nambu-Goldstone bosons at the chosen locus of the Higgs branch. The only subtlety is that the contributions of these free hypermultiplets are graded according to the charges of the Higgs branch chiral ring operator that becomes the field operator for the Nambu-Goldstone boson in the IR, so we have
\begin{equation}
\label{eq:NGB_index}
\II_{\rm NGB}({\bf x}_{\rm UV};q)\ceq \PE\left(\sum_{\OO_i}\frac{q^{R_{\OO_i}}{\bf x}^{f_{\OO_i}}}{1-q}\right)~.
\end{equation}
The reason that Eqn.~\eqref{eq:schematic_index_reduction} involves a limit is that the index will have a pole at the specialized values of the fugacities. It is easy to see that this will be the case because operators that acquire expectation values in the Higgs branch vacuum of interest will always be uncharged under all of the fugacities appearing in the specialized index. This invariably leads to a divergence in the index.

Now let us return to the specific case of interest: the reduction of punctures in class $\SS$ theories. All of the conditions listed above are met. The only accidental symmetries are those that act only on the decoupled Nambu-Goldstone bosons arising from the spontaneous breaking of global and scale symmetries. The Cartan generator of the low energy $\suf(2)_R$ (when restricted to act in the interacting sector) was identified in Eqn.~\eqref{eq:new_R_cartan}. Finally, we know precisely which operators in the UV theory will become the field operators for the Nambu-Goldstone bosons (\cf\ Eqn.~\eqref{eq:mu_expansion}). Consequently we know how these decoupling operators are acted upon by the UV symmetries. 

Describing the index of the (interacting part) of the IR theory resulting from the Higgsing associated to an embedding $\Lambda$ in terms of the theory with maximal punctures is now a simple exercise. The relevant specialization is accomplished by redefining the $\suf(2)_R$ Cartan in the index according to Eqn.~\eqref{eq:new_R_cartan}, which leads to the replacement rule ${\bf x}\mapsto\mathrm{fug}_{\Lambda}(\mathbf{x}_{\Lambda};q)$. The character $\chi_\Rf^\gf$ is regular under this specialization. To check that we obtain the expected wavefunction for the reduced puncture given in Eqns. \eqref{eq:schur_wave_functions} and \eqref{eq:K-factor} it only remains to verify that the $K$-factors behave in the expected manner. The fugacity replacement in the $K$-factor of the maximal puncture leads to the following rewriting,
\begin{equation}
\label{eq:K_factor_fugacity_replacement_1}
K_{\mathrm{max}}(\mathbf{a};q) = \PE \left[ \frac{q}{1-q} \chi_{\mathrm{adj}_{\gf}}(\mathbf{a}) \right] ~\longrightarrow~ \PE \left[ \frac{q}{1-q} \sum_j \chi_{\RR_j^{(\mathrm{adj})}}^{\hhf_{\Lambda}}(\mathbf{a}_{\Lambda}) \ \chi_{V_j}^{\suf(2)}(q^{\frac{1}{2}},q^{-\frac{1}{2}})\right]~,
\end{equation}
and upon expanding out the character $\chi_{V_j}^{\suf(2)}(q^{\frac{1}{2}},q^{-\frac{1}{2}}) = \sum_{m=-j}^{+j} q^m$, we find the expression
\begin{equation}
\label{eq:K_factor_fugacity_replacement_2}
K_{\mathrm{max}}(\mathbf{a};q) \rightarrow \PE\left[ \sum_j \frac{q^{j+1}}{1-q} \chi_{\RR_j^{(\mathrm{adj})}}^{\hhf_{\Lambda}}(\mathbf{a}_{\Lambda}) \right] \PE\left[ \frac{q}{1-q} \sum_j \goodchi_{\RR_j^{(\mathrm{adj})}}^{\hhf_{\Lambda}}(\mathbf{a}_{\Lambda}) \ \sum_{m=-j}^{+j-1} q^m \right]~.
\end{equation}
The first factor here reproduces the $K$-factor of the reduced flavor puncture given in Eqn.~\eqref{eq:K-factor}. The second factor is strictly divergent because there are constant terms in the plethystic exponent. However it is precisely this second factor that is cancelled by the denominator in Eqn.~\eqref{eq:schematic_index_reduction}. We have been a little careless in this treatment by making a formal fugacity replacement and then cancelling an infinite factor. A more rigorous treatment proceeds via the limiting procedure described above, and produces the same result.

%% file: sections/Section_3/S3.tex

\section{Chiral algebras of class \texorpdfstring{$\SS$}{S}}
\label{sec:TQFT}

The organization of class $\SS$ theories in terms of two-dimensional conformal geometry has important implications for observables of these theories. In particular, any observable that is independent of exactly marginal couplings should give rise to a (generalized) topological quantum field theory upon identifying a given theory with its UV curve. As reviewed above, this insight was originally exploited in the study of the superconformal index \cite{Gaiotto:2012xa,Gadde:2009kb,Gadde:2010te,Gadde:2011ik}. Subsequently the strategy was formalized and extended to the case of the (maximal) Higgs branch in \cite{Moore:2011ee}. There it was emphasized that this approach has the additional benefit of providing a way to study the superstructure of class $\SS$ with some degree of mathematical rigor, evading problems associated with the definition of interacting quantum field theories. The basic idea is summarized in the following commutative diagram. 
\begin{equation}
\label{eq:category_diagram}
\begin{tikzcd}[column sep=small]
& \TT\,[\,\CC_{g,s};\gf;\{\Lambda_i\}\,] \arrow{dr}{\Pb} 	& \\
\left\{\CC_{g,s};\{\Lambda_i\}\right\} \arrow{ur}{\TT_{\gf}}\arrow{rr}{\Pb\;\circ\;\TT_{\gf}}	& 	& {\Pb}\,[\,\TT\,[\,\CC_{g,s};\gf;\{\Lambda_i\}\,]\,]
\end{tikzcd}
\end{equation}
For some protected observable $\Pb$ that can be defined for an $\NN=2$ SCFT, one defines the composition $\Pb\,\circ\,\TT_{\gf}$ that associates the observable in question directly to a UV curve. When the observable is something relatively simple -- like the holomorphic symplectic manifolds studied in \cite{Moore:2011ee} -- one should be able to define this composition in a rigorous fashion without having to define the more complicated $\TT_{\gf}$-functor at all.

In the present work we take as our ``observable'' the protected chiral algebra, which is indeed independent of marginal couplings. The composition $\goodchi\,\circ\,\TT_{\gf}$ has as its image the chiral algebras of class $\SS$, which are labelled by Riemann surfaces whose punctures are decorated by embeddings $\Lambda:\slf(2)\hookrightarrow\slf(n)$. This class of chiral algebras has the form of a generalized topological quantum field theory.

The aim of this section is to develop a basic picture of the structure of this TQFT and to characterize it to the extent possible. In the first subsection, we make some general statements about the implications of the TQFT structure from a physicist's point of view. We also make a modest attempt to formalize the predicted structure in a language closer to that employed in the mathematics literature. In the second subsection, we discuss the basic building blocks of the TQFT for the $\suf(2)$ and $\suf(3)$ cases. We also make a conjecture about the general case. In the last subsection we make some comments about the constraints of associativity and possible approaches to solving for the class $\SS$ chiral algebras at various levels of generality.

\bigskip
\input{./sections/Section_3/S3_1}
\bigskip
\input{./sections/Section_3/S3_2}
\bigskip
\input{./sections/Section_3/S3_3}
\bigskip\bigskip

%% file: sections/Section_3/S3_1.tex

\subsection{A TQFT valued in chiral algebras}
\label{subsec:generalized_tqft}

\begin{figure}[t!]
\centering
\raisebox{-0.5\height}{\subfloat[Cap]{
\includegraphicsif{scale=.45}{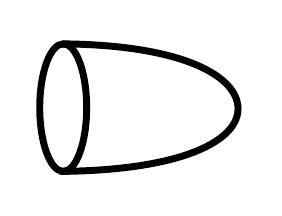}
\label{fig:subfig1}}}
\qquad\qquad
\raisebox{-0.5\height}{\subfloat[Cylinder]{
\includegraphicsif{scale=.45}{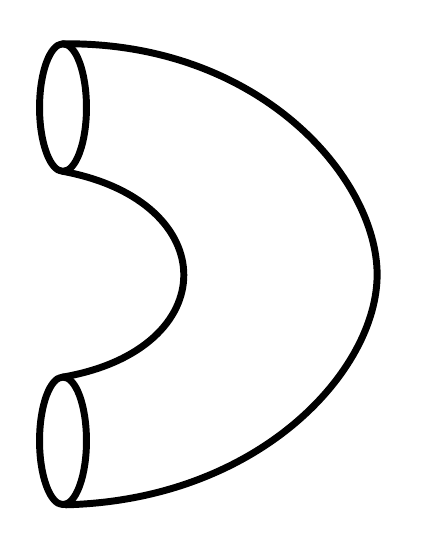}
\label{fig:subfig2}}}
\qquad\qquad
\raisebox{-0.5\height}{\subfloat[Trinion]{
\includegraphicsif{scale=.45}{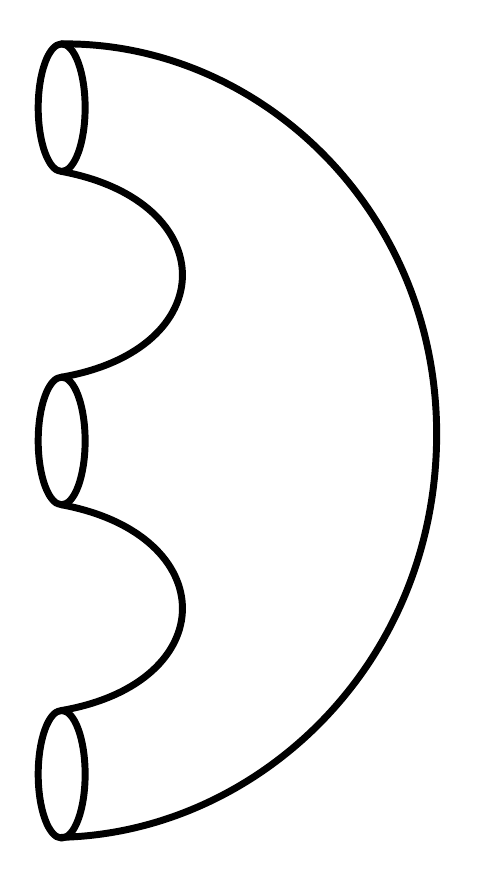}
\label{fig:subfig3}}}
\caption{Elementary building blocks of a two-dimensional TQFT.}
\label{fig:globfig}
\end{figure}

\begin{figure}[t!]
\centering
\includegraphicsif{scale=.4}{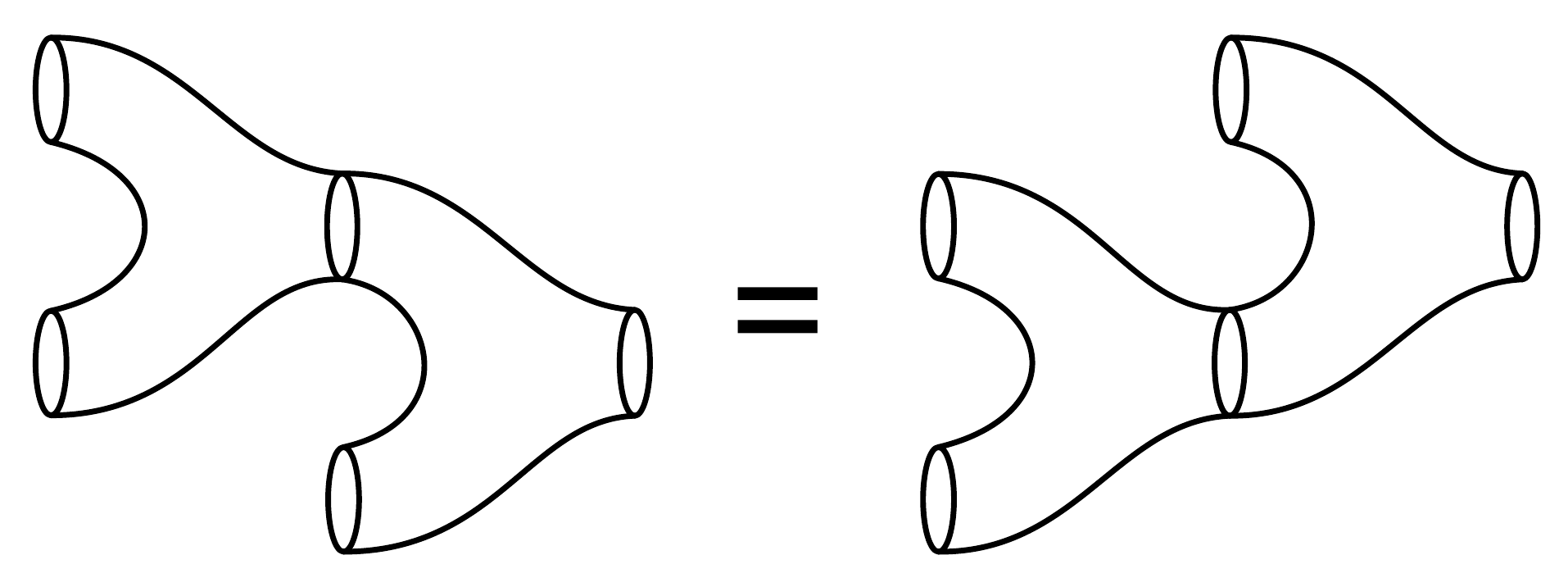}
\caption{Associativity of composition of $T_n$ chiral algebras.}
\label{fig:tqft_topological_invariance}
\end{figure}

In a physicist's language, the type of generalized TQFT we have in mind is specified by associating a chiral algebra with each of a small number of (topological) Riemann surfaces with boundary, namely the genus zero surface with one, two, or three boundary circles (see Fig.\,\ref{fig:globfig}).\footnote{Strictly speaking, this is a redundant amount of information because composing a trinion with a cap produces a cylinder. In anticipating the fact that the chiral algebra for the cap is somewhat difficult to understand, we are considering them independently.} We must further give a meaning to the procedure of gluing Riemann surfaces along common boundaries at the level of the chiral algebra. Self-consistency of the generalized TQFT then requires that the resulting structure be associative in that it reflects the equivalence of Fig.\,\ref{fig:tqft_topological_invariance}.

\begin{figure}[ht!]
\centering
\raisebox{-0.5\height}{\subfloat[Decorated cap]{~~~
\includegraphicsif{scale=.45}{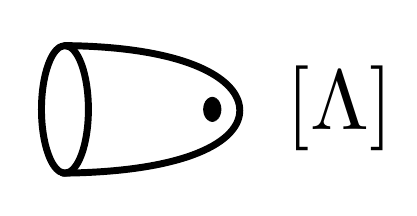}~~~
\label{fig:dec_cap}}}
\qquad\qquad
\raisebox{-0.5\height}{\subfloat[Decorated cylinder]{~~~
\includegraphicsif{scale=.45}{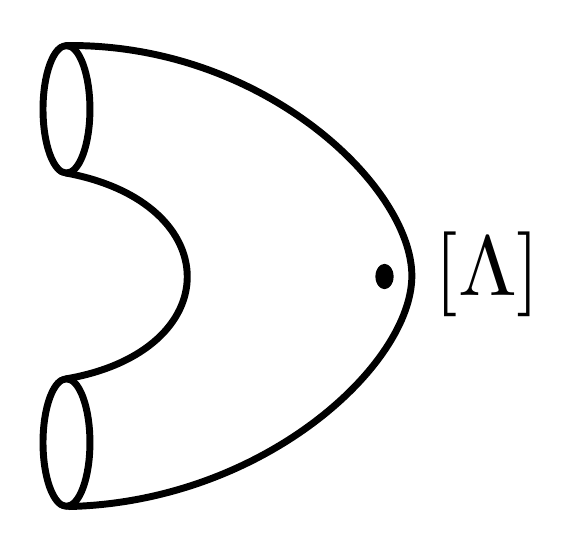}~~~
\label{fig:dec_cyl}}}
\caption{Additional building blocks of a class $\SS$ TQFT.}
\label{fig:decorated_fixtures}
\end{figure}

The full class $\SS$ structure is more complicated than can be captured by this basic version of a generalized TQFT due to the possibility of choosing nontrivial embeddings to decorate the punctures. We can partially introduce this additional structure by allowing the decorated objects illustrated in Fig.\,\ref{fig:decorated_fixtures}. For our purposes these will be thought of as decorated versions of the cap and cylinder. In choosing this interpretation, we are ignoring the fact that in class $\SS$ one can in certain cases glue along a non-maximal puncture. This fact plays an important role already in the basic example of Argyres-Seiberg duality interpreted as a class $\SS$ duality. These decorated fixtures will also be required to satisfy certain obvious associativity conditions.

We can define the gluing operation for chiral algebras associated to these elementary surfaces by knowing a few of the general features of these chiral algebras. Namely, it is guaranteed that the chiral algebras associated to these surfaces will include affine current subalgebras associated to their boundary circles. Indeed, to every full puncture in a class $\SS$ theory of type $\suf(n)$ there is associated an $\suf(n)$ global symmetry with central charge $k_{4d}=2n$. Correspondingly, the associated chiral algebra will have an $\widehat{\suf(n)}_{-n}$ affine current subalgebra. Knowing this, the composition rule for chiral algebras follows more or less immediately from the rules for gauging reviewed in Sec. \ref{subsec:chiral_review}. Two legs with maximal punctures can be glued by introducing $(b,c)$ ghosts transforming in the adjoint of $\suf(n)$ and passing to the cohomology of a BRST operator formed with the diagonal combination of the two affine current algebras.

Given the fairly involved nature of this gluing operation, associativity for the TQFT as illustrated in Fig. \ref{fig:tqft_topological_invariance} is an extremely nontrivial property. Indeed, it is the reflection of generalized $S$-duality of the four-dimensional SCFTs of class $\SS$ at the level of chiral algebras. It is not \emph{a priori} obvious that it should even be possible to find chiral algebras for which this gluing will satisfy the associativity conditions, and the existence of such a family of chiral algebras is an interesting prediction that follows from the existence of the class $\SS$ landscape.

\bigskip\medskip

For the sake of the mathematically inclined reader, we can now formalize this structure a bit more to bring the definition of this generalized TQFT into line with the standard mathematical description. This type of a formalization has also been presented by Yuji Tachikawa \cite{Tachikawa:2013un} in a lecture shortly following the completion of \cite{Beem:2013sza}. The structure in question is a strict symmetric monoidal functor between two symmetric monoidal categories that we outline now.

\subsubsection*{The source category}

\begin{figure}[t!]
\centering
\includegraphicsif{scale=.47}{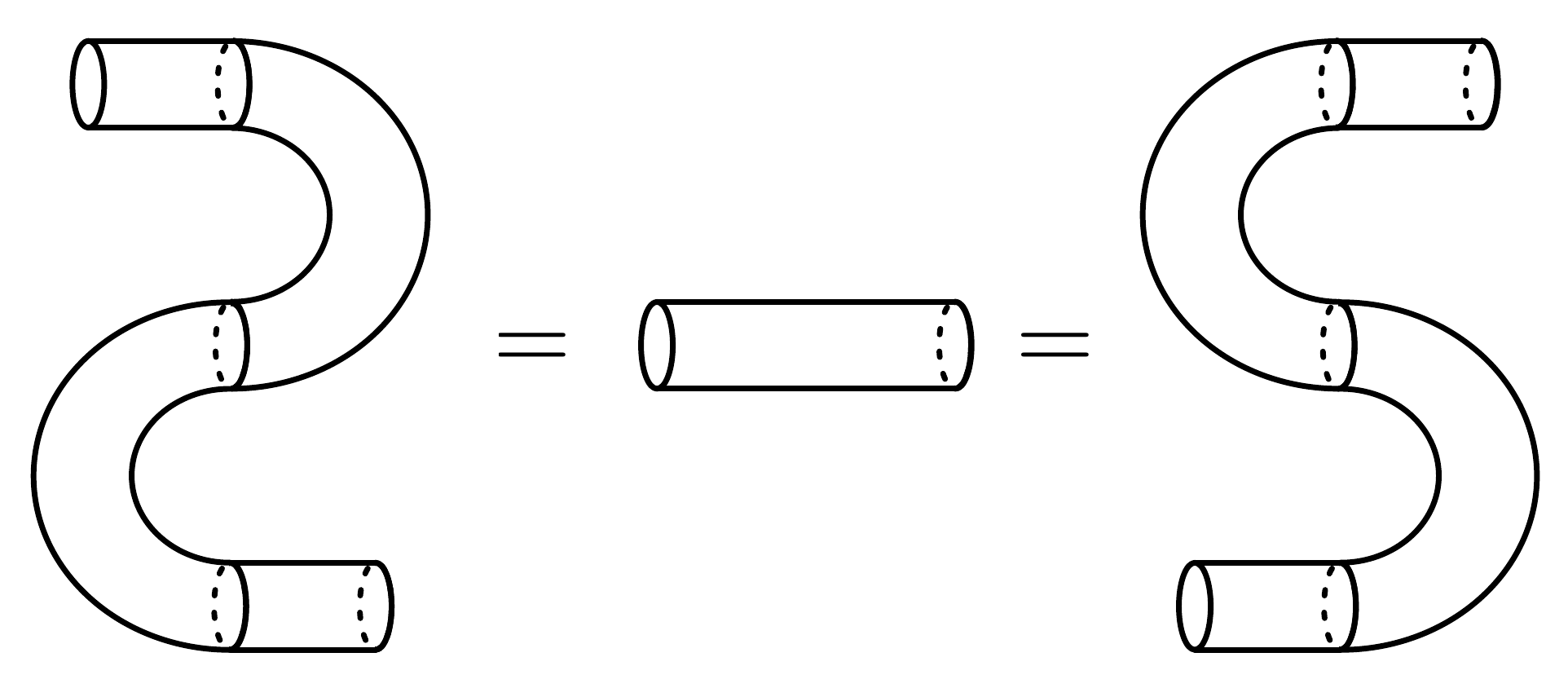}
\caption{Duality and the S-diagram.}
\label{fig:S_cobordism}
\end{figure}

The source category is a decorated version of the usual bordism category $\Bo_2$. It has previously appeared in \cite{Moore:2011ee} for the same purpose. In fact, there is a separate such category for each simply laced Lie algebra $\gf$ (which for us will always be $\suf(n)$ for some $n$), and we will denote it as $\Bo_2^{(\gf)}$. The category has the following structure:
\begin{itemize}
\item[$\bullet$] The objects of $\Bo_2^{(\gf)}$ are the same as for the $\Bo_2$ -- they are closed oriented one-manifolds (\ie, disjoint unions of circles).
\item[$\bullet$] A morphism in $\Bo_2$ between two objects $B_1$ and $B_2$ is a two-dimensional oriented manifold $B$ that is a bordism from $B_1$ to $B_2$. A morphism in $\Bo_2^{(\gf)}$ is a morphism from $\Bo_2$ that is additionally decorated by an arbitrary finite number of marked points $\{s_i\}$, each of which is labelled by an embedding $\Lambda_i:\suf(2)\hookrightarrow\gf$.
\item[$\bullet$] Composition is the usual composition of bordisms by gluing along boundaries.
\item[$\bullet$] The symmetric monoidal structure is given by taking disjoint unions.
\item[$\bullet$] This category has duality, which follows from the existence of left- and right-facing cylinders for which the S-bordisms of Fig. \ref{fig:S_cobordism} are equivalent to the identity.
\end{itemize}

\subsubsection*{The target category}	

The target category is a certain category of chiral algebras that we will call $\Cb\Ab_{\gf}$. We define it as follows
\begin{itemize}
\item[$\bullet$] The objects are finite tensor powers of the $\gf$ affine current algebra at the critical level. This includes the case where the power is zero, which corresponds to the trivial chiral algebra for which only the identity operator is present.
$$
{\bf Obj}(\Cb\Ab_{\gf})=\prod_{n=0}^{\infty}\left(\otimes^n \hat{\gf}_{-h^{\vee}}\right)~.
$$ 
\item[$\bullet$] Given two objects $\of_1=\otimes^{n_1}\hat\gf_{-h^\vee}$ and $\of_2=\otimes^{n_2}\hat\gf_{-h^\vee}$, the morphisms ${\bf Hom}(\of_1,\of_2)$ are \emph{conformal} chiral algebras containing $\of_1\otimes\of_2$ as a subalgebra. Note that this precludes a morphism which is just equal to several copies of the critical affine Lie algebra, since there would be no stress tensor.
\item[$\bullet$] For $\goodchi_1\in{\bf Hom}(\of_1,\of_2)$ and $\goodchi_2\in{\bf Hom}(\of_2,\of_3)$, the composition $\goodchi_2\circ\goodchi_1\in{\bf Hom}(\of_1,\of_3)$ is obtained by the BRST construction of Sec. \ref{subsec:chiral_review}. That is, one first introduces $\dim\gf$ copies of the $(1,0)$ ghost system and then passes to the cohomology of the nilpotent BRST operator relative to the $b$-ghost zero modes,
$$
\goodchi_2\circ\goodchi_1=H_{\rm BRST}^*(\psi\in\goodchi_1\otimes\goodchi_2\otimes\goodchi_{(b,c)_{\gf}}\,\restr{}{}\,b_0\psi=0)~.
$$
It is straightforward to show that this composition rule is associative.
\item[$\bullet$] The symmetric monoidal structure is given by taking tensor products of chiral algebras.
\item[$\bullet$] The duality structure in this category is somewhat complicated and involves the precise form of the chiral algebra that is the image of the cylinder in ${\bf Hom}(S^1\sqcup S^1,\varnothing)$. We delay discussion of this chiral algebra until Sec. \ref{subsec:cylinder}. For now, we define a weaker version of duality -- namely that there exists a certain action of $(\Zb_2)^r$ on the collection $\coprod_{p+q=r}{\bf Hom}((\hat\gf_{-h^\vee})^p,(\hat\gf_{-h^\vee})^q)$ that corresponds to the action of changing external legs of a bordism from ingoing to outgoing and vice versa. This action is simple to describe. Note that a chiral algebra belonging to the above collection of ${\bf Hom}$ spaces can be described as $r$ copies of the critical $\hat\gf$ current algebra along with (possibly infinitely many) additional generators transforming as modules. The primary states of each such module with respect to the affine current algebras will transform in some representation $\Rf_1\otimes\cdots\otimes\Rf_r$ of the global $\suf(n)^r$ symmetry. The duality action associated to flipping the $i$'th leg of a bordism then acts as $\Rf_i\mapsto\Rf_i^*$, and this action lifts to the full chiral algebra in the obvious way.
\end{itemize}

\subsubsection*{The functor}

A chiral algebra-valued TQFT of type $\gf$ can now be defined as a functor that realizes the horizontal arrow in diagram \eqref{eq:category_diagram},
$$
\goodchi\circ\TT:\Bo_2^{(\gf)}\rightarrow\mathbb{CA}_{\gf}~.
$$
The image of such a functor in $\Cb\Ab_{\gf}$ defines a very interesting set of chiral algebras. The necessary ingredients to define this functor are those outlined in the previous discussion. Namely, we need to specify the images of the basic topological Riemann surfaces in Fig. \ref{fig:globfig} and the decorated versions in Fig. \ref{fig:decorated_fixtures}. In order for this to be a functor, the composition of Riemann surfaces with three boundary components must be associative in the sense of Fig. \ref{fig:tqft_topological_invariance}. A similar associativity condition is obtained by replacing any of the boundary components in Fig. \ref{fig:tqft_topological_invariance} with a general decoration $\Lambda$.

The problem of including decorations can be self-consistently ignored in order to focus on the subproblem in which the source category is the more traditional bordism category $\Bo_2$. In the remainder of this section we will address the problem of understanding this more basic version of the TQFT, sometimes with the addition of simple punctures, but not the most general case. The addition of arbitrary decorations will be discussed in Sec. \ref{sec:reducing}.

%% file: sections/Section_3/S3_2.tex

\subsection{Lagrangian class \texorpdfstring{$\SS$}{S} building blocks}
\label{subsec:lagrangian_building_blocks}

The basic building blocks of class $\SS$ SCFTs are the theories associated to spheres with three punctures. Of these, the simplest case is the theory with two maximal punctures and one minimal puncture. This is the only regular configuration which gives rise to a Lagrangian theory for arbitrary choice of ADE algebra. For the $\suf(n)$ theory, it is the theory of $n^2$ free hypermultiplets, so the associated chiral algebra is the theory of $n^2$ symplectic boson pairs \cite{Beem:2013sza}. Though this chiral algebra has a full $\uspf(2n^2)$ symmetry, it is natural to use a basis which makes manifest the $\suf(n)_1\times\suf(n)_2\times\uf(1)$ symmetry associated to the punctures,
\begin{equation}
\label{eq:free_hyper_OPE}
q^i_a(z)\tilde q_j^b(0)\sim\frac{\delta^i_j\delta_a^b}{z}~,\qquad i=1,\ldots,n~\quad a=1,\ldots,n~.
\end{equation}
The currents generating the puncture symmetries are the chiral algebra relatives of the moment map operators in the free hypermultiplet theory,
\begin{equation}
\label{eq:free_hyper_currents}
\begin{split}
\makebox[14ex][l]{$(J_{\suf(n)_{1}})^i_j(z)$}&\ceq~ -(q^i_a\tilde q_j^a)(z)+\tfrac{1}{n}\delta_i^j(q^k_a\tilde q_k^a)(z)~,\\
\makebox[14ex][l]{($J_{\suf(n)_{2}})_a^b(z)$}&\ceq~ -(q^i_a\tilde q_i^b)(z)+\tfrac{1}{n}\delta_a^b(q^i_c\tilde q_i^c)(z)~,\\
\makebox[14ex][l]{\ph{(}$J_{ \uf(1)}(z)$}  &\ceq~ -(q^i_a\tilde q_i^a)(z)~.
\end{split}
\end{equation}
The $\suf(n)$ current algebras are each at level $k_{\suf(n)}=-n$. Additionally, the canonical stress tensor for this chiral algebra descends from the $\suf(2)_R$ current of the free hypermultiplet theory
\begin{equation}
\label{eq:free_hyper_stress_tensor}
T(z)\ceq (q_i^a\partial\tilde q_a^i)(z)-(\tilde q_a^i\partial q_i^a)(z)~.
\end{equation}
The central charge of the Virasoro symmetry generated by this operator is given by $c=-n^2$.

These Lagrangian building blocks can be used to build up the chiral algebras associated to any of the Lagrangian class $\SS$ theories, \ie, to those theories constructed from linear or circular quivers. For example, the chiral algebras for $\NN=2$ superconformal QCD were studied in \cite{Beem:2013sza}, and these theories are constructed from a pair of these free field trinions by gauging a single $\suf(n)$ symmetry. The TQFT structure associated to these Lagrangian theories is already quite interesting, but we will not dwell on the subject here since these Lagrangian constructions are only the tip of the iceberg for class $\SS$. Indeed, from an abstract point of view there is a different set of theories that are the most natural starting point for an investigation of class $\SS$ chiral algebras. These are the chiral algebras associated to spheres with three maximal punctures.

\medskip
\subsection{Trinion chiral algebras}
\label{subsec:building_blocks}

Our first order of business should then be to understand the elementary building blocks for class $\SS$ chiral algebras of type $\gf=\suf(n)$, which are the trinion chiral algebras $\goodchi[T_n]$. In this section we will try to outline the general properties of these chiral algebras. It is possible that these properties will actually make it possible to fix the chiral algebras completely. It is a hard problem to characterize these algebras for arbitrary $n$. Doing so implicitly involves fixing an infinite amount of CFT data (\ie, operator dimensions and OPE coefficients) for the $T_n$ SCFTs, and this data is apparently inaccessible to the usual techniques used to study these theories. Nevertheless, many properties for these chiral algebras can be deduced from the structure of the $\goodchi$ map and from generalized $S$-duality.

\subsubsection*{Central charge}

From the general results reviewed in Section \ref{subsec:chiral_review}, we know that the chiral algebra of any $T_n$ theory should include a Virasoro subalgebra, the central charge of which is determined by the $c$-type Weyl anomaly coefficient of the parent theory according to the relation $c_{2d}=-12c_{4d}$. The central charges of the $T_n$ theories have been computed in \cite{Gaiotto:2009gz}, and from those results we conclude that the corresponding chiral algebras will have Virasoro central charges given by
\begin{equation}
\label{eq:T_N_central_charge}
c_{2d}(\chi[\TT_{n}])=-2n^3+3n^2+n-2~.
\end{equation}
For any value of $n$ the Virasoro central charge predicted by this equation is an even negative integer. These chiral algebras will necessarily be non-unitary, as is always the case for the protected chiral algebras of four-dimensional theories. For reference, we display the Virasoro central charges for $\goodchi[T_n]$ for low values of $n$ in Table \ref{tab:vir_central_charges}.

\begin{table}[t!]
\centering
\begin{tabular}{c|rrrrrr}
  $n$ 		& $\ph{-3}2\ph{3}$ & $\ph{-3}3\ph{3}$ & $\ph{-3}4\ph{3}$ & $\ph{-33}5$ & $\ph{-33}6$ & $\ph{-33}7$ \\ \hline
  $c_{2d}$	& $\ph{3}-4\ph{3}$ & $-26\ph{3}$ & $-78\ph{3}$ & $-172$ & $-320$ & $-534$\\
\end{tabular}
\caption{Central charges of the chiral algebras $\goodchi[T_n]$ for small values of $n$.}
\label{tab:vir_central_charges}
\end{table}

\subsubsection*{Affine current subalgebras}

Global symmetries of the $T_n$ theories imply the presence of chiral subalgebras that are isomorphic to the affine current algebras for the same symmetry algebra. The levels $k_{2d}$ of these affine current algebras are fixed in terms of the four-dimensional flavor central charges $k_{4d}$ according to $k_{2d}=-\frac12 k_{4d}$. The $T_n$ theories have $\suf(n)^3$ global symmetry with each $\suf(n)$ factor associated to one of the punctures on the UV curve. The flavor central charge for each $\suf(n)$ is given by $k_{4d}=2n$. Consequently, the chiral algebras $\goodchi[T_n]$ will have affine current subalgebras of the form 
\begin{equation}
\label{eq:affine_subalgebra}
\widehat{\mf su(n)}_{-n}\times\widehat{\mf su(n)}_{-n}\times\widehat{\mf su(n)}_{-n}\subset \goodchi[T_n]~.
\end{equation} 
Note that $k_{2d}=-n$ is the \emph{critical level} for an $\widehat{\suf(n)}$ current algebra, which means that the Sugawara construction of a stress tensor fails to be normalizable. The chiral algebras $\goodchi[T_n]$ will still have perfectly good stress tensors, but they will not be given by the Sugawara construction. Precisely the critical affine current algebra $\suf(n)_{-n}$ has been argued in \cite{Beem:2014kka} to describe the protected chiral algebra that lives on maximal codimension two defects of the six-dimensional $(2,0)$ theory in flat six dimensional space. Its reappearance as a subalgebra of the class $\SS$ chiral algebra is then quite natural. It would be interesting to develop a better first-principles understanding of the relationship between BPS local operators supported on codimension two defects in six dimensions and local operators in the class $\SS$ theories obtained by compactification in the presence of said defects.

\subsubsection*{Chiral algebra generators from the Higgs branch}

A definitive characterization of the generators of the protected chiral algebra in terms of the operator spectrum of the parent theory is presently lacking. However, as we reviewed in Section \ref{subsec:chiral_review}, any generator of the Hall-Littlewood chiral ring is guaranteed to be a generator of the chiral algebra. For the $T_n$ theories, the Hall-Littlewood chiral ring is actually the same thing as the Higgs branch chiral ring due to the absence of $\DD$ and $\bar\DD$ multiplets in genus zero class $\SS$ theories. The list of generators of the Higgs branch chiral ring is known for the $T_n$ theories, so we have a natural first guess for the list of generators of these chiral algebras.

In the interacting theories (all but the $T_2$ case), the moment map operators for the flavor symmetry acting on the Higgs branch are chiral ring generators. The corresponding chiral algebra generators are the affine currents described above. There are additional generators of the form \cite{Gadde:2013fma}
\begin{equation}
\label{eq:extra_higgs_generators}
Q^{\II_1\II_2\II_3}_{(\ell)}~,\qquad \ell=1,\cdots, n-1~.
\end{equation}
These operators are scalars of dimension $\Delta=\ell(n-\ell)$ that transform in the $\wedge^\ell$ representation (the $\ell$-fold antisymmetric tensor) of each of the $\suf(n)$ flavor symmetries. There must therefore be \emph{at least} this many additional chiral algebra generators. We may denote these chiral algebra generators as
\begin{equation}
\label{eq:extra_chiral_algebra_generators}
W_{(\ell)}^{\II_1\II_2\II_3}(z)~, \qquad \II=[i_1 \cdots i_\ell]~,\quad  i_*=1,\ldots,n~.
\end{equation} 
These operators will have dimension $h_\ell=\frac12 \ell(n-\ell)$, so for $n>3$ we are guaranteed to have non-linear chiral algebras.

For $n>3$ the stress tensor must be an independent generator of the chiral algebra. This is because the stress tensor can only be a composite of other chiral algebra operators with dimension $h\leqslant1$. For an interacting theory there can be no chiral algebra operators of dimension $h=1/2$, so the only possibility is that the stress tensor is a Sugawara stress tensor built as a composite of affine currents. This can only happen if the $\suf(n)^3$ symmetry is enhanced, since as we have seen above the affine currents associated to the $\suf(n)$ symmetries are at the critical level and therefore do not admit a normalizable Sugawara stress tensor. Such an enhancement of the flavor symmetry only happens for the $n=3$ case, as will be discussed in greater detail below.

Let us now consider the two simplest cases of trinion chiral algebras: $n=2$ and $n=3$. These are both exceptional in some sense compared to our expectations for generic $n$, which will ultimately make them easier to work with in our examples.

\medskip
\subsubsection{The \texorpdfstring{$\goodchi[T_2]$}{Chi[T2]} chiral algebra}
\label{subsubsec:t2_chiral_algebra}

In the rank one case, the trinion SCFT is a theory of free hypermultiplets. This case is exceptional compared to the general free hypermultiplets discussed in Section \ref{subsec:lagrangian_building_blocks} because for $\suf(2)$ the maximal puncture and minimal puncture are the same, so the minimal puncture also carries an $\suf(2)$ flavor symmetry, and instead of $n^2$ hypermultiplets transforming in the bifundamental of $\suf(n)\times\suf(n)$, one instead describes the free fields as $2^3=8$ half hypermultiplets transforming in the trifundamental representation of $\suf(2)^3$. Consequently the symplectic bosons describing this theory are organized into a trifundamental field $q_{abc}(z)$ with $a,b,c=1,2$, with OPE given by
\begin{equation}
\label{eq:trifundamental_OPE}
q_{abc}(z)q_{a'b'c'}(w)\sim\frac{\epsilon_{aa'}\epsilon_{bb'}\epsilon_{cc'}}{z-w}~.
\end{equation}
Each of the three $\suf(2)$ subalgebras has a corresponding $\wh{\suf(2)}_{-2}$ affine current algebra in the chiral algebra. For example, the currents associated to the first puncture are given by
\begin{align}\label{eq:T2_affine_currents}
\begin{split}
J_1^{+}(z)  &~\ceq~ \frac{1}{2}\epsilon^{bb'}\epsilon^{cc'}(q_{1bc}q_{1b'c'})(z)~,\\
J_1^{-}(z)  &~\ceq~ \frac{1}{2}\epsilon^{bb'}\epsilon^{cc'}(q_{2bc}q_{2b'c'})(z)~,\\
J_1^{\,0}(z)&~\ceq~ \frac{1}{4}\epsilon^{bb'}\epsilon^{cc'}\Big[(q_{1bc}q_{2b'c'})(z)+(q_{2bc}q_{1b'c'})(z)\Big]~.
\end{split}\end{align}
The currents associated to the second and third punctures are constructed analogously. The stress tensor is now given by
\begin{equation}
\label{eq:T2_stress_tensor}
T(z)\ceq \e^{aa'}\e^{bb'}\e^{cc'}(q_{abc}\partial q_{a'b'c'})(z)~,
\end{equation}
with corresponding Virasoro central charge given by $c_{2d}=-4$.

In this simple case it is easy to explicitly compare the Schur superconformal index for the $T_2$ theory with the vacuum character of the chiral algebra. The Schur index has appeared explicitly in, \eg, \cite{Gadde:2009kb}. It is given by a single plethystic exponential,
\begin{equation}
\label{eq:T2_index}
\II(q;{\bf a},{\bf b},{\bf c})=\PE\left[\frac{q^{\frac12}}{1-q}\goodchi_\square(\bf a)\goodchi_\square(\bf b)\goodchi_\square(\bf c)\right]~.
\end{equation}
This is easily recognized as the vacuum character of the symplectic boson system defined here. The only comment that needs to be made is that there are no null states that have to be removed from the freely generated character of the symplectic boson algebra. In the next example this simplifying characteristic will be absent.

Crossing symmetry, or associativity of gluing, was investigated for this chiral algebra in \cite{Beem:2013sza}. There it was proposed that the complete chiral algebra obtained when gluing two copies of $\goodchi[T_2]$ is the $\widehat{\sof(8)}$ affine current algebra at level $k_{\sof(8)}=-2$, and this proposal was checked up to level $h=5$. If the chiral algebra of the four-punctured sphere is precisely this current algebra, then the crossing symmetry relation is implied immediately. This is because the $\sof(8)$ current algebra has an automorphism as a consequence of triality that exchanges the $\suf(2)$ subalgebras in accordance with Figure \ref{fig:tqft_topological_invariance}. If one could prove that the solution to the BRST problem for this gluing is the $\wh{\sof(8)}$ current algebra, one would therefore have a proof of generalized $S$-duality at the level of the chiral algebra for all rank one theories of class $\SS$. We hope that such a proof will turn out to be attainable in the future.

\medskip
\subsubsection{The \texorpdfstring{$\goodchi[T_3]$}{Chi[T3]} chiral algebra}
\label{subsubsec:t3_chiral_algebra}

The $T_3$ theory is the rank-one $\ef_6$ theory of Minahan and Nemeschanksky \cite{Minahan:1996fg}. Before describing its chiral algebra, let us list a number of known properties of this theory.
\begin{enumerate}
\item[$\bullet$]The $a$ and $c_{4d}$ anomaly coefficients are known to be given by $a=\frac{41}{24}$ and $c_{4d}=\frac{13}{6}$.
\item[$\bullet$]The global symmetry is $\ef_6$, for which the flavor central charge is $k_{\ef_6}=6$. This is an enhancement of the $\suf(3)^3$ symmetry associated with the punctures. It can be understood as a consequence of the fact that the extra Higgs branch generators have dimension two in this case, which means that they behave as moment maps for additional symmetry generators.
\item[$\bullet$]The Higgs branch of this theory is the $\ef_6$ one-instanton moduli space, which is the same thing as the minimal nilpotent orbit of $\ef_6$. This property follows immediately from the realization of this theory as a single D3 brane probing an $\ef_6$ singularity in F-theory.
\item[$\bullet$]A corollary of this characterization of the Higgs branch is that the Higgs branch chiral ring is finitely generated by the moment map operators $\mu_A$ for $A=1,\ldots,78$, subject to the \emph{Joseph relations} (see {\it e.g.} \cite{Gaiotto:2008nz}),
$$\restr{(\mu\otimes\mu)}{\bOn\oplus\mathbf{650}}=0~.$$
\item[$\bullet$]The superconformal index of the $T_3$ theory was computed in \cite{Gadde:2010te}. This leads to a formula for the Schur limit of the index given by 
\begin{equation}\begin{split}
\II_{T_3}(q)=1\ &+\ q\ \chi_{\mathbf{[0,0,0,0,0,1]}}\ \\&+\ q^2\ (\chi_{\mathbf{[0,0,0,0,0,2]}}+\chi_{\mathbf{[0,0,0,0,0,1]}}+1)\ \\ 
&+\ q^3\ (\chi_{\mathbf{[0,0,0,0,0,3]}}+\chi_{\mathbf{[0,0,0,0,0,2]}}+\chi_{\mathbf{[0,0,1,0,0,0]}}+2\ \chi_{\mathbf{[0,0,0,0,0,1]}}+1)\ \\
&+\ q^4\ (\chi_{\mathbf{[0,0,0,0,0,4]}} + \chi_{\mathbf{[0,0,0,0,0,3]}} + \chi_{\mathbf{[0,0,1,0,0,1]}} + 3\ \chi_{\mathbf{[0,0,0,0,0,2]}}\\
&\ph{+\ q^4\ ( }\   + \chi_{\mathbf{[0,0,1,0,0,0]}}+ \chi_{\mathbf{[1,0,0,0,1,0]}} + 3\ \chi_{\mathbf{[0,0,0,0,0,1]}} +2)\ \notag\\
& + \ldots
\end{split}\end{equation}
where we denoted the $\mf{e}_6$ representations by their Dynkin labels and suppressed the fugacity-dependence.
\end{enumerate}

The only chiral algebra generators that are guaranteed to be present on general grounds are the seventy-eight affine currents that descend from the four-dimensional moment map operators. The level of the affine current algebra generated by these operators will be $k=-3$. Note that this is \emph{not} the critical level for $\ef_6$. The $\suf(3)^3$ symmetry associated to the punctures is enhanced, and criticality of the subalgebras does not imply criticality of the enhanced symmetry algebra. For this reason, it is possible to construct a Sugawara stress tensor for the current algebra that is properly normalized, and indeed the correct value of the central charge is given by
\begin{equation}
\label{eq:sugawara_central_charge_check}
c_{2d}=-26=\frac{-3\dim(\ef_6)}{-3+h^\vee_{\ef_6}}=c_{\rm Sugawara}~.
\end{equation}
One then suspects that the chiral algebra does not have an independent stress tensor as a generator, but instead the Sugawara construction yields the true stress tensor. Indeed, this was proven in \cite{Beem:2013sza} to follow from the saturation of certain unitarity bounds by the central charges of this theory.

This leads to a natural proposal for the $\goodchi[T_3]$ chiral algebra that was already put forward in \cite{Beem:2013sza}. The proposal is that the correct chiral algebra is simply the $\widehat{\ef}_6$ affine current algebra at level $k=-3$. The singular OPEs of the seventy-eight affine currents are fixed to the canonical form,\footnote{Our conventions are that the roots of $\ef_6$ have squared length equal to two.}
\begin{equation}
\label{eq:e6_OPE}
J_A(z)J_B(0) ~\sim~ \frac{-3\,\delta_{AB}}{z^2} + \frac{f_{AB}^{\ph{AB}C}\,J_C(0)}{z}~.
\end{equation}
It is natural to consider the subalgebra $\suf(3)^3\subset \ef_6$ associated to the three punctures on the UV curve and to decompose the currents accordingly. The adjoint representation of $\ef_6$ decomposes as
\begin{equation}
\label{eq:e6_adjoint_decomposition}
\mathbf{78}~\longrightarrow~(\mathbf{8,1,1})+(\mathbf{1,8,1})+(\mathbf{1,1,8})+(\mathbf{3,3,3})+(\mathbf{\bar{3},\bar 3,\bar 3})~.
\end{equation}
The affine currents are therefore rearranged into three sets of $\suf(3)$ affine currents along with one tri-fundamental and one tri-antifundamental set of dimension one currents,
\begin{equation}
\label{eq:e6_decomposed_current_list}
J_A(z)~\longrightarrow~\left\{(J^{1})_{a}^{\,a'}(z)~,~(J^{2})_{b}^{\,b'}(z)~,~(J^{3})_{c}^{\,c'}(z)~,~W_{abc}(z)~,~{\wt W}^{abc}(z)\right\}~.
\end{equation}
The singular OPEs for this basis of generators are listed in Appendix \ref{app:level_by_level}. It is perhaps interesting to note that given this list of generators and the requirement that the $\suf(3)$ current algebras are all at the critical level, the only solution to crossing symmetry for the chiral algebra that includes no additional generators is the $\wh{\ef}_6$ current algebra with $k=-3$. So the chiral algebra is completely inflexible once the generators and their symmetry properties are specified.

A nice check of the whole story is that the Joseph relations are reproduced automatically by the chiral algebra. For the non-singlet relation, this follows in a simple way from the presence of a set of null states in the chiral algebra.
\begin{equation}
\label{eq: e6 nullprediction in 650}
\restr{}{}\!\restr{}{}P^{AB}_{\bf 650}(J_AJ_B)(z)\restr{}{}\!\restr{}{}^2=0 \quad\Longleftrightarrow\quad \restr{(\mu\otimes\mu)}{\bf 650}=0~,
\end{equation}
where $P^{AB}_{\bf 650}$ is a projector onto the ${\bf 650}$ representation. These states are only null at this particular value of the level, so we see a close relationship between the flavor central charge and the geometry of the Higgs branch. Similarly, the singlet relation follows from the identification of the Sugawara stress tensor with the true stress tensor of the chiral algebra,
\begin{equation}
T(z)=\frac{1}{-3+h^\vee}(J_AJ_A)(z) \quad\Longleftrightarrow\quad \restr{(\mu\otimes\mu)}{\bf 1}=0~.
\end{equation}
So in this relation we see that the geometry of the Higgs branch is further tied in with the value of the $c$-type central charge in four dimensions.

Note that these successes at the level of reproducing the Higgs branch chiral ring relations follow entirely from the existence of an $\wh{\ef}_6$ current algebra at level $k=-3$ in the chiral algebra. However what is not necessarily implied is the absence of additional chiral algebra generators transforming as some module of the affine Lie algebra. We can test the claim that there are no additional generators by comparing the partition function of the current algebra to the Schur limit of the superconformal index for $T_3$ (\cf\ \cite{Gadde:2010te}).\footnote{Because the current algebra is entirely bosonic, the $\Zb_2$ graded vacuum character is the same as the ungraded vacuum character. Indeed, it is a prediction of our identification of the $\goodchi[T_3]$ chiral algebra that there are no cancellations in the Schur index between operators that individually contribute.} This comparison is made somewhat difficult by the fact that affine Lie algebras at negative integer dimension have complicated sets of null states in their vacuum module, and these must be subtracted to produce the correct index. The upshot is that up to level four, the vacuum character does indeed match the superconformal index. In order for this match to work, it is crucial that the $\wh{\ef}_6$ current algebra has certain null states at the special value $k=-3$. In Table \ref{Tab:T3_index}, we show the operator content up to level four of a generic $\wh{\ef}_6$ current algebra along with the subtractions that occur at this particular value of the level. It is only after making these subtractions that the vacuum character matches the Schur index. Thus we conclude that if there are any additional generators of the $\goodchi[T_3]$ chiral algebra, they must have dimension greater than or equal to five.

\begin{table}\small
\centering
\begin{tabular}{cl}
\hline
\hline
dimension & $\ef_6$ representations with multiplicities $m_{\rm generic}\blue{/m_{k=-3}}$\\
\hline
$0$ &   $1\ph{/0}\times\mathbf{[0,0,0,0,0,0]}$\\
$1$ &   $1\ph{/0}\times\mathbf{[0,0,0,0,0,1]}$\\
$2$ &   $1\ph{/0}\times\mathbf{[0,0,0,0,0,2]}$,~ $1\blue{/0}\times\mathbf{[1,0,0,0,1,0]}$,~   $1\ph{/0}\times\mathbf{[0,0,0,0,0,1]}$,~   $1\ph{/0}\times\mathbf{[0,0,0,0,0,0]}$\\
$3$ &   $1\ph{/0}\times\mathbf{[0,0,0,0,0,3]}$,~ $1\blue{/0}\times\mathbf{[1,0,0,0,1,1]}$,~   $1\ph{/0}\times\mathbf{[0,0,0,0,0,2]}$,~ $2\blue{/1}\times\mathbf{[0,0,1,0,0,0]}$,\\
	& $2\blue{/0}\times\mathbf{[1,0,0,0,1,0]}$,~ $3\blue{/2}\times\mathbf{[0,0,0,0,0,1]}$,~   $1\ph{/0}\times\mathbf{[0,0,0,0,0,0]}$\\
$4$ &   $1\ph{/0}\times\mathbf{[0,0,0,0,0,4]}$,~ $1\blue{/0}\times\mathbf{[1,0,0,0,1,2]}$,~ $1\blue{/0}\times\mathbf{[2,0,0,0,2,0]}$,~   $1\ph{/0}\times\mathbf{[0,0,0,0,0,3]}$,\\ 
	& $2\blue{/1}\times\mathbf{[0,0,1,0,0,1]}$,~ $1\blue{/0}\times\mathbf{[0,1,0,1,0,0]}$,~ $3\blue{/0}\times\mathbf{[1,0,0,0,1,1]}$,~ $2\blue{/0}\times\mathbf{[1,1,0,0,0,0]}$,\\ 
	& $2\blue{/0}\times\mathbf{[0,0,0,1,1,0]}$,~ $5\blue{/3}\times\mathbf{[0,0,0,0,0,2]}$,~ $3\blue{/1}\times\mathbf{[0,0,1,0,0,0]}$,~ $6\blue{/1}\times\mathbf{[1,0,0,0,1,0]}$,\\
	& $6\blue{/3}\times\mathbf{[0,0,0,0,0,1]}$,~ $3\blue{/2}\times\mathbf{[0,0,0,0,0,0]}$\\
\hline
\end{tabular}
\caption{\label{Tab:T3_index}The operator content of the $\ef_6$ current algebra up to dimension four. The first multiplicity is valid for generic values of the level, \ie, any value of $k$ where null states are completely absent. The second multiplicity is valid for $k=-3$, and if no second multiplicity is given then the original multiplicity is also the correct one for $k=-3$. These latter multiplicities precisely reproduce the coefficients appearing in the Schur superconformal index for the $T_3$ theory.}
\end{table}

A more refined test of our identification of the $\goodchi[T_3]$ chiral algebra comes from the requirement of compatibility with Argyres-Seiberg duality \cite{Argyres:2007cn}. The meaning of Argyres-Seiberg duality at the level of the chiral algebra is as follows. Introduce a pair of symplectic bosons transforming in the fundamental representation of an $\suf(2)$ flavor symmetry,
\begin{equation}
\label{eq:AS_sb_OPE}
q_\alpha(z) \tilde q^\beta (0) \sim  \frac{\delta_\a^{\ph{a}\beta}}{z}~,\qquad \alpha,\beta=1,2~.
\end{equation}
In this symplectic boson algebra one can construct an $\suf(2)$ current algebra at level $k=-1$. Now take the $\ef_6$ current algebra and consider an $\suf(2)\times\suf(6)\subset\ef_6$ maximal subalgebra. The $\suf(2)$ current algebra coming from this subalgebra has level $k=-3$. Thus the combined level of the symplectic-boson-plus-$\goodchi[T_3]$ system is $k_{tot}=-4$, and consequently this current algebra can be gauged in the manner described in Section \ref{subsec:chiral_review} by introducing a $(b,c)$ ghost system in the adjoint of $\suf(2)$ and passing to the cohomology of the appropriate BRST operator. The resulting chiral algebra should be \emph{identical} to the chiral algebra obtained by taking two copies of the $n=3$ free hypermultiplet chiral algebra of Section \ref{subsec:lagrangian_building_blocks} and gauging a diagonal $\suf(3)$ current algebra. This comparison is detailed in Appendix \ref{app:level_by_level}. 

Although we have not been able to completely prove the equivalence of these two chiral algebras (the BRST problem for this type of gauging is not easy to solve), we do find the following. On each side of the duality, we are able to determine the generators of dimensions $h=1$ and $h=3/2$ which amount to a $\widehat{\uf(6)}_{-6}$ current algebra in addition to a pair of dimension $h=\frac32$ generators transforming in the tri-fundamental and tri-antifundamental representations of $\uf(6)$, with singular OPEs given by
\begin{equation}
b_{i_1i_2i_3}(z)\tilde b^{j_1j_2j_3}(0) \sim \frac{36\,\delta_{[i_1}^{[j_1} \delta_{\ph{[}\!i_2}^{\ph{[}\!j_2} \delta_{i_3]}^{j_3]}}{z^3} - \frac{36\,  \delta_{[i_1}^{[j_1} \delta_{\ph{[}\!i_2}^{\ph{[}\!j_2}\hat J_{i_3]}^{j_3]}(0)}{z^2}+\frac{18\, 	\delta_{[i_1}^{[j_1} 	  \hat J_{\ph{[}\!i_2}^{\ph{[}\!j_2}\hat J_{i_3]}^{j_3]}(0) -  18\,\delta_{[i_1}^{[j_1} \delta_{\ph{[}\!i_2}^{\ph{[}\!j_2} \del \hat J_{i_3]}^{j_3]}(0)}{z}~.
\end{equation}
Thus these operators in addition to the $\uf(6)$ currents form a closed $\WW$-algebra which is common to both sides of the duality. We expect that these $\WW$-algebras are in fact the entire chiral algebras in question. However, it should be noted that the existence of this $\WW$-algebra actually follows from what we have established about the $\goodchi[T_3]$ chiral algebra without any additional assumptions. That is to say, the possible addition of generators of dimension greater than four could not disrupt the presence of this $\WW$-algebra. In this sense, common appearance of this algebra can be taken as a check of Argyres-Seiberg duality that goes well beyond the check of \cite{Gaiotto:2008nz} at the level of the Higgs branch chiral ring. It not only implies a match of a much larger set of operators than just those appearing in the chiral ring, but it also amounts to a match of the three-point functions for those operators, which include the Higgs branch chiral ring operators.

Finally, let us mention one last consistency check on the identification of $\goodchi[T_3]$ to which we will return in Section \ref{subsec:examples}. When one of the three maximal punctures of the $T_3$ theory is reduced to a minimal puncture by Higgsing, the resulting theory is simply that of nine free hypermultiplets transforming in the bifundamental representation of the remaining $\suf(3)\times\suf(3)$ flavor symmetry (along with a $\uf(1)$ baryon number symmetry associated to the minimal puncture). Therefore if we have correctly identified the $\goodchi[T_3]$ chiral algebra, then it should have the property that when the corresponding reduction procedure is carried out, the result is the symplectic boson chiral algebra of Section \ref{subsec:lagrangian_building_blocks}. The proposal we have given will indeed pass this check, but we postpone the discussion until after we present the reduction procedure in Section \ref{sec:reducing}.

\medskip
\subsubsection{A proposal for \texorpdfstring{$\goodchi[T_n]$}{Chi[Tn]}}
\label{subsubsec:tn_chiral_algebra}

We have seen above that for ranks one and two, the trinion chiral algebras are finitely generated (in the chiral algebra sense) by currents that descend from four-dimensional generators of the Higgs branch chiral ring. We know from the results of \cite{Beem:2013sza} that this cannot be a characterization that holds true for the chiral algebra of an \emph{arbitrary} $\NN=2$ SCFT. Moreover, in an interacting theory where the $\suf(n)^3$ symmetry is not enhanced to a larger global symmetry algebra, the chiral algebra stress tensor cannot be the Sugawara stress tensor of the dimension one currents. This follows from the fact that the $\suf(n)$ current algebras are at the critical level, so the Sugawara construction fails to produce an appropriate stress tensor. Therefore there must be at least an additional generator corresponding to the stress tensor. The results of \cite{Lemos:2014lua} further suggest that there should be additional generators in one-to-one correspondence with the generators of the $\WW_n$ algebra -- \ie, generators of dimensions $3,\ldots,n-1$. Aside from that, however, there is room to hope that there will be no additional generators for the trinion chiral algebras. One piece of partial evidence in favor of this suggestion is the absence of additional HL chiral ring generators on top of those generating the Higgs branch chiral ring. This follows from the fact that the $T_n$ theories have genus zero UV curves. Taking this as sufficient reason to formulate a conjecture, we propose the following:\footnote{This conjecture is different from the one that appeared in the earliest version of this paper. It has been changed to reflect the results of \cite{Lemos:2014lua}, where the original version of the conjecture was ruled out and replaced by the modified version that appears here.}

\begin{conj}[$T_{n\geqslant3}$ chiral algebras]
\label{conj:Tn}
The protected chiral algebra of the $T_n$ SCFT for any $n\geqslant3$ is a $\WW$-algebra with the following generators
\begin{enumerate}
\item[$\bullet$] Three sets of $\suf(n)^3$ affine currents at the critical level $k=-n$. 
\item[$\bullet$] One current of dimension $\frac12\ell(n-\ell)$ transforming in the $(\wedge^{\ell},\wedge^{\ell},\wedge^{\ell})$ representation of $\suf(n)^3$ for each $\ell=1,\ldots,n-1$.
\item[$\bullet$] Operators $W_i$, $i=1,\ldots,n-1$ of dimension $i+1$ that are $\suf(n)^3$ singlets. The dimension two operator is identified as a stress tensor $W_1(z)\equiv T(z)$ with Virasoro central charge equal to $c_{2d}=-2n^3+3n^2+n-2$. In special cases some of these operators may be redundant.
\end{enumerate}
\end{conj}

\noindent At any $n\geqslant4$, the very existence of such a $\WW$-algebra is quite nontrivial, since for a randomly chosen set of generators one doesn't expect to be able to solve the associated Jacobi identities. In fact if the singular OPEs of such a $\WW$-algebra can be chosen so that the algebra is associative, it seems likely that the requirements of associativity will \emph{completely fix} the structure constants, rendering the chiral algebra unique. It is worth observing that precisely such uniqueness occurs in the case of the $T_3$ chiral algebra. The characterization given by the conjecture above for $n=3$ doesn't explicitly imply $\ef_6$ symmetry enhancement, but the unique chiral algebra satisfying the requirements listed is precisely the $\ef_6$ current algebra at the appropriate level. A similar uniqueness result is currently under investigation for the $T_4$ chiral algebra \cite{Lemos:2014lua}.

Before moving on, let us extrapolate a bit from Conjecture \ref{conj:Tn} to make a further conjecture that, while not extremely well-supported, is consistent with everything we know at this time.

\begin{conj}[Genus zero chiral algebras]
\label{conj:genus_zero}
The protected chiral algebra of any class $\SS$ SCFT of type $A_n$ whose UV curve has genus zero is a $\WW$-algebra with singlet generators $W_i$, $i=1,\ldots, n$ of dimension $i+1$ and additional currents associated to Higgs branch chiral ring generators of the four-dimensional theory. In special cases some of the $W_i$ may be related to composites -- in particular when the central charge is equal to its Sugawara value with respect to the affine currents, then the stress tensor $W_1(z)$ is a composite.
\end{conj}

\noindent The modest evidence in favor of this proposal is that genus zero theories have honest Higgs branches with no residual $U(1)$ gauge fields in the IR, so they don't have any of the additional $\NN=1$ chiral ring generators discussed in Section \ref{subsec:chiral_review}. Additionally the examples of \cite{Beem:2013sza} for which there were chiral algebra generators unrelated to four-dimensional chiral ring generators was a genus one and two theories. It would be interesting to explore this conjecture further, even in the Lagrangian case.

%% file: sections/Section_3/S3_3.tex

\subsection{A theory space bootstrap?}
\label{subsec:gluing_and_associativity}

\begin{figure}[t!]
\centering
\includegraphicsif{scale=.45}{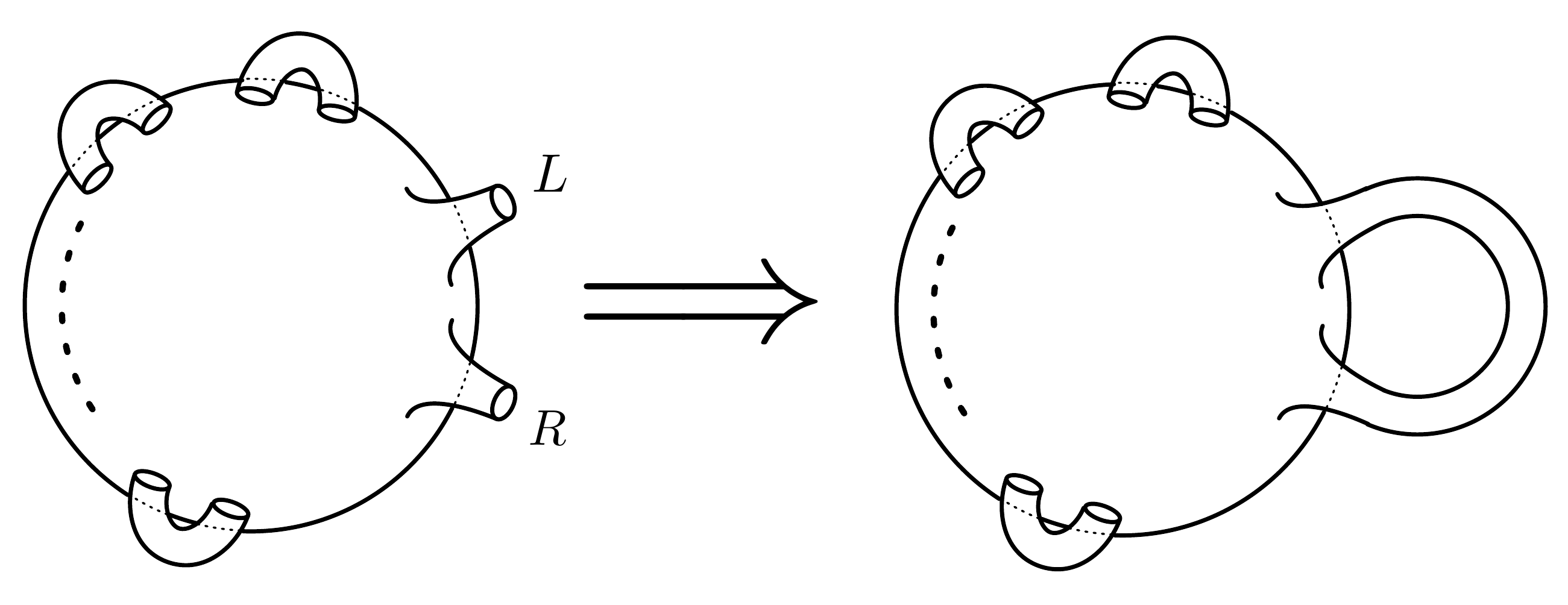}
\caption{Gluing together maximal punctures.}
\label{fig:gluing_punctures}
\end{figure}

Chiral algebras of general theories with maximal punctures can be constructed from the $T_n$ chiral algebra by means of the BRST procedure reviewed in Sec.\;\ref{subsec:chiral_review}. Namely, let us suppose that we are handed the chiral algebra $\TT$ associated to some (possibly disconnected) UV curve with at least two maximal punctures, that we will label $L$ and $R$. The chiral algebra associated to the UV curve where these two punctures are glued together, which we will call $\TT_{\rm glued}$ is obtained in two steps. We first introduce a system of $n^2-1$ $(b,c)$ ghost pairs of dimensions $(1,0)$,
\begin{equation}
\label{eq:adjoint_ghosts_OPE}
b_A(z)c^B(w)\sim\frac{\delta_A^{\ph{A}B}}{z-w}~,\qquad A,B=1,\ldots,n^2-1~.
\end{equation}
These are taken to transform in the adjoint representation of $\suf(n)$, and we can construct the $\suf(n)$ affine currents for that symmetry accordingly,
\begin{equation}
\label{eq:adjoint_ghosts_current}
J^{bc}_A(z) \colonequals -f_{AB}^{\ph{AB}C}(c^B b_C)(z)~.
\end{equation}
The ghost current algebra has level $k_{2d}^{(b,c)}=2h^\vee=2n$. The chiral algebra of the glued configuration is now defined in terms of the ghosts and the chiral algebra of the original system by the BRST procedure of Sec.\;\ref{subsec:chiral_review}. In addition to $\suf(n)$ currents coming from the ghost sector, there will be two more $\suf(n)$ currents $J_A^{L}(z)$ and $J_A^{R}(z)$ associated to the two punctures being glued. A nilpotent BRST operator is defined using these various $\suf(n)$ currents,
\begin{equation}
\label{eq:BRST_op_def}
Q_{\rm BRST}\colonequals\oint\frac{dz}{2\pi i}j_{\rm BRST}(z)~,\qquad j_{\rm BRST}(z)\colonequals (c^A J_A^{L})(z)+(c^A J_A^{R})(z)+\frac{1}{2}(c^A J^{bc}_A)(z)~.
\end{equation}
The nilpotency of $Q_{\rm BRST}$ requires that the sum of the levels of the two matter sector affine currents be given by $k_{L}+k_{R}=-2h^{\vee}$. As usual, this is a reflection of the requirement that the beta function for the newly introduced four-dimensional gauge coupling vanishes. The new chiral algebra is given by
\begin{equation}
\label{eq:gauging_BRST_def}
\goodchi[\TT_{\rm glued}]=H^*_{\rm BRST}\left[\psi\in\goodchi[\TT]\otimes\goodchi_{(b,c)}~|~b_{0}\psi=0\right]~.
\end{equation}
Using this gluing procedure, one may start with a collection of disconnected $\goodchi[T_n]$ chiral algebras and build up the chiral algebra for an arbitrary generalized quiver diagram with maximal punctures. 

The deepest property of the chiral algebras obtained in this manner, which is also the principal condition that must be imposed in order for the map described in the previous section to be a functor, is that they depend only on the topology of the generalized quiver. Of course this is the chiral algebra reflection of generalized $S$-duality in four dimensions, and follows from the more elementary requirement that the gluing described here is associative (alternatively, crossing-symmetric) in the manner represented pictorially in Fig.\;\ref{fig:tqft_topological_invariance}. This is a very strict requirement, and it is conceivable that the $\goodchi[T_n]$ chiral algebras might be the unique possible choices for the image of the trinion in $\Cb\Ab_{\suf(n)}$ that satisfy this condition. Indeed, this requirement of theory-space crossing symmetry imposes a strong constraint on any proposal for the $\goodchi[T_n]$ chiral algebras. For the $\goodchi[T_3]$ theory, where we have a proposal for the chiral algebra, it would be interesting to investigate this associativity condition. For the general case, it is interesting to ask whether this constraint might help to \emph{determine} the appropriate trinion chiral algebras. At present, we see no obvious strategies that would utilize this direct approach.

\begin{figure}[t!]
\centering
\includegraphicsif{scale=.5}{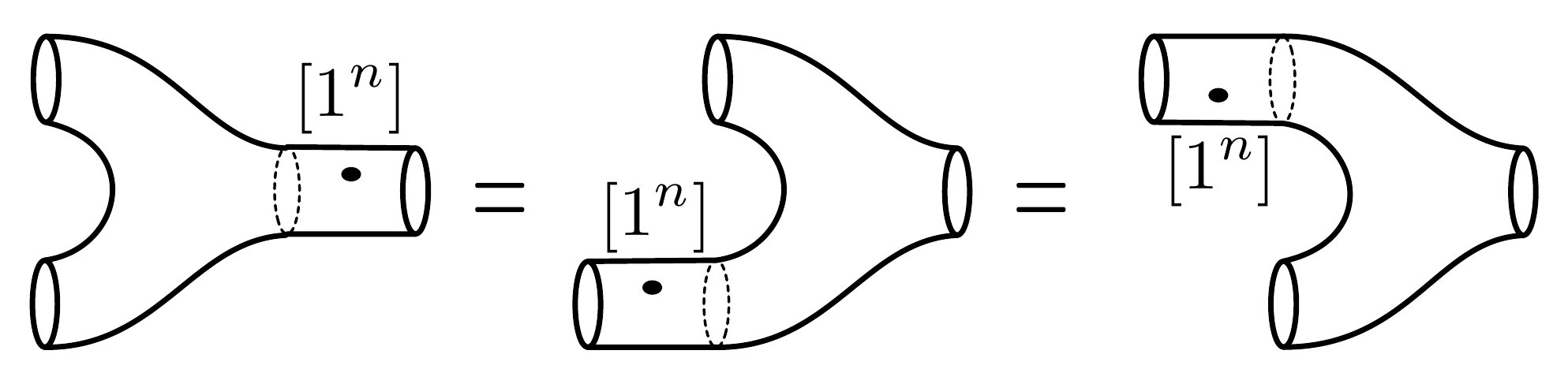}
\caption{Associativity with respect to gluing in free hypermultiplets.}
\label{fig:free_hyper_associativity}
\end{figure}

Although we will have more to say about reduced punctures in Sec.\;\ref{sec:reducing}, we should point out that the associativity conditions described here apply equally well to the case when not all punctures are maximal. A particularly interesting case that we can consider immediately is when one puncture is minimal. In this case, the requirement of associativity is the one illustrated in Fig.\;\ref{fig:free_hyper_associativity}. This relation is interesting because the theory with two maximal punctures and one minimal puncture is a known quantity -- the free hypermultiplet chiral algebra of Sec.\;\ref{subsec:lagrangian_building_blocks} -- and so the relation amounts to probing the unknown trinion chiral algebra by coupling it to a known theory. One may hope that this is a sufficient condition in place of the full $T_n$ associativity from which to try to bootstrap the class $\SS$ chiral algebras. In fact, as we will see in the next section, this condition does follow directly from the full puncture condition, though the converse is not obvious.

Leaving direct approaches to the theory space bootstrap as an open problem, let us note that associativity combined with the conjectures of the previous subsection provide a very constraining framework within which we can attempt to characterize various class $\SS$ chiral algebras. Namely, Conjecture \ref{conj:genus_zero} suggests a list of generators for an arbitrary genus zero chiral algebra, and the requirement of associativity implies the presence of an automorphism that acts as permutations on the $\suf(n)$ subalgebras associated to the various punctures. This permutation symmetry vastly constrains the possible OPE coefficients of the aforementioned generators, which leads to a straightforward problem of solving the Jacobi identities for such a chiral algebra.

As a simple example of this approach, let us consider the rank one chiral algebra associated to the sphere with five punctures. In this case, the chiral algebra generators associated to Higgs branch chiral ring generators are five sets of $\suf(2)$ affine currents at level $k=-2$ 
along with a single additional generator of dimension $h=3/2$ with a fundamental index with respect to each $\suf(2)$ symmetry. Since this is a generic case, the stress tensor will be an independent generator. If Conjecture \ref{conj:genus_zero} is correct, then there should be a $\WW$-algebra with precisely these generators that, due to associativity, has an $S_5$ automorphism group that acts as permutations on the five $\suf(2)$ subalgebras. Consequently, the number of independent parameters in the singular OPE of the $\WW$-algebra generators is quite small. The only singular OPE not fixed by flavor symmetries and Virasoro symmetry is that of two copies of the quinfundamental field,
\begin{multline}
\label{eq:five_puncture_OPE}
Q_{abcde}(z)Q_{a'b'c'd'e'}(w)~\sim~\frac{\e_{aa'}\e_{bb'}\e_{cc'}\e_{dd'}\e_{ee'}}{(z-w)^3}+\\
\frac{\alpha (J_{aa'}\e_{bb'}\e_{cc'}\e_{dd'}\e_{ee'}+\text{permutations})}{(z-w)^2}+\frac{(\beta\,T+\gamma(\e^{f f'}\e^{f'' f'''}J_{f f''}J_{f' f'''}+\text{4 more}))\e_{aa'}\e_{bb'}\e_{cc'}\e_{dd'}\e_{ee'}}{(z-w)}+\\
\frac{\zeta(\partial J_{aa'}\e_{bb'}\e_{cc'}\e_{dd'}\e_{ee'}+\text{permutations})}{z-w}+\frac{\eta(J_{aa'}J_{bb'}\e_{cc'}\e_{dd'}\e_{ee'}+\text{permutations})}{(z-w)}~.
\end{multline}
The parameters $\alpha$, $\beta$ and $\zeta$ are constrained in terms of the central charges $c=-24$ and $k=-2$ by comparing with the $\langle QQT\rangle$ and $\langle QQJ\rangle$ three-point functions:
\begin{equation}
\label{eq:five_puncture_coeffs}
-8\,\beta + 20\,\gamma = 1~,\qquad
\zeta=\frac{1}{4}~,\qquad
\alpha=\frac{1}{2}~.
\end{equation}
This leaves a total of two adjustable parameters, which we may take to be $\{\gamma,\eta\}$. It is a highly nontrivial fact then that the Jacobi identities for this $\WW$-algebra can indeed be solved for a unique choice of these parameters,
\begin{equation}
\label{eq:five_puncture_coeffs_2}
\gamma=-\frac{1}{20}~,\quad\eta=\frac14~.
\end{equation}
Interestingly, this solution of crossing symmetry is special to the $\suf(2)$ level taking the critical value $k=-2$ and the Virasoro central charge taking the expected value $c_{2d}=-24$. Had we not fixed them by hand, we could have derived them from crossing symmetry here.

We consider the existence and uniqueness of this solution as strong evidence in favor of the validity of Conjecture \ref{conj:genus_zero} in this instance, seeing as the existence of such a $\WW$-algebra would otherwise be somewhat unexpected. Indeed, this characterization of the class $\SS$ chiral algebras becomes all the more invaluable for non-Lagrangian theories. See \cite{Lemos:2014lua} for a discussion of the case of $\goodchi[T_4]$.

%% file: sections/Section_4/S4.tex

\section{Reduced punctures}
\label{sec:reducing}

The $T_n$ building blocks outlined in Sec.\;\ref{subsec:building_blocks} only allow us to construct class $\SS$ chiral algebras associated to undecorated UV curves, while the inclusion of the free hypermultiplet chiral algebras of Sec.\;\ref{subsec:lagrangian_building_blocks} allow for decoration by minimal punctures only. The purpose of this section is to develop the tools necessary to describe theories that correspond to UV curves with general non-trivial embeddings decorating some of their punctures.

From the TQFT perspective, the most natural way to introduce the necessary additional ingredients is to find a chiral algebra associated to the decorated cap of Fig. \ref{fig:dec_cap}. This turns out not to be the most obvious approach from a physical perspective since the cap doesn't correspond to any four-dimensional SCFT.%
\footnote{It does however correspond to a true compactification of the six-dimensional $(2,0)$ theory \cite{Gaiotto:2011xs}. We will return to the notion of such a decorated cap in Sec.\;\ref{subsec:decorated_cap}.}
Rather, it is more natural to develop a procedure for reducing a maximal puncture to a non-maximal that mimics the Higgsing procedure reviewed in Sec.\;\ref{subsec:class_S_review}. Naively, the four-dimensional Higgsing prescription need not lead to a simple recipe for producing the chiral algebra of the Higgsed theory in terms of that of the original theory. This is because the Higgsing spontaneously breaks the superconformal symmetry that is used to argue for the very existence of a chiral algebra, with the theory only recovering superconformal invariance in the low energy limit. Consequently one could imagine that the Higgsing procedure irrecoverably requires that we abandon the chiral algebraic language until reaching the far infrared.

Nevertheless, it turns out that the chiral algebra does admit its own Higgsing procedure that has the desired result. Such a procedure cannot literally amount to Higgsing in the chiral algebra, because quantum mechanically in two dimensions there are no continuous moduli spaces of vacua. The best that we can do is to try to impose a quantum-mechanical \emph{constraint} on the chiral algebra. A natural expectation for the constraint is that it should fix to a non-zero value the chiral algebra operator that corresponds to the Higgs branch chiral ring operator that gets an expectation value. This means imposing the constraint
\begin{equation}
\label{eq:positive_current_constraint}
J_{\alpha_-}(z)=A~,
\end{equation}
where $T_{\alpha_-}=\Lambda(t_-)$. Here $A$ is a dimensionful constant that will be irrelevant to the final answer as long as it is nonzero. We might also expect that we should constrain some of the remaining currents to vanish. A motivation for such additional constraints is that when expanded around the new vacuum on the Higgs branch, many of the moment map operators become field operators for the Nambu-Goldstone bosons of spontaneously broken flavor symmetry, and we want to ignore those and focus on the chiral algebra associated to just the interacting part of the reduced theory.

There happens to be a natural conjecture for the full set of constraints that should be imposed. This conjecture, which was already foreshadowed in \cite{Beem:2013sza}, is as follows:
\begin{conj}\label{conj:qDS}
The chiral algebra associated to a class $\SS$ theory with a puncture of type $\Lambda$ is obtained by performing quantum Drinfeld-Sokolov (qDS) reduction with respect to the embedding $\Lambda$ on the chiral algebra for the theory where the same puncture is maximal.
\end{conj}
Quantum Drinfeld-Sokolov in its most basic form is a procedure by which one obtains a new chiral algebra by imposing constraints on an affine Lie algebra $\hat\gf$, with the constraints being specified by an embedding $\Lambda:\suf(2)\hookrightarrow\gf$. In the case of interest to us, the chiral algebra on which we will impose these constraints is generally larger than just an affine Lie algebra. Nevertheless, these constraints can still be consistently imposed in the same manner. This conjecture therefore amounts to a choice of the additional constraints beyond \eqref{eq:positive_current_constraint} that should be imposed in order to reduce a puncture. It is interesting to note that the right set of constraints will turn out to fix only \emph{half} of the currents that are expected to become Nambu-Goldstone bosons. We will see that the removal of the remaining Nambu-Goldstone bosons occurs in a more subtle manner.

Before delving into the details, we should make the observation that this answer is not unexpected in light of the pre-existing connections between non-maximal defects in the $(2,0)$ theory and qDS reduction \cite{Alday:2010vg,Chacaltana:2012zy}. Though a sharp connection between the AGT story and the protected chiral algebra construction is still lacking, we take this as a positive indication that such a connection is there and remains to be clarified. We now turn to a more precise description of qDS reduction for chiral algebras with affine symmetry. We will first develop the general machinery for performing such a reduction in the cases of interest, whereafter we will perform a number of tests of the claim that this is the correct procedure for reducing the ranks of punctures in class $\SS$ chiral algebras.

\bigskip
\input{./sections/Section_4/S4_1}
\bigskip
\input{./sections/Section_4/S4_2}
\bigskip
\input{./sections/Section_4/S4_3}
\bigskip
\input{./sections/Section_4/S4_4}
\bigskip
\bigskip

%% file: sections/Section_4/S4_1.tex

\subsection{Quantum Drinfeld-Sokolov for modules}
\label{subsubsec:qDSspecseq}

Quantum Drinfeld-Sokolov reduction is a procedure for imposing a set of constraints given below in Eqn. \eqref{eq:qDSconstraints} at the quantum level for an affine Lie algebra $\hat\gf$ at any level. In the following discussion, we will closely follow the analysis of \cite{deBoer:1993iz} (see also \cite{deBoer:1992sy} for a similar discussion for finite dimensional algebras). Although traditionally the starting point for this procedure is a pure affine Lie algebra, our interest is in the case of a more general chiral algebra with an affine Lie subalgebra at the critical level. Said differently, we are interested in performing qDS reduction for nontrivial $\hat{\gf}_{-h^\vee}$ modules. We will utilize essentially the same spectral sequence argument as was used in \cite{deBoer:1993iz}. Some basic facts about spectral sequences are collected in Appendix \ref{app:spectral_sequences} for the convenience of the reader.

The general setup with which we are concerned is the following. We begin with a chiral algebra (for simplicity we take it to be finitely generated) with an $\widehat{\suf(n)}_{k}$ affine subalgebra. We denote the generating currents of the affine subalgebra as $J_{A}(z)$, while the additional generators of the chiral algebra will be denoted as $\{\phi^i(z)\}$, each of which transforms in some representation $\Rf_i$ of $\suf(n)$.

We now choose some embedding $\Lambda:\suf(2)\hookrightarrow\suf(N)$, for which the images of the $\suf(2)$ generators $\{t_0,t_+,t_-\}$ will be denoted by $\{\Lambda(t_0),\Lambda(t_+),\Lambda(t_-\}$. The embedded Cartan then defines a grading on the Lie-algebra,
\begin{equation}
\label{eq:cartan_grading_def}
\gf = \bigoplus_{m\in\frac12 \Zb}\gf_m~,\qquad \gf_m\ceq\left\{T_A \in \gf~\vert~{\rm ad}_{\Lambda(t_0)}T_A = m\,T_A \right\}~.
\end{equation}
When the embedded Cartan is chosen such that some of the currents have half-integral grading, then some of the associated constraints are second-class and cannot be enforced by a straightforward BRST procedure. Fortunately, it has been shown that one may circumvent this problem by selecting an alternative Cartan generator $\delta$ which exhibits integer grading and imposing the corresponding first class constraints \cite{Feher:1992ed,deBoer:1992sy,deBoer:1993iz}. We will adopt the convention that an index $\alpha$ ($\bar\alpha$) runs over all roots with negative (non-negative) grading with respect to $\delta$, while Latin indices run over all roots. The first-class constraints to be imposed are then as follows,
\begin{equation}
\label{eq:qDSconstraints}
J_{\alpha} = A\,\delta_{\alpha\alpha_-}~,
\end{equation}
where $\Lambda(t_-) = T_{\alpha_-}$. These constraints are imposed \`a la BRST by introducing dimension $(1,0)$ ghost pairs $(c^\alpha,b_\alpha)$ in one-to-one correspondence with the generators $T_{\alpha}$. These ghosts have the usual singular OPE
\begin{equation}
\label{eq:BRST_ghost_OPE}
c^\alpha(z) b_\beta(0)\sim \frac{\delta^\alpha_{\ph{\alpha}\beta}}{z}~,
\end{equation}
and allow us to define a BRST current
\begin{equation}
\label{eq:BRSTcurrent}
d(z) = \left(J_{\alpha}(z) - A\,\delta_{\alpha\alpha_-}\right)c^{\alpha}(z) - \frac12 f_{\alpha\beta}^{\ph{\alpha\beta}\gamma}(b_\gamma(c^\alpha c^\beta))(z)~.
\end{equation}
The reduced chiral algebra is defined to be the BRST-cohomology of the combined ghost/matter system. Note that this definition is perfectly reasonable for the case where we are reducing not just the affine current algebra, but a module thereof. The presence of the module doesn't modify the system of constraints of the BRST differential, but as we shall see, the operators in the modules will be modified in a nontrivial way in the constrained theory.

This cohomological problem can be solved via a modest generalization of the approach of \cite{Feigin:1990pn,deBoer:1993iz}. We first split the BRST current into a sum of two terms,
\begin{align}\begin{split}
d_0(z) &~=~ \left(-A\,\delta_{\alpha\alpha_-}\right) c^{\alpha}(z)~,\\
d_1(z) &~=~ J_{\alpha}(z)c^{\alpha}(z)-\frac12 f_{\alpha\beta}^{\ph{\alpha\beta}\gamma}(b_\gamma(c^\alpha c^\beta))(z)~.\label{eq:differential0}
\end{split}\end{align}
We now introduce a bi-grading for the currents and ghosts so that the differentials $(d_0,d_1)$ have bi-grades $(1,0)$ and $(0,1)$, respectively,
\begin{alignat}{3}
&\text{deg }(J_A(z)) 	  &~=~& (m,-m)~,		\qquad &&T_A\in \gf_m~,\notag\\
&\text{deg }(c^\alpha(z)) &~=~& (-m,1+m)~,  	\qquad &&T_\alpha\in \gf_m~,\label{eq:bigrading}\\
&\text{deg }(b_\alpha(z)) &~=~& (m,-m-1)~, 		\qquad &&T_\alpha\in \gf_m\notag~.
\end{alignat}
This bi-grading can also be extended to the additional generators $\phi^i$. We decompose each such generator into weight vectors of $\suf(n)$ according to
\begin{equation}
\label{eq:phi_decomposition}
\phi^i=\phi^i_It_I^{(\Rf_i)}~,\qquad I=1,\ldots,\dim\Rf_i~,
\end{equation}
where the $t_I^{(\Rf_i)}$ form a weight basis for the representation $\Rf_i$ with weights defined according to 
\begin{equation}
\label{eq:general_rep_weight_basis}
H_\alpha\cdot t_I^{(\Rf_i)} = \lambda^{(\Rf_i)}_{I,\alpha}\, t_I^{(\Rf_i)}~,
\end{equation}
where $H_\alpha$ is an element of the Cartan subalgebra of $\suf(n)$. Given the element $\delta$ in terms of which our grading is defined, the bi-grading of the extra generators can be defined according to
\begin{equation}
\label{eq:bigradingextrafields}
\text{deg }(\phi^i_I) = (\delta\cdot t_I^{(\Rf_i)},\,-\delta\cdot t_I^{(\Rf_i)})~.
\end{equation}
The differentials $(d_0,d_1)$ are each differentials in their own right, that is, they satisfy
\begin{equation}
\label{eq:double_sequence_differentials}
d_0^2=d_1^2=d_0d_1+d_1d_0=0~.
\end{equation}
Therefore they define a double complex on the Hilbert space of the ghost/matter chiral algebra, which is the starting point for a spectral sequence computation of the cohomology.

It turns out that a simplification occurs if instead of trying to compute the cohomology of the double complex straight off, we first introduce ``hatted currents'' \cite{Feigin:1990pn,deBoer:1993iz},
\begin{equation}
\label{eq:hattedcurrents}
\hat{J}_A(z) = J_A(z) + f_{A\beta}^{\ph{a\beta}\gamma}(b^{\ph{a}}_\gamma c_{\ph{a}}^{\,\beta})(z)~.
\end{equation}
Let us denote by $\Ab_1$ the subalgebra generated by $b_\alpha(z)$ and $\hat{J}_\alpha(z)$, and by $\Ab_2$ the subalgebra produced by the remaining generators $c^\alpha(z)$, $\hat{J}_{\bar\alpha}(z)$, and $\phi^i(z)$. One then finds that $d(\Ab_1)\subseteq\Ab_1$ and $d(\Ab_2)\subseteq\Ab_2$, with the generators of $\Ab_1$ additionally obeying
\begin{equation}
\label{eq:trivial_cohomology}
d(b_{\alpha}(z))= \hat{J}_{\alpha}(z)-A\delta_{\alpha\alpha_-}~,\qquad d(\hat{J}_{\alpha}(z))=0~.
\end{equation}
It follows that the BRST cohomology of $\Ab_1$ is trivial: $H^*(\Ab_1,d)=\Cb$. From the K\"unneth formula (see Appendix \ref{app:spectral_sequences}), it follows that the BRST cohomology of the chiral algebra is isomorphic to the cohomology of the smaller algebra $\Ab_2$,
\begin{equation}
\label{eq:cohomology_simplification}
H^*(\Ab,d) \cong H^*(\Ab_2,d)~.
\end{equation}
Our task then simplifies: we need only compute the cohomology of $\Ab_2$. We will address this smaller problem by means of a spectral sequence for the double complex $(\Ab_2,d_0,d_1)$.

The first step in the spectral sequence computation is to compute the cohomology $H^*(\Ab_2,d_0)$. The only nontrivial part of this computation is the same as in the case without modules. This is because the additional generators $\phi^i_{I}(z)$ have vanishing singular OPE with the $c$-ghosts, rendering them $d_0$-closed. Moreover, they can never be $d_0$-exact because the $b$-ghosts are absent from $\Ab_2$. For the currents and ghosts, one first computes
\begin{equation}
\label{eq:d0_of_Jhat}
d_0(\hat{J}_{\bar{\alpha}}(z)) = -A f_{\bar{\alpha}\beta}^{\ph{\alpha\beta}\gamma} \delta^{\ph{a}}_{\gamma\alpha_-}c^{\,\beta}(z) = - \Tr\left({\rm ad}_{\Lambda(t_+)}T_{\bar{\alpha}}\cdot T_\beta\right)c^{\,\beta}(z)~.
\end{equation}
It follows that $d_0(\hat{J}_{\bar\alpha}(z))=0$ if and only if $T_{\bar{\alpha}}\in \ker({\rm ad}_{\Lambda(t_+)})$. The same equation implies that the $c^\alpha(z)$ ghosts are $d_0$-exact for any $\alpha$. Because the $d_0$-cohomology thus computed is supported entirely at ghost number zero, the spectral sequence terminates at the first step. At the level of vector spaces we find
\begin{equation}
\label{eq:vector_space_cohomology}
H^*(\Ab,d) \cong H^*(\Ab_2,d_0)~,
\end{equation}
with $H^*(\Ab_2,d_0)$ being generated by the $\phi^i_I(z)$ and by $J_{\bar\alpha}(z)$ for $T_{\bar\alpha}\in\ker({\rm ad}_{\Lambda(t_+)})$.

In order to improve this result to produce the vertex operator algebra structure on this vector space, we can construct representatives of these with the correct OPEs using the tic-tac-toe procedure. Letting $\psi(z)$ be a generator satisfying $d_0(\psi(z))=0$, the corresponding chiral algebra generator $\Psi(z)$ is given by
\begin{equation}
\label{eq:tic-tac-toed_generator}
\Psi(z) = \sum_l (-1)^l \psi_l(z)~,
\end{equation}
where $\psi_l(z)$ is fixed by the condition
\begin{equation}
\label{eq:tic-tac-toe_condition}
\psi_0(z) \ceq \psi(z)~,\quad d_1(\psi_l(z)) = d_0(\psi_{l+1}(z))~.
\end{equation}
At the end, this procedure will give a collection of generators of the qDS reduced theory along with their singular OPEs and it would seem that we are finished. However, it is important to realize that this may not be a minimal set of generators, in that some of the generators may be expressible as composites of lower dimension generators due to null states. The existence of null relations of this type is very sensitive to the detailed structure of the original chiral algebra. For example, the level of the current algebra being reduced plays an important role. In practice, we will find for the class $\SS$ chiral algebras, \emph{most} of the generators $\Psi(z)$ produced by the above construction do in fact participate in such null relations.

Some null states of the reduced theory can be deduced from the presence of null states in the starting chiral algebra. This can be an efficient way to generate redundancies amongst the naive generators of the qDS reduced theory like the ones described above. Abstractly, we can understand this phenomenon as follows. Consider a null operator $N^K(z)$ that is present in the original $\WW$-algebra, and that transforms in some representation $\mathfrak{R}$ of the symmetry algebra that is being reduced. Given an embedding $\Lambda,$ the representation $\mathfrak{R}$ decomposes as in \eqref{eq:generaldecomposition} under $\mathfrak{g}_{\Lambda}\oplus \Lambda(\mathfrak{su}(2)).$ We can thus split the index $K$ accordingly and obtain $\{N^{k_j,m_j}(z)\}_{j\geqslant 0},$ where $k_j$ is an index labeling the representation $\mathcal{R}_j^{(\mathfrak{R})}$ and $m_j$ labels the Cartan of the spin $j$ representation $V_j.$ For fixed values of the index $m_j$ we find an operator that will have proper dimension with respect to the new stress tensor \eqref{eq:qDSstresstensor}. Moreover, since introducing a set of free ghost pairs naturally preserves the null property of the original operator and restricting oneself to the BRST cohomology does not spoil it either, we find that this operator is null in the qDS reduced theory. In practice, for each value of $m_j$ one chooses a representative of the BRST class $N^{k_j,m_j}(z) + d(\ldots)$ that only involves the generators of the qDS reduced theory.

There are a couple of features of the qDS reduced theory that can be deduced without studying the full procedure in specific examples. These features provide us with the most general test of the conjecture that qDS reduction is the correct way to reduce the ranks of punctures in the chiral algebra. The first of these features is the Virasoro central charge of the reduced theory, a subject to which we turn presently.

%% file: sections/Section_4/S4_2.tex

\subsection{Virasoro central charge and the reduced stress tensor}
\label{subsec:reduced_central_charges}

A useful feature of qDS reduction is that the stress tensor of a qDS reduced chiral algebra takes a canonical form (up to BRST-exact terms) in which it is written as a shift of the stress tensor of the unreduced theory,
\begin{equation}
\label{eq:qDSstresstensor}
T = T_{\star} - \partial J_0 + \partial b_\alpha c^\alpha - (1+\lambda_\alpha) \partial (b_\alpha c^\alpha)~.
\end{equation}
Here $T_{\star}$ is the stress tensor of the unreduced theory, $J_0$ is the affine current of the $U(1)$ symmetry corresponding to $\Lambda(t_0)$, and $\lambda_\alpha$ is the weight for $T_\alpha$ with respect to $\Lambda(t_0)$ as defined by Eqn. \eqref{eq:general_rep_weight_basis}.%
\footnote{Note that in the case of half-integral gradings, the weights $\lambda_\alpha$ are defined with respect to $\Lambda(t_0)$ and \emph{not} with respect to the alternate Cartan element $\delta$.}
The dimensions of the ghosts measured by this new stress tensor are $h_{b_\alpha} = 1+\lambda_\alpha$ and $h_{c^\alpha} = -\lambda_\alpha$. Meanwhile the dimensions of all remaining fields are simply shifted by their $J_0$ charge.

The central charge of the reduced theory can be read off from the most singular term in the self-OPE of the reduced stress tensor. The result is given by \cite{Feher:1992ed}
\begin{align}\label{eq:central_charge_shift}
\begin{split}
c-c_{\star}	&~= \left(\dim\gf_0-\frac12\dim\gf_{\frac12}-12\left|\sqrt{k+h^\vee}\Lambda(t_0)-\frac{\rho}{\sqrt{k+h^\vee}}\right|^2\right)-\left(\frac{k\dim\gf}{k+h^\vee}\right)~,\\
			&~= \dim\gf_0-\frac12\dim \gf_{\frac{1}{2}}-12(k+h^\vee)\left|\Lambda(t_0)\right|^2+24\Lambda(t_0)\cdot\rho-\dim\gf~.
\end{split}
\end{align}
Here $\rho$ is the Weyl vector of $\suf(n)$, and in passing to the second line, we have used the Freudenthal-de Vries strange formula $|\rho|^2 = \frac{h^\vee}{12}\dim\gf$. In the cases of interest the level of the current algebra is always given by $k=-h^\vee$ and there is a further simplification,
\begin{equation}
\label{eq:change2dcentralcharge}
c=c_{\star}+\dim\gf_0-\frac12\dim\gf_{\frac12}+24\Lambda(t_0)\cdot\rho-\dim\gf~.
\end{equation}

This shift of two-dimensional central charge can be compared to our expectations based on the four-dimensional results in Eqns. \eqref{eq:4dcentralcharges}-\eqref{eq:nv_nh_defs_2}. The change of the four-dimensional central charge that occurs upon reducing a maximal puncture down to a smaller puncture labelled by the embedding $\Lambda$ is given by
\begin{align}\begin{split}
\label{eq:change4dcentralcharge}
-12(c_{4d} - c_{4d,\text{orig.}} ) &~=~ 2(n_v(\text{max.}) - n_v(\Lambda)) + (n_h(\text{max.}) - n_h(\Lambda))~,\\
									&~=~ \dim \gf_0 - \frac{1}{2}\dim \gf_{\frac{1}{2}}  +24  \Lambda(t_0)\cdot \rho  -\dim\gf~.
\end{split}\end{align}
Thus we see precise agreement with the change in two-dimensional central charge induced by qDS reduction and that of the four-dimensional charge induced by Higgsing after accounting for the relation  $c_{2d} = -12 c_{4d}$. We take this as a strong indication the the qDS prescription for reducing chiral algebras is indeed the correct one.

%% file: sections/Section_4/S4_3.tex

\subsection{Reduction of the superconformal index}
\label{subsec:reduced_index}

We can now check that the qDS reduction procedure has an effect on the (graded) partition function of the chiral algebra that mimics the prescription for reducing the Schur superconformal index described in Sec.\,\ref{subsec:class_S_review}. As was reviewed above, the Schur limit of the superconformal index is equivalent to a graded partition function of the corresponding chiral algebra,
\begin{equation}
\label{eq:chiral_character}
\II_{\chi}(q; {\bf x}) ~\ceq~ \Tr_{\HH_{\chi}}\,(-1)^F q^{L_0} ~=~ \II^{\rm Schur}(q; {\bf x})~.
\end{equation}
Computing this graded partition function is straightforward for the qDS reduced theory owing to the fact that the BRST differential commutes with all of the fugacities ${\bf x}$ that may appear in the index and has odd fermion number. This means that we can ignore the cohomological aspect of the reduction and simply compute the partition function of the larger Hilbert space obtained by tensoring the unreduced chiral algebra with the appropriate ghosts system.\footnote{There is a caveat to this argument, which is that if there are null states in the reduced theory that do not originate as null states in the parent theory, then their subtraction will not necessarily be accomplished by this procedure. We operate under the assumption that such spurious null states do not appear. This assumption appears to be confirmed by the coherence between this procedure and that discussed in Sec.\,\ref{subsec:class_S_review}.}

This simpler problem of computing the partition function of the larger Hilbert space parallels the index computation described in Sec.\,\ref{subsec:class_S_review}. There are again two steps -- the inclusion of the ghosts, and the specialization of fugacities to reflect the symmetries preserved by the BRST differential. Including the ghosts in the partition function before specializing the fugacities requires us to assign them charges with respect to the UV symmetries. This can be done in a canonical fashion so that upon specializing the fugacities the BRST current will be neutral with respect to IR symmetries and have conformal dimension one. 

Recall that the ghost sector involves one pair of ghosts $(b_\alpha,c^\alpha)$ for each generator $T_{\alpha}$ that is negatively graded with respect to $\delta$. The charge assignments are then the obvious ones -- namely the charges of $b_\alpha$ are the same as those of $T_\alpha$ (let us call them $f_\alpha$), while those of $c^{\alpha}$ are minus those of $b_{\alpha}$. With these charge assignments, the graded partition function of the reduced chiral algebra can be obtained as a specialization that mimics that which led to the superconformal index,
\begin{equation}
\label{eq:chiral_index_specialization}
\II_{\chi_{\Lambda}}(q;{\bf x}_{\Lambda})=\lim_{{\bf x} \to {\bf x}_\Lambda}\II_{\chi}(q;{\bf x})\,\II_{(b,c)_{\Lambda}}(q;{\bf x})~,\qquad \II_{(b,c)_{\Lambda}}\ceq\PE\left[-\!\!\!\!\sum_{T_{\alpha}\in\gf_{<0}}\left(\frac{q\,{\bf x}^{f_{\alpha}}}{1-q}+\frac{{\bf x}^{-f_{\alpha}}}{1-q}\right)\right].
\end{equation}
As in the discussion of the index in Sec.\,\ref{subsec:reduced_index}, we can formally perform the specialization ignoring divergences that occur in both the numerator and the denominator as a consequence of constant terms in the plethystic exponent. In doing this, the flavor fugacities are replaced by fugacities for the Cartan generators of $\hhf_\Lambda$, while the $q$-grading is shifted by the Cartan element of the embedded $\suf(2)$. This leads to the following formal expression for the contribution of the ghosts,\footnote{For simplicity, we write the expression here for the case where $\Lambda(t_0)$ provides an integral grading so there is no auxiliary $\delta$. The case of half-interal grading can be treated with modest modifications.}
\begin{equation}
\label{eq:formal_character_specialized_ghosts}
\II_{(b,c)_{\Lambda}}~~ ``=" ~~\PE\left[ \frac{-q}{1-q} \sum_j \goodchi_{\RR_j^{(\mathrm{adj})}}^{\hhf_{\Lambda}}(\mathbf{a}_{\Lambda}) \sum_{i=-j}^{-1}q^i -\frac{1}{1-q}\sum_j \goodchi_{\RR_j^{(\mathrm{adj})}}^{\hhf_{\Lambda}}(\mathbf{a}^{-1}_{\Lambda})\sum_{i=1}^{j}q^i \right]~.
\end{equation}
After a small amount of rearrangement and the recognition that the representations $\RR_j^{(\mathrm{adj})}$ are pseudoreal, one finds that this exactly reproduces the formal denominator in Eqn.\,\eqref{eq:K_factor_fugacity_replacement_2}. Again, when the limit in Eqn.\,\eqref{eq:chiral_index_specialization} is taken carefully, the divergences in this formal denominator cancel against equivalent terms in the $K$-factors of the numerator to produce a finite result. It is interesting that in spite of the asymmetry between $b$ and $c$ ghosts in this procedure, they ultimately play the same role from the point of view of four-dimensional physics -- each ghost is responsible for cancelling the effect of a single Nambu-Goldstone boson from the index.

Before moving on to examples, we recall that in \cite{Beem:2014kka} it was observed that the $K$-factor for a maximal puncture matches the character of the corresponding affine Lie algebra at the critical level, and it was conjectured that a similar statement would be true for reduced punctures. That is to say, the $K$-factor associated to the reduction of type $\Lambda$ should be the character of the qDS reduction of type $\Lambda$ of the critical affine Lie algebra. Given the analysis to this point, this statement becomes almost a triviality. The qDS reduction of the affine current algebra proceeds by introducing the same collection of ghosts as we have used here, and so the effect on the graded partition function is the introduction of the same ghost term given in Eqn.\,\eqref{eq:formal_character_specialized_ghosts} and the same specialization of fugacities. Thus, the identification of the $K$-factors given in Eqn.\,\eqref{eq:K-factor} with the character of the qDS reduction of the critical affine Lie algebra depends only on our ability to equate the index (\ie, the partition function graded by $(-1)^F$) with the ungraded vacuum character. This is a simple consequence of the fact that the starting current algebra consists of all bosonic operators and the spectral sequence calculation of Sec.\,\ref{subsubsec:qDSspecseq} only found BRST cohomology elements at ghost number zero.

%% file: sections/Section_4/S4_4.tex

\subsection{Simple examples}
\label{subsec:examples}

In light of the analysis in Section \ref{subsubsec:qDSspecseq}, the reduction problem admits an algorithmic solution subject to two conditions. (A) the starting chiral algebra should be finitely generated, \ie, it admits a description as a $\WW$-algebra. (B) the $L_0$ operator of the reduced theory should have a positive definite spectrum. The latter condition must hold for any reductions where the endpoint corresponds to a physical class $\SS$ theory, while the former conditions is conjectured to be true for general class $\SS$ theories but is more certainly true in some simple examples. Given these conditions, the procedure is as follows:
\begin{enumerate}
\item[$\bullet$] List the (possibly redundant) generators of the qDS reduced chiral algebra at the level of vector spaces. These are given by the hatted currents $\hat J_{\bar \alpha}$ for which $T_{\bar{\alpha}}\in \ker({\rm ad}_{\Lambda(t_+)})$, along with all of the additional generators $\{\phi_i\}$.
\item[$\bullet$] Apply the tic-tac-toe algorithm to construct genuine generators of the chiral algebra. The OPEs of these reduced chiral algebra generators can be computed directly using the OPEs of the original, unreduced fields.
\item[$\bullet$] Compute the null states at each level up to that of the highest-dimensional generator in order to check for redundancy. Remove any redundant generators. What remains is a description of the reduced chiral algebra as a $\WW$-algebra.
\end{enumerate}
This procedure is still morally a correct one when the two conditions listed above fail to be met, but in those cases the algorithm will not necessarily terminate in finite time. In the examples discussed in this subsection, both conditions above will indeed be satisfied, so this algorithm will be sufficient to determine the answer entirely.

\begin{figure}[t!]
\centering
\includegraphics[scale=.5]{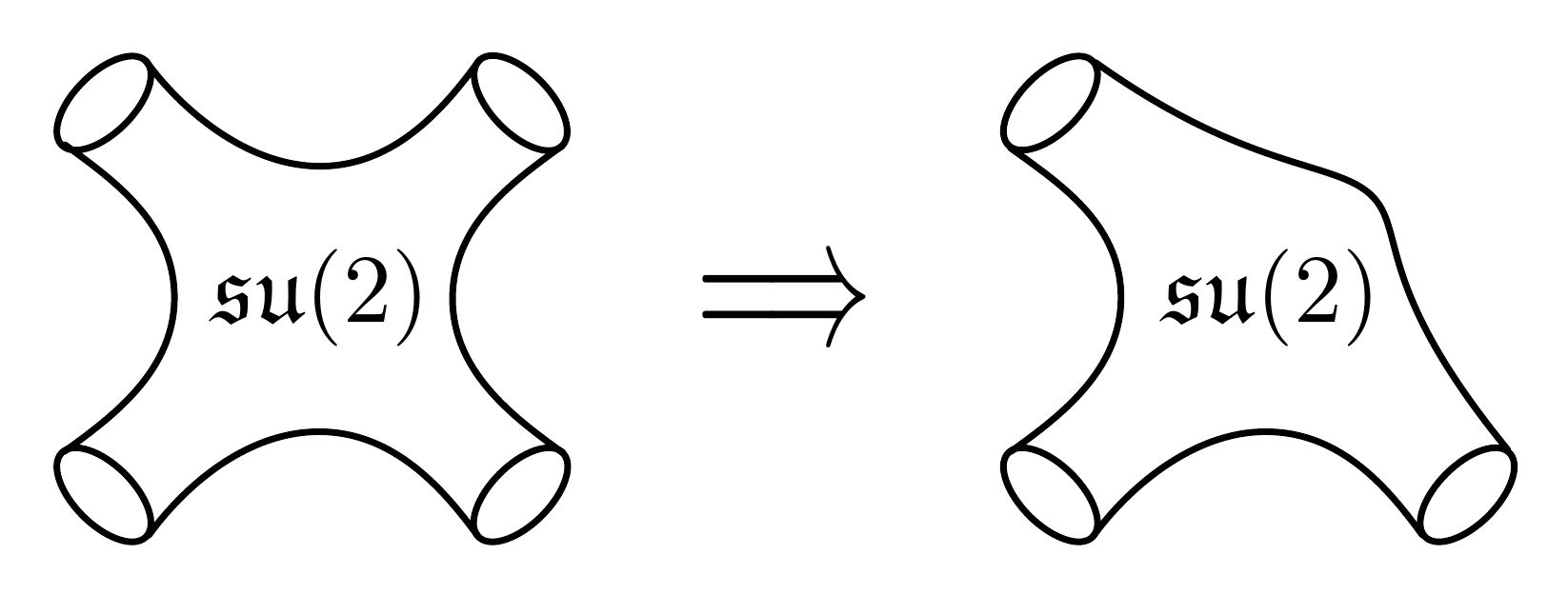}
\caption{Reduction from the $\sof(8)$ theory to $T_2$.}
\label{fig:4_to_3_reduction}
\end{figure}

We now consider a pair of simple cases in which the reduction can be performed quite explicitly. Our first example will be the complete closure of a single puncture in the rank one theory of a four-punctured sphere, which as we reviewed above has as its chiral algebra the affine Lie algebra $\wh{\sof(8)}_{-2}$. The result of this closure is expected to be the $T_2$ theory (see Figure \ref{fig:4_to_3_reduction}). The second example will be the partial reduction (corresponding to the semi-regular embedding) of one puncture in the $T_3$ theory to produce a theory of free bifundamental hypermultiplets, which should correspond to free symplectic bosons at the level of the chiral algebra. Details of the second calculation beyond what is included in this summary can be found in Appendix \ref{subapp:e6_to_free}.

\bigskip
\subsubsection*{Reducing $\wh{\sof(8)}_{-2}$ to $\goodchi[T_2]$}

The starting point for our first reduction is the affine Lie algebra $\widehat{\sof(8)}_{-2}$. We first introduce a basis for the affine currents that is appropriate for class $\SS$ and for the reduction we aim to perform. The adjoint of $\sof(8)$ decomposes into irreps of the $\suf(2)^{(1)}\times\suf(2)^{(2)}\times\suf(2)^{(3)}\times\suf(2)^{(4)}$ symmetries associated to punctures according to
\begin{equation}
\label{eq:SO8currentalgebra}
\rep{28}_{\sof(8)} ~\to~
(\rep3,\rep1,\rep1,\rep1)\oplus
(\rep1,\rep3,\rep1,\rep1)\oplus
(\rep1,\rep1,\rep3,\rep1)\oplus
(\rep1,\rep1,\rep1,\rep3)\oplus
(\rep2,\rep2,\rep2,\rep2)~.
\end{equation}
Accordingly, we assemble the twenty-eight affine currents into these irreps,
\begin{equation}
\label{eq:so8_relabelled_currents}
J_A(z)~\to~ \{J^{(1)}_{(a_1b_1)}(z)~,~J^{(2)}_{(a_2b_2)}(z)~,~J^{(3)}_{(a_3b_3)}(z)~,~J^{(4)}_{(a_4b_4)}(z)~,~J_{a_1a_2a_3a_4}(z)\}~,
\end{equation}
where $a_I,b_I$ are fundamental indices of $\suf(2)^{(I)}$. In this basis, the OPEs of the affine Lie algebra are given by
{\small
\begin{align}\label{eq:so8algebra}
\begin{split}
\makebox[1in][r]{$J^{(I)}_{ab}(z) J^{(J)}_{cd} (w)$}	&~\sim~ \frac{- k (\e_{ac} \e_{bd} + \e_{a d} \e_{bc}) \delta^{IJ}}{2(z-w)^2} + \frac{f^{ef}_{ab ;cd} J^{(I)}_{ef} \delta^{IJ}}{z-w}~,\\
\makebox[1in][r]{$J^{(1)}_{ab} (z) J_{c d e f}(w)$} 	&~\sim~ \frac{ \e_{a c} J_{b d e f} + \e_{b c} J_{a de f}}{2(z-w)}~,\\
\makebox[1in][r]{$J_{a b c d}(z) J_{efgh}(w)$} 			&~\sim~ \frac{k \e_{ae} \e_{bf} \e_{cg} \e_{dh}}{(z-w)^2} + \frac{J^{(1)}_{ae} \e_{bf} \e_{cg} \e_{dh} + \e_{ae} J^{(2)}_{bf} \e_{cg} \e_{dh} + \e_{ae} \e_{bf} J^{(3)}_{cg} \e_{dh}+ \e_{ae} \e_{bf} \e_{cg} J^{(4)}_{dh}}{z-w}~,
\end{split}\end{align}}%
and similarly for the other $J^{(I)}$. Here the $\suf(2)$ structure constants are given by $f^{ef}_{ab;cd} = \hf (\e_{a c} \dt_b^e \dt_d^f + \e_{bc} \dt_a^e \dt_d^f + \e_{ad} \dt_b^e \dt_c^f + \e_{bd} \dt_a^e \dt_c^f)$, and for our case of interest level is fixed to $k=-2$. 

We will choose the first puncture to close, meaning we will perform qDS reduction on the current algebra generated by $J^{(1)}_{(ab)}$ with respect to the principal embedding,
\begin{equation}
\label{eq:so8_reduction_embedding}
\Lambda(t_+)= -T_{11}~,\qquad\Lambda(t_-)= T_{22}~,\qquad \Lambda(t_0)= -T_{(12)}~.
\end{equation}
The grading provided by $\Lambda(t_0)$ is integral, so we can proceed without introducing any auxiliary grading. The only constraint to be imposed in this case is $J_{22}^{(1)}(z) = 1$. This is accomplished with the help of a single ghost pair $(c^{22},b_{22})$, in terms of which the BRST operator is given by
\begin{equation}
\label{eq:so8_brst_operator}
d(z)=c^{22}(J_{22}-1)(z)~.
\end{equation}
The remaining three sets of $\suf(2)$ affine currents can be thought of as trivial modules of the reduced currents, while the quadrilinear currents provide a nontrivial module. In the language of the previous subsection we have\footnote{We should note that there is something slightly unconventional about the reduction procedure here. In this example the entire starting chiral algebra is an affine current algebra, so one could in principle perform qDS reduction in the entirely standard manner. This is \emph{not} what our prescription tells us to do. Instead, we treat a single $\suf(2)$ subalgebra as the target of the reduction, and the rest as modules. The two procedures are naively inequivalent, although we have not checked in detail to make sure that the results don't turn out the same.}
\begin{equation}
\label{eq:so8_extra_generators}
\{\phi^i\} = \{J^{(2)}_{(a_2b_2)}~,J^{(3)}_{(a_3b_3)}~,J^{(4)}_{(a_4b_4)}~, J_{a_1a_2a_3a_4}\}~.
\end{equation}

The reduced generators of step one are simply the hatted current $\hat{J}^{(1)}_{{11}} = J^{(1)}_{{11}}$ along with the additional generators in \eqref{eq:so8_extra_generators}. Applying the tic-tac-toe procedure produces true generators of the reduced chiral algebra,
\begin{align}\label{eq:finalgeneratorsSO8}
\begin{split}
\makebox[11ex][l]{$\hat{\JJ}^{(1)}_{{11}}$} 		&~\colonequals~	\hat J^{(1)}_{{11}} - \hat J^{(1)}_{{12}} \hat J^{(1)}_{{12}} - (k+1) \partial(\hat J^{(1)}_{{12}})~,\\
\makebox[11ex][l]{$(\JJ_1)_{a_2a_3a_4}$}			&~\colonequals~	J_{{1}a_2a_3a_4} - \hat J^{(1)}_{{12}} J_{{2}a_2a_3a_4}~,\\
\makebox[11ex][l]{$(\JJ_2)_{a_2a_3a_4}$}			&~\colonequals~	J_{{2}a_2a_3a_4}~,\\
\makebox[11ex][l]{$\JJ^{(I=\{2,3,4\})}_{a_Ib_I}$}	&~\colonequals~	J^{(I=\{2,3,4\})}_{a_Ib_I}~,
\end{split}\end{align}
where $\hat J^{(1)}_{{12}} \colonequals J^{(1)}_{{12}} + b_{{22}}c^{{2 2}}$. 

The stress tensor of the reduced algebra takes the form given in Eqn. \eqref{eq:qDSstresstensor}, where the original stress tensor was the Sugawara stress tensor of $\widehat{\sof(8)}_{-2}$ and the generator of the embedded Cartan is $J_0 = - J^{(1)}_{{12}}$. We can then compute the conformal dimensions of the new generators and we find
\begin{align}\begin{split}
\makebox[1.1in][l]{$[\hat{\JJ}^{(1)}_{{11}}]=2$~,}\qquad &[\JJ^{(I)}_{a_Ib_I}]=1~,\\
\makebox[1.1in][l]{$[(\JJ_1)_{a_2a_3a_4}]=3/2$~,} \qquad &[(\JJ_2)_{a_2a_3a_4}]=1/2~.
\end{split}\end{align}
The currents $\JJ^{(I)}_{a_Ib_I}$ persist as affine currents of $\suf(2)$ subalgebras, so all of their singular OPEs with other generators are determined by the symmetry properties of the latter. Explicit calculation determines the OPEs that are not fixed by symmetry to take the following form,
{\small
\begin{align}\begin{split}
\hat{\JJ}^{(1)}_{{11}}(z)\hat{\JJ}^{(1)}_{{11}}(0)	&~\sim~  -\frac{1}{2}\frac{(2+k)(1+2k)(4+3k)}{z^4} - \frac{2(2+k)\hat{\JJ}^{(1)}_{{11}}(0)}{z^2}-\frac{(2+k)\partial \hat{\JJ}^{(1)}_{{11}}(0)}{z}\\
\hat{\JJ}^{(1)}_{{11}}(z) (\JJ_1)_{a_2a_3a_4}(0) 	&~\sim~  -\frac{1}{2}\frac{(2+k)(1+2k)(\JJ_2)_{a_2a_3a_4}(0)}{z^3}-\frac{1}{4}\frac{(7+2k)(\JJ_1)_{a_2a_3a_4}(0)}{z^2} - \frac{(\hat{\JJ}^{(1)}_{{11}}(\JJ_2)_{a_2a_3a_4})(0)}{z}\\
\hat{\JJ}^{(1)}_{{11}}(z) (\JJ_2)_{a_2a_3a_4}(0) 	&~\sim~  \ph{-}\frac{1}{4}\frac{(1+2k)(\JJ_2)_{a_2a_3a_4}(0)}{z^2} + \frac{(\JJ_1)_{a_2a_3a_4}(0)}{z}\\
(\JJ_1)_{a_2a_3a_4}(z) (\JJ_2)_{b_2b_3b_4}(0) 		&~\sim~  -\frac{1}{2}\frac{(1+2k)\epsilon_{a_2b_2}\epsilon_{a_3b_3}\epsilon_{a_4b_4}}{z^2}+\frac{-\frac{1}{2}((\JJ_2)_{a_2a_3a_4} (\JJ_2)_{b_2b_3b_4})(0) + \mf{J}_{a_2a_3a_4;b_2b_3b_4}(0)}{z}\\
(\JJ_2)_{a_2a_3a_4}(z) (\JJ_2)_{b_2b_3b_4}(0) 		&~\sim~ \ph{-}\frac{\epsilon_{a_2b_2}\epsilon_{a_3b_3}\epsilon_{a_4b_4}}{z}\\
(\JJ_1)_{a_2a_3a_4}(z) (\JJ_1)_{b_2b_3b_4}(0) 		&~\sim~ \ph{-}\frac{3}{4}\frac{(1+2k)\epsilon_{a_2b_2}\epsilon_{a_3b_3}\epsilon_{a_4b_4}}{z^3}+\frac{\frac{1}{4}(3+2k)((\JJ_2)_{a_2a_3a_4} (\JJ_2)_{b_2b_3b_4})(0) -\mf{J}_{a_2a_3a_4;b_2b_3b_4}(0)}{z^2}\\
&\qquad\qquad+\frac{\frac{1}{4}((\JJ_2)_{a_2a_3a_4} \partial(\JJ_2)_{b_2b_3b_4})(0)+\frac{1}{2}(1+k)(\partial(\JJ_2)_{a_2a_3a_4} (\JJ_2)_{b_2b_3b_4})(0)}{z}\\
&\qquad\qquad\qquad-\frac{1}{2}\frac{ \partial\mf{J}_{a_2a_3a_4;b_2b_3b_4}(0) }{z}~,
\end{split}\end{align}
}
where 
\begin{equation}
\mf{J}_{a_2a_3a_4;b_2b_3b_4}(z)=\JJ^{(2)}_{a_2b_2}(z)\epsilon_{a_3b_3}\epsilon_{a_4b_4} + \JJ^{(3)}_{a_3b_3}(z)\epsilon_{a_2b_2}\epsilon_{a_4b_4}+\JJ^{(4)}_{a_4b_4}(z)\epsilon_{a_2b_2}\epsilon_{a_3b_3}~,
\end{equation}
and we have removed $d$-exact terms.

We expect the result of this reduction procedure to be the trifundamental symplectic boson algebra $\goodchi[T_2]$, and $(\JJ_2)_{a_2a_3a_4}(z)$ has the correct dimension and OPE to be identified with the trifundamental generator $q_{a_2a_3a_4}$. In order to complete the argument, we need all of the remaining reduced generators to be expressible as composites of this basic generator. Indeed it turns out to be a straightforward exercise to compute the null states in the reduced algebra at dimensions $h=1,\frac32,2$ and to verify that null relations allow all the other generators to be written as normal ordered products of (derivatives of) $(\JJ_2)_{a_2a_3a_4}(z)$. For example, we should expect that the $\suf(2)$ affine currents should be equivalent to the bilinears currents of Eqn. \eqref{eq:T2_affine_currents}, and indeed there are null relations (only for $k=-2$) that allow us to declare such an equivalence,
\begin{align}\label{eq:null_relation_free_currents}
\begin{split}
\tfrac12(\JJ_2)_{abc}(\JJ_2)_{a'b'c'}\e^{bb'}\e^{cc'} &~=~ \JJ^{(2)}_{aa'}~,\\ 
\tfrac12(\JJ_2)_{abc}(\JJ_2)_{a'b'c'}\e^{aa'}\e^{cc'} &~=~ \JJ^{(3)}_{bb'}~,\\
\tfrac12(\JJ_2)_{abc}(\JJ_2)_{a'b'c'}\e^{aa'}\e^{bb'} &~=~ \JJ^{(4)}_{cc'}~,
\end{split}\end{align}
At dimensions $h=3/2$ and $h=2$ there are additional null states for our special value of the level,
\begin{align}
(\JJ_1)_{bcd}			&~=~ -\tfrac{3}{2}\partial (\JJ_2)_{bcd} + \tfrac{2}{3}(\JJ_2)_{(b_1(c_1(d_1}(\JJ_2)_{b_2)c_2)d_2)}(\JJ_2)_{b_3c_3d_3} \epsilon^{b_2b_3}\epsilon^{c_2c_3}\epsilon^{d_2d_3}~,\\
\begin{split}
\hat{\JJ}^{(1)}_{{11}}  &~=~ -\tfrac{3}{4} (\JJ_2)_{b_1c_1d_1} \partial (\JJ_2)_{b_2c_2d_2} \e^{b_1b_2}\e^{c_1c_2}\e^{d_1d_2}\\ 
						&~\ph{=}~ -\tfrac{1}{6}(\JJ_2)_{b_1c_1d_1}(\JJ_2)_{(b_2(c_2(d_2}(\JJ_2)_{b_3)c_3)d_3)}(\JJ_2)_{b_4c_4d_4}\e^{b_1b_2}\e^{c_1c_2}\e^{d_1d_2}\e^{b_3b_4}\e^{c_3c_4}\e^{d_3d_4}~.
\end{split}
\end{align}
Thus all of the additional generators are realized as composites of the basic field $(\JJ_2)_{abc}(z)$, and we have reproduced the $\goodchi[T_2]$ chiral algebra from qDS reduction of the $\sof(8)$ affine current algebra at level $k=-2$. We should re-emphasize that the redundancy amongst generators due to null states depends crucially on the precise value of the level. This is another instance of a general lesson that we have learned: the protected chiral algebras of $\NN=2$ SCFTs realize very special values of their central charges and levels at which nontrivial cancellations tend to take place. We will see more of this phenomenon in the next example.

\bigskip
\subsubsection*{Reducing $(\,\wh{\ef}_6\,)_{-3}$ to symplectic bosons}

In this case, our starting point is again an affine Lie algebra, this time $(\wh{\ef}_6)_{-3}$. Also we are again led to decompose the adjoint representation of $\ef_6$ under the maximal $\suf(3)_1\times\suf(3)_2\times\suf(3)_3$ subalgebra associated to the punctures on the UV curve as was done in \eqref{eq:e6_adjoint_decomposition}, leading to a basis of currents given by \eqref{eq:e6_decomposed_current_list} subject to singular OPEs given by Eqn.~\eqref{eq:e6_decomposed_current_OPEs}. Our aim is now to perform a partial reduction of the first puncture. Accordingly, we divide the generating currents as usual,
\begin{equation}
\label{eq:e6_generators_reduction_basis}
(J^{1})_{a}^{\ph{a}a'}~, \qquad\qquad \{\phi_i\} = \{(J^{2})_{b}^{\ph{b}b'}~, (J^{3})_{c}^{\ph{c}c'}~, W_{abc}~, \tilde W^{abc} \}~,
\end{equation}
where now $a,b,c$ are fundamental indices of $\suf(3)_{1,2,3}$, and the adjoint representation is represented by a fundamental and antifundamental index subject to a tracelessness condition.

The partial closing down to a minimal puncture is accomplished by means of the subregular embedding,
\begin{equation}
\Lambda(t_0) = \frac12(T_1^{\ph{1}1} - T_3^{\ph{1}3})~,\qquad
\Lambda(t_-) = T_3^{\ph{1}1}~,\qquad
\Lambda(t_+) = T_1^{\ph{1}3}~.
\end{equation}
The grading induced by the embedded Cartan turns out to be half-integral in this case and must therefore be supplanted by the integral $\delta$ grading. Under this grading the generators $\Lambda(t_-) = T_3^{\ph{1}1}$ and  $T_{3}^{\ph{1}2}$ are negative and of grade minus one. The relevant constraints are thus $\left(J^{1}\right)_3^{\ 1} = 1$ and $\left(J^{1}\right)_3^{\ 2} = 0.$ The implementation of these constraints via the BRST procedure introduces two ghost pairs $b_3^{\ 1}, c_1^{\ 3} $ and $b_3^{\ 2},c_2^{\ 3}.$

\begin{figure}[t!]
\centering
\includegraphics[scale=.5]{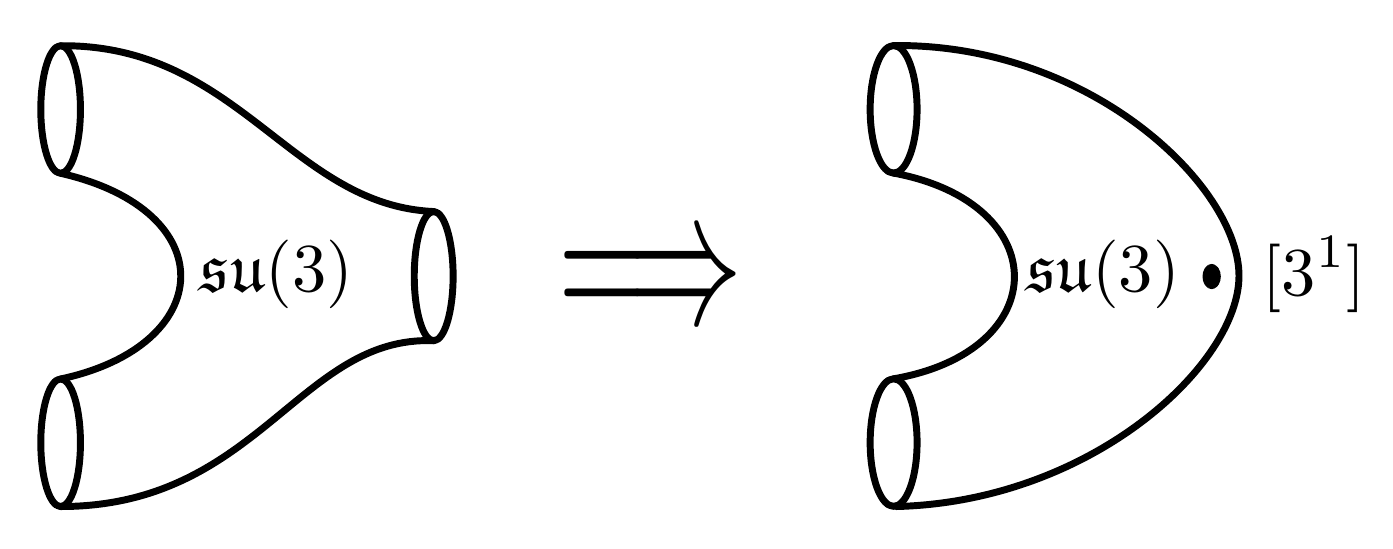}
\caption{Reduction from the $\ef_6$ theory to free hypermultiplets.}
\label{fig:3_to_hyper_reduction}
\end{figure}

In the reduction of $\goodchi[T_3],$ one finds that the currents $(\hat J^{1})_{\bar \alpha}$ such that  $T_{\bar{\alpha}}\in \ker(ad(\Lambda(t_+))),$ are given by $(\hat J^{1})_1^{\ 2},(\hat J^{1})_1^{\ 3},(\hat J^{1})_2^{\ 3},$ and the current generating the reduced $\uf(1)$ symmetry
\begin{equation}
\JJ_{\uf(1)} = (\hat J^{1})_1^{\ 1}-2(\hat J^{1})_2^{\ 2}+(\hat J^{1})_3^{\ 3}\;.
\end{equation}
Together with the additional generators in \eqref{eq:e6_generators_reduction_basis}, these constitute the generators of the cohomology at the level of vector spaces. The tic-tac-toe procedure produces honest chiral algebra generators, which we denote by the calligraphic version of the same letter as the vector space generator. The quantum numbers of these redundant generators are summarized in Table \ref{tab:T3_reduced_generators}. Their precise expressions can be found in Appendix \ref{subapp:e6_to_free}.
\begin{table}[ht!]
\begin{center}
\begin{tabular}{ c|c|c|c|c }
\hline\hline
~~Generator~~ & Dimension & $U(1)$ & $SU(3)_2$ & ~~$SU(3)_3$~~ \\
\hline
$\JJ_{\uf(1)} $ 					& $1$ 		&$\ph{-}0$ 		 	& $\bOn$ 		& $\bOn$	 	\\
$(\hat {\JJ}^{1})_{1\ph{\hf}}^{\ph{1}2}$ 	& $\frac32$ & $\ph{-}3$		 	& $\bOn$ 		& $\bOn$	 	\\
$(\hat {\JJ}^{1})_{1\ph{\hf}}^{\ph1 3} $ 	& $2$ 		& $\ph{-}0$ 	 	& $\bOn$ 		& $\bOn$		\\
$(\hat {\JJ}^{1})_{2\ph{\hf}}^{\ph1 3} $ 	& $\frac32$ & $-3$ 			 	& $\bOn$ 		& $\bOn$		\\
${\WW}_{1bc}$ 								& $\frac32$ & $\ph{-}1$     	& $\bTh$ 		& $\bTh$		\\
${\WW}_{2bc}$ 								& $1$ 		& $-2$	 	        & $\bTh$ 		& $\bTh$		\\
${\WW}_{3bc}$ 								& $\frac12$ & $\ph{-}1$	        & $\bTh$ 		& $\bTh$		\\
$\tilde{\WW}^{1bc}$ 						& $\frac12$ & $-1$ 	 	        & $\bar{\bTh}$ 	& $\bar{\bTh}$	\\
$\tilde{\WW}^{2bc}$ 						& $1$ 		& $\ph{-}2$	        & $\bar{\bTh}$ 	& $\bar{\bTh}$	\\
$\tilde{\WW}^{3bc}$ 						& $\frac32$ & $-1$	 	        & $\bar{\bTh}$ 	& $\bar{\bTh}$	\\
$(\JJ^{2})_{b}^{\ph{b}b'}$ 				& $1$		& $\ph{-}0$		 	& $\mathbf{8}$ 	& $\bOn$		\\
$(\JJ^{3})_{c}^{\ph{c}c'}$ 				& $1$		& $\ph{-}0$		 	& $\bOn$ 		& $\mathbf{8}$	\\
\end{tabular}
\end{center}
\caption{The quantum numbers of redundant generators of the reduced $T_3$ chiral algebra.\label{tab:T3_reduced_generators}}
\end{table}

Again, we see that there are dimension one half generators $(\WW_3)_{bc}=W_{3bc}$ and $(\tilde\WW^1)^{bc}=\tilde W^{1bc}$ that one naturally expects should be identified as the symplectic bosons of the reduced theory. Indeed, up to $d$-exact terms, the OPE for these generators is exactly what we expect from the desired symplectic bosons,
\begin{equation}
\label{eq:reduced_symplectic_boson_OPE}
(\WW_3)_{bc}(z) (\tilde{\WW}^1)^{b'c'}(0)\sim  \frac{\delta_{b}^{\ph{b}b'}\delta_{c}^{\ph{c}c'}}{z}~.
\end{equation}
These generators thus have the correct dimension, charges and OPE to be identified with the expected hypermultiplet generators. Again, by studying the null relations of the reduced chiral algebra at levels $h=1,\frac32,2$ one finds that precisely when the level $k=-3$, all of the higher dimensional generators in Table \ref{tab:T3_reduced_generators} are related to composites of $(\WW_3)_{bc}$ and $(\tilde {\WW}^1)^{bc}$ (see Appendix \ref{subapp:e6_to_free}). In particular, one can verify that the $\uf(1)\oplus \suf(3)_2 \oplus \suf(3)_3$ currents are equal to their usual free field expression modulo null states.

%% file: sections/Section_5/S5.tex

\section{Cylinders and Caps}
\label{sec:cyl_and_cap}

The procedure we have introduced for reducing punctures is sufficiently general that there is no obstacle to formally defining chiral algebras associated to unphysical curves such as the cylinder and (decorated) cap. These are unphysical curves from the point of view of class $\SS$ SCFTs, although they have a physical interpretation in terms of theories perturbed by irrelevant operators that correspond to assigning a finite area to the UV curve \cite{Gaiotto:2011xs}. It would be interesting to interpret the chiral algebras associated with these curves in terms of those constructions, although naively extrapolating away from conformal fixed points seems impossible. (There are other unphysical curves, such as a thrice-punctured sphere with two minimal punctures and one maximal puncture, and the chiral algebras for these can also be defined. We focus on cylinders and caps in this section as they are particularly natural objects in the TQFT.) 

The chiral algebra associated to a cylinder is a particularly natural object to consider from the TQFT perspective because it corresponds to the identity morphism (when taken with one ingoing and one outgoing leg). When taken with two ingoing or two outgoing legs, it is the chiral algebra avatar of the evaluation and coevaluation maps, respectively, of an ordinary two-dimensional TQFT. Similarly, the chiral algebra of the undecorated cap is the chiral algebra version of the trace map.

On the whole, we have not been able to systematically solve the BRST problem for these theories in the general case. This is because, as we shall see, the chiral algebras involve dimension zero (or negative dimension) operators, which prevent us from applying the simple algorithm set forth in Sec. \ref{sec:reducing}. Nevertheless, we are able to develop a compelling picture of the mechanics of the cylinder chiral algebra. It would be interesting from a purely vertex operator algebra point of view to construct these algebras rigorously.

\bigskip
\input{./sections/Section_5/S5_1}
\bigskip
\input{./sections/Section_5/S5_2}
\bigskip\bigskip

%% file: sections/Section_5/S5_1.tex

\subsection{The cylinder chiral algebra}
\label{subsec:cylinder}

\begin{figure}[t!]
\centering
\includegraphics[scale=.5]{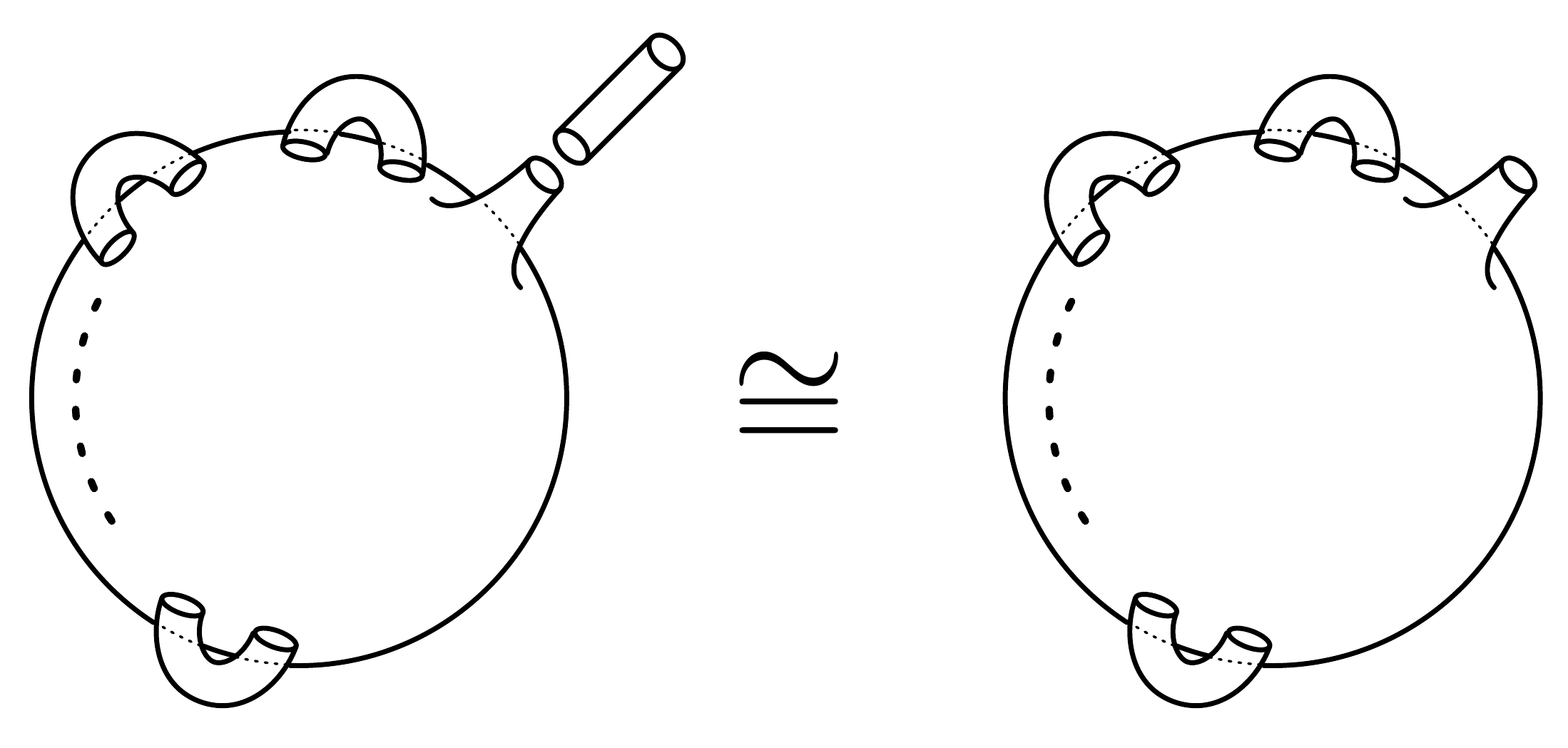}
\caption{Characteristic property of the identity morphism.}
\label{fig:tqft_identity}
\end{figure}

The chiral algebra associated to a cylinder should be obtained by performing a complete qDS reduction on one puncture of the trinion chiral algebra $\goodchi[T_n]$. In the generalized TQFT, the cylinder chiral algebra plays the role of the identity morphism for a single copy of the affine Lie algebra, ${\rm Id}:\wh{\suf(n)}_{-n}\mapsto\wh{\suf(n)}_{-n}$. The essential property associated with an identity morphism is illustrated in Figure \ref{fig:tqft_identity}.
As a statement about chiral algebras, the identity property is quite interesting. It means that the chiral algebra should have the property that when tensored with another class $\SS$ chiral algebra $\goodchi[\TT]$ along with the usual $(b,c)$ ghosts, restriction to the appropriate BRST cohomology produces a chiral algebra that is isomorphic to the original class $\SS$ chiral algebra,
\begin{equation}
H^*_{\rm BRST}\left(\psi\in\goodchi_{\rm Id}\otimes\goodchi[\TT]\otimes\goodchi_{bc}\,\restr{}{}\,b_0\psi=0\right)\cong\goodchi[\TT]~.
\end{equation}

As stated above, the qDS reduction problem in this case is substantially complicated by the fact that amongst the list of naive generators of the reduced chiral algebra, there will always be dimension zero currents. Consequently, a systematic solution of the BRST problem that removes redundancies from the list of generators is difficult even in the case of the $\goodchi[T_2]$ and $\goodchi[T_3]$ theories, for which the starting point of the reduction is known. A somewhat detailed analysis of the $\suf(3)$ case can be found in Appendix \ref{app:cylinders_and_caps}.

Although we don't have a general first principles solution, the general structure of the reduction and our intuition gained from other examples suggests a simple characterization of the cylinder chiral algebra. We state this here as a conjecture.
\begin{conj}[Cylinder chiral algebra]\label{conj_cylinder}
The chiral algebra associated to a cylinder of type $\suf(n)$ is finitely generated by an $\widehat{\suf(n)}_{-n}$ affine current algebra $\{(\JJ_{L})_A(z)$, $A=1,\ldots,n^2-1\}$, along with dimension zero currents $\{g_{ab}(z),~a,b=1,\ldots,n\}$ that are acted upon on the left by the affine currents. These dimension zero currents further obey a determinant condition $\det g=1$, \ie, they form a matrix that belongs to $SL(n,\Cb)$.
\end{conj}
This turns out to be a surprisingly interesting chiral algebra. Let us mention a few of its properties.

The first key property -- one which is not completely obvious from the description -- is that this chiral algebra actually has two commuting $\widehat{\suf(n)}_{-n}$ current algebras. The second set of affine currents are defined as follows
\begin{equation}\label{eqn: Rightcurrent_cylinder}
\left(\JJ_{R}\right)_c^{\ c^\prime}(z)\colonequals \left(\JJ_{L}\right)_{b^\prime}^{\ b}\  g_{b c}\  g^{b^\prime c^\prime} + n\left(g_{bc}\ \partial g^{bc^\prime} - \frac{1}{n} \delta_c^{c^\prime} g_{bd}\ \partial g^{bd} \right)~,
\end{equation}
where we have traded the adjoint index for a fundamental and antifundamental index satisfying a tracelessness condition, and we've also introduced the shorthand
\begin{equation}\label{eqn: determinantconstraint}
g^{ab}(z) = \frac{1}{n!}\epsilon^{a a_2\ldots a_n}\epsilon^{b b_2\ldots b_n} \left(g_{a_2b_2}\ldots g_{a_nb_n}\right)(z)~.
\end{equation}
Because of the determinant condition, this can be thought of as the inverse of $g_{ab}(z)$. The currents $(\JJ_R)_A(z)$ act on the dimension zero currents on the right. The $\JJ_R$ currents and the $\JJ_L$ currents have nonsingular OPE with one another, so they generate commuting affine symmetries. These are the symmetries associated with the two full punctures of the cylinder.

The key feature of this chiral algebra should be its behavior as the identity under gluing to other class $\SS$ chiral algebras. Let us thus consider a chiral algebra associated to a UV curve $\CC_{g,s\geq1}$ with at least one maximal puncture. Let us consider a general operator in this theory which will take the form  $X_{a_1 a_2\ldots a_p}^{b_1b_2\ldots b_q}$, with $p$ fundamental indices and $q$ antifundamental indices (possibly subject to (anti)symmetrizations and tracelessness conditions) of the flavor symmetry associated to the maximal puncture and with its transformation properties under other flavor symmetries suppressed. Then our expectations is that after gluing in the cylinder, there will be a new operator of the same dimension of the same form, but where its transformation under the symmetry of the original maximal puncture has been replaced with a transformation under the symmetry at the unglued end of the cylinder.

We can see how this might come about. Gluing a cylinder to the maximal puncture means tensoring the original chiral algebra with the chiral algebra of conjecture \ref{conj_cylinder} in addition to the usual adjoint $(b,c)$ system of dimensions $(1,0)$. We then restrict ourselves to the BRST cohomology (relative to the $b$-ghost zero modes) of the nilpotent operator
\begin{equation}\label{eqn: BRSTcurrentcylinder}
Q_{\rm BRST} = \oint dz\,c^{A} ((\JJ_L)_{A} + J^{\TT}_{A} + \frac12 J^{\text{gh}}_{A})~,
\end{equation}
where $J^{\TT}_{A}$ is the current for the symmetry associated to the puncture on $\CC_{g,s\geq1}$ that is being glued. Our original operator, which was charged under the $\suf(n)$ that is being gauged and therefore does not survive the passage to BRST cohomology, has a related \emph{transferred operator} of the following form
\begin{equation}
\hat X^{c_1 c_2\ldots c_p}_{d_1d_2\ldots d_q} =  X_{a_1 a_2\ldots a_p}^{b_1b_2\ldots b_q}\  g^{a_1c_1}\ g^{a_2c_2}\ldots g^{a_pc_p}\  g_{b_1d_1}\ g_{b_2d_2}\ldots g_{b_qd_q}~.
\end{equation}
This operator \emph{is} gauge invariant, since the gauged symmetry acts on $g_{ab}, g^{ab}$ on the left. In this sense the $g_{ab}$ fields effectively transfer and conjugate the symmetry from one end of the cylinder to the other. Notice that the transferred operators have the same dimension as before, because the $g_{ab}(z)$ have dimension zero. What's more, by virtue of the unit determinant condition on $g_{ab}$, we see that the OPE of the transferred fields ends up being exactly the conjugate of the OPE of the original fields. It therefore seems likely that we recover precisely the same chiral algebra on the other end of the cylinder (up to conjugation of $\suf(n)$ representations). Of course, for this construction to work we have to assume that the spectrum of physical operators will consist only of the transferred operators. It would be interesting to prove this conjecture.

Finally, one can't help but notice the similarities between this description of the cylinder chiral algebra and the discussions of \cite{Moore:2011ee} regarding the holomorphic symplectic manifold associated with the cylinder in the Higgs branch TQFT. In that work, the hyperk\"ahler manifold $T^* G_{\Cb}$ was associated to the cylinder. It is interesting to note that the chiral algebra we have described in Conjecture \ref{conj_cylinder} seems to be precisely what one obtains from studying the half-twisted $(0,2)$ supersymmetric sigma model on $G_{\Cb}$ \cite{Witten:2005px,Kapustin:2005pt}. Alternatively, it describes the global sections of the sheaf of chiral differential operators on $G_{\Cb}$ as defined in \cite{Malikov:cdr1,Malikov:cdr2,Malikov:gerbes1,Malikov:gerbes2,Malikov:gerbes3}. This connection is exciting, but remains mostly mysterious to the authors at present.

%% file: sections/Section_5/S5_2.tex

\subsection{The (decorated) cap chiral algebra}
\label{subsec:decorated_cap}

The chiral algebra associated to a decorated cap can be defined by partially reducing one puncture of the cylinder chiral algebra. The resulting chiral algebra should have the interesting property that if you glue it to another class $\SS$ chiral algebra using the standard gauging BRST construction, it effectively performs the appropriate qDS reduction on the original chiral algebra.

In trying to characterize this chiral algebra, one immediately encounters the problem that it includes operators of negative dimension. Namely, consider the first steps of the general reduction procedure as applied to the cylinder chiral algebra. The (potentially redundant) generators for the decorated cap labeled by an embedding $\Lambda$ include the usual currents $\hat J_{\bar\alpha}$ for $T_{\bar\alpha} \in \ker( ad_{{\Lambda}(t_+)})$, the dimensions of which are shifted by their ${\Lambda}(t_0)$ weight. However, there are additional generators coming from the dimension zero bifundamental fields $g_{ab}$ of the cylinder theory. In terms of the reduced symmetry associated with the decoration, these fields are reorganized as follows: for each irrep of $\suf(2)$ in the decomposition \eqref{eq:generaldecomposition} of the fundamental representation there are $2j+1$ generators transforming in representation $\ff\otimes\RR_j^{(\ff)}$ with dimensions $-j,-j+1,\ldots,j$. The dimension zero null relation corresponding to the determinant condition in the cylinder theory of the cylinder theory is expected to descend to the cap theory. The superconformal index (see App. \ref{subapp:cylinder_cap_index}) supports this expectation, and further suggests that there may be no additional redundancies.

The existence of negative dimension operators makes this a rather exotic chiral algebra, and we will not explore it much further. Nevertheless, let us offer a couple of brief comments. In the description of the cap chiral algebra given in the previous paragraph, it is not immediately clear that an affine current algebra associated to the maximal puncture survives. However, one finds that the necessary dimension one currents can be constructed using the above fields in a manner similar to \eqref{eqn: Rightcurrent_cylinder}, using only those elements of the left current algebra that survive in the cap chiral algebra. When gluing the cap to another theory $\TT$, this current algebra will enter in the BRST current \eqref{eqn: BRSTcurrentcylinder}. As in the case of the cylinder, the Gauss law constraint can be solved by constructing transferred fields, which thanks to nonzero conformal dimension of the various components of $g_{ab}$ end up with their dimensions shifted correctly. It remains to verify that restricting to the BRST cohomology removes the transferred versions of the currents $J^{\TT}_{A}$ for $T_{A} \not\in \ker( ad_{{\Lambda}(t_+)})$.

%% file: sections/acknowledge.tex

\acknowledgments
The authors have greatly benefited from discussions with Tudor Dimofte, Davide Gaiotto, Andy Neitzke, David Simmons-Duffin, and Edward Witten.
We would also like to express our particular gratitude to Shlomo Razamat and Yuji Tachikawa for many helpful conversations.
The work of C.B. is supported in part by NSF grant PHY-1314311.
L. R. gratefully acknowledges the generous support of the Simons Foundation and of the Solomon Guggenheim Foundation
and the wonderful hospitality of the IAS, Princeton and of the KITP, Santa Barbara during his sabbatical leave. 
The work of W.P. and L.R. is supported in part by NSF Grant PHY-1316617.

%% file: appendices/app_A.tex
\section{Details for rank two theories}
\label{app:level_by_level}

This appendix includes details regarding a number of calculations having to do with operations on the $\goodchi[T_3]$ chiral algebra described in Sec.~\ref{subsubsec:t3_chiral_algebra}. For all of these calculations, it is useful to have the realization of the $\goodchi[T_3]$ chiral algebra, which coincides with the affine $\ef_6$ current algebra at level $k=-3$, in the basis relevant for class $\SS$ given in Eqn. \eqref{eq:e6_decomposed_current_list}. In this basis, the singular OPEs are as follows,
\begin{equation}
\label{eq:e6_decomposed_current_OPEs}
\begin{split}
(J^{1})_{a}^{\,a'}\!(z)(J^{1})_{\tilde a}^{\,\tilde a'}\!(0)&~\sim~ 
\frac{ k (\delta_{a}^{\,\tilde a'}\delta_{\tilde a}^{\,a'} 
- \frac{1}{3} \delta_{a}^{\,a'}\delta_{\tilde a}^{\,\tilde a'} )}{z^2} 
+ \frac{ \delta_{\tilde a}^{\,a'}\ (J^{1})_{a}^{\,\tilde a'} -\delta_{a}^{\,\tilde a'}\ (J^{1})_{\tilde a}^{\,a'}}{z}~,\\
(J^{1})_{a}^{\,a'}\!(z)W_{a''bc}(0) &~\sim~ \ph{-}\frac{1}{z}\left(\delta_{a''}^{\,a'} W_{abc} -\frac{1}{3} \delta_{a}^{\,a'} W_{a''bc}\right)~,\\
(J^{1})_{a}^{\,a'}\!(z)\wt W^{a''bc}(0) &~\sim~ -\frac{1}{z}\left(\delta_{a}^{\,a''} \wt W^{a'bc} -\frac{1}{3} \delta_{a}^{\,a'} \wt W^{a''bc}\right)~,\\
W_{abc}(z)W_{a'b'c'}(0) &~\sim~ -\frac{1}{z}\epsilon_{aa'a''}\epsilon_{bb'b''}\epsilon_{cc'c''}  \wt W^{a''b''c''}~,\\
\wt W^{abc}(z)\wt W^{a'b'c'}(0) &~\sim~ \frac{1}{z}\epsilon^{aa'a''}\epsilon^{bb'b''}\epsilon^{cc'c''} W_{a''b''c''}~,\\
W_{abc}(z)\wt W^{a'b'c'}(0) &~\sim~ \frac{ k}{z^2} \delta_{a}^{\,a'}\delta_{b}^{\,b'}\delta_{c}^{\,c'} +\frac{1}{z}\left((J^{1})_{a}^{\,a'}\delta_{b}^{\,b'}\delta_{c}^{\,c'} + \delta_{a}^{\,a'}(J^{2})_{b}^{\,b'}\delta_{c}^{\,c'} + \delta_{a}^{\,a'}\delta_{b}^{\,b'}(J^{3})_{c}^{\,c'}\right)~,
\end{split}
\end{equation}
and similarly for $(J^{2})$ and $(J^{3})$.

We saw illustrated in Table \ref{Tab:T3_index} that there exists for $k=-3$ a null relation in the ${\bf 650}$-dimensional representation of $\ef_6$. This is in agreement with the Higgs branch relation of Eqn. \eqref{eq: e6 nullprediction in 650}. It will prove useful to have the explicit expression for the components of this null vector upon decomposition in terms of $\oplus_{I=1}^3 \suf(3)_I$ representations. Group theoretically, the decomposition in question is given by
\begin{multline*}
\mathbf{650} ~\rightarrow~ 2\times(\mathbf{1,1,1}) + (\mathbf{8,1,1})+(\mathbf{1,8,1})+(\mathbf{1,1,8})+(\mathbf{8,8,1})+(\mathbf{1,8,8})+(\mathbf{8,1,8})+\\2\times(\mathbf{3,3,3})+2\times(\mathbf{\bar{3},\bar{3},\bar{3}})+(\mathbf{6,\bar{3},\bar{3}})+(\mathbf{\bar{6},3,3})+(\mathbf{\bar{3},6,\bar{3}})+(\mathbf{3,\bar{6},3})+(\mathbf{\bar{3},\bar{3},6})+(\mathbf{3,3,\bar{6}})~.
\end{multline*} 
The corresponding null vectors arise from the relations summarized in Table \ref{tab:relation_reductions}.

\renewcommand{\arraystretch}{1.75}
\begin{table}[t!]
\centering
\begin{tabular}{|l|p{10cm}|}
\hline\hline
Representation & Null relation \\
\hline
$(\mathbf{1,1,1})$ & $(J^{1})_{a_1}^{\ a_2} (J^{1})_{a_2}^{\ a_1}= (J^{2})_{b_1}^{\ b_2} (J^{2})_{b_2}^{\ b_1} = (J^{3})_{c_1}^{\ c_2} (J^{3})_{c_2}^{\ c_1}$\\
\hline 
$(\mathbf{8,1,1})+(\mathbf{1,8,1})+(\mathbf{1,1,8})$ & $\frac{1}{3} \left(W_{a_1bc}\tilde W^{a_2bc}-\frac{1}{3}\delta_{a_1}^{a_2}W_{abc}\tilde W^{abc} \right) +\newline{}\qquad\qquad~ \left((J^{1})_{a_1}^{\ a} (J^{1})_{a}^{\ a_2}- \frac{1}{3}\delta_{a_1}^{a_2} (J^{1})_{\alpha}^{\ a}(J^{1})_{a}^{\ \alpha} \right)-3 \partial (J^{1})_{a_1}^{\ a_2}=0$\\
\hline
$(\mathbf{8,8,1})+(\mathbf{1,8,8})+(\mathbf{8,1,8})$ & $W_{a_1b_1c}\tilde W^{a_2b_2c} - \frac{1}{3}\delta_{a_1}^{a_2}W_{ab_1c}\tilde W^{ab_2c} - \frac{1}{3}\delta_{b_1}^{b_2}\left(W_{a_1bc}\tilde W^{a_2bc}\right)+\newline
{}\qquad\qquad\qquad\qquad\quad\frac{1}{9}\delta_{a_1}^{a_2}\delta_{b_1}^{b_2}\left(W_{abc}\tilde W^{abc}\right) + (J^{1})_{a_1}^{\ a_2} (J^{2})_{b_1}^{\ b_2}=0$\\
\hline
$(\mathbf{3,3,3})$ & 
$(J^{1})_{a}^{\ \alpha}W_{\alpha bc} = (J^{2})_{b}^{\ \beta}W_{a \beta c} = (J^{3})_{c}^{\ \gamma}W_{a b\gamma}$\\
\hline
$(\mathbf{\bar{3},\bar{3},\bar{3}})$ &
$(J^{1})_{\alpha}^{\ a}\tilde W^{\alpha bc} = (J^{2})_{\beta}^{\ b}\tilde W^{a \beta c} = (J^{3})_{\gamma}^{\ c}\tilde W^{a bc}$\\
\hline
$(\mathbf{\bar{6},3,3})+(\mathbf{3,\bar{6},3})+(\mathbf{3,3,\bar{6}})$ &
$2(J^{1})_{\alpha_1}^{\ (a_1|} W_{\alpha_2bc}\epsilon^{\alpha_1\alpha_2|a_2)} + \tilde W^{(a_1b_1c_1}\tilde W^{a_2)b_2c_2} \epsilon_{bb_1b_2}\epsilon_{cc_1c_2}=0$\\
\hline
$(\mathbf{6,\bar{3},\bar{3}})+(\mathbf{\bar{3},6,\bar{3}})+(\mathbf{\bar{3},\bar{3},6})$ &
$2(J^{1})_{(a_1|}^{\ \ \alpha_1} \tilde W^{\alpha_2bc}\epsilon_{\alpha_1\alpha_2|a_2)} +  W_{(a_1b_1c_1} W_{a_2)b_2c_2} \epsilon^{bb_1b_2}\epsilon^{cc_1c_2}=0$\\
\hline
\end{tabular}
\caption{\label{tab:relation_reductions}Null state relations at level two in the $\goodchi[T_3]$ chiral algebra.} 
\end{table}

\input{./appendices/app_A1}
\bigskip
\input{./appendices/app_A2}
\bigskip

%% file: appendices/app_A1.tex

\subsection{Argyres-Seiberg duality}
\label{subapp:Argyres_Seiberg}

First we describe in detail the check of Argyres-Seiberg duality at the level of chiral algebras described in Section \ref{subsubsec:t3_chiral_algebra}. The first duality frame is that of SQCD, the chiral algebra of which was described in \cite{Beem:2013sza}. There the generators of the chiral algebra were found to include a $\widehat{\suf(6)}_{-3}\times\widehat{\uf(1)}$ affine current algebra with currents $J_i^j$ and $J$, along with baryonic and anti-baryonic operators $\{b_{ijk},\tilde b^{ijk}\}$ of dimension $\Delta=\frac32$. The singular OPEs for these generators are as follows,
{\allowdisplaybreaks
\begin{equation}
\label{eq:scqcdopes}
\begin{alignedat}{3}
&& J_i^j(z) J_k^l(0) 	                      &~~\sim~&       &-\frac{3 (\delta_i^l \delta^j_k - \text{trace})}{z^2} ~+~ \frac{\delta_k^j J_i^l(z)-\delta_i^l J_k^j(z)}{z}~,\\
&& J(z) J(0) 			                          &~~\sim~&       &-\frac{18}{z^2}~,\\
&& J_i^j(z) b_{k_1 k_2 k_3}(0)              &~~\sim~&       &\ph{-}\frac{3 \delta_{[k_1|}^j b_{i |k_2 k_3]}(0) - \hf \delta_i^j b_{k_1 k_2 k_3}(0)}{z}~,\\
&& J(z) b_{k_1 k_2 k_3}(0)                  &~~\sim~&       &\ph{-}\frac{3b_{k_1 k_2 k_3}(0)}{z}~,\\
&& J(z) b^{k_1 k_2 k_3}(0)                  &~~\sim~&       &-\frac{3b^{k_1 k_2 k_3}(0)}{z}~,\\
&& b_{i_1i_2i_3}(z)\tilde b^{j_1j_2j_3}(0)  &~~\sim~&       &\ph{-}\frac{36\,\delta_{[i_1}^{[j_1} \delta_{\ph{[}\!i_2}^{\ph{[}\!j_2} \delta_{i_3]}^{j_3]}}{z^3} - \frac{36\,  \delta_{[i_1}^{[j_1} \delta_{\ph{[}\!i_2}^{\ph{[}\!j_2}\hat J_{i_3]}^{j_3]}(0)}{z^2} \\
&&                                          &~~\ph{\sim}~&  &\ph{-}\qquad +\frac{18\, 	\delta_{[i_1}^{[j_1} 	  \hat J_{\ph{[}\!i_2}^{\ph{[}\!j_2} 	 \hat J_{i_3]}^{j_3]}(0) -  18\,\delta_{[i_1}^{[j_1} \delta_{\ph{[}\!i_2}^{\ph{[}\!j_2} \del \hat J_{i_3]}^{j_3]}(0)}{z}~.
\end{alignedat}
\end{equation}}%
Antisymmetrizations are performed with weight one, and lower (upper) indices $i,j,\ldots$ transform in the fundamental (antifundamental) representation of $\suf(6)$. In the last line we have introduced the $\uf(6)$ current $\hat J^i_j \colonequals J^i_j + \frac{1}{6} \delta^i_j J$. 
It was conjectured in \cite{Beem:2013sza} that the SQCD chiral algebra is a $\WW$-algebra with just these generators. This proposal passed a few simple checks. All the generators of the Hall-Littlewood chiral ring have been accounted for and the OPE closes. There is no additional stress tensor as a generator because the Sugawara stress tensor of the $\uf(6)$ current algebra turns out to do the job (this again implies a relation in the Higgs branch chiral ring of SQCD). The spectrum of the chiral algebra generated by these operators also correctly reproduces the low-order expansion of the superconformal index.

Our aim in the remainder of this appendix is to reproduce this chiral algebra from the `exceptional side' of the duality using our proposal that the chiral algebra $\goodchi[T_3]$ is the current algebra $(\,\widehat{\ef}_6\,)_{-3}$. The two free hypermultiplets contribute symplectic bosons $q_{\alpha}$ and $\tilde q^{\alpha}$ with $\alpha=1,2$ with singular OPE given by
\begin{equation}
\label{eq:app_symplectic_boson_OPE}
q_\alpha(z) \tilde q^\beta (0) \sim  \frac{\delta_\a^\b}{z}~.
\end{equation}
The $\goodchi[T_3]$ chiral algebra should be re-expressed in terms of an $\suf(6)\times\suf(2)$ maximal subalgebra, in terms of which the affine currents split as
\begin{equation}
\label{eq:e6_as_relabeling}
\{J_{A=1,\ldots,78}\} ~~\Longrightarrow~~ \{X^i_j,~ Y^{[ijk]}_\a,~ Z_\a^\b\}~.
\end{equation}
The operators $X^i_j$ and $Z_\a^\b$ are the affine currents of $\suf(6)$ and $\suf(2)$, respectively, with $X^i_i = Z_\a^\a = 0$. The additional operators $Y^{ijk}_\a$ transform in the $(\mathbf{20},\mathbf{2})$ of $\suf(6)\times\suf(2)$. The nonvanishing OPEs amongst these operators are simply a rewriting of the $\wh{\ef}_6$ current algebra,
\begin{eqnarray}
\label{eq:e6_ope_decomposition}
X_i^j(z) X_k^l(0) &~\sim~& - \frac{3 (\delta_i^l \delta^j_k - \text{trace})}{z^2} + \frac{\delta_k^j X_i^l(0) - \delta_i^l X_k^j(0)}{z} \nn\\
Z_\a^\b(z) Z_\g^\delta(0) &~\sim~& - \frac{3 (\delta_\a^\delta \delta^\b_\g - \text{trace})}{z^2} + \frac{\delta_\g^\b Z_\a^\d(0) - \delta_\a^\delta Z_\g^\b(0)}{z}\nn\\
X_i^j(z) Y^{klm}_\a (0) &~\sim~& - \frac{3  \delta_i^{[k} Y_\a^{lm]j}(0)}{z} - \text{trace}\\
Z_\a^\b(z) Y^{ijk}_\g (0) &~\sim~& \ph{-}\frac{\delta_\g^\b Y^{ijk}_{\a} (0)}{z} - \text{trace}\nn\\
Y^{ijk}_\a (z) Y^{lmn}_\b (0) &~\sim~& \ph{-}\frac{\e_{\a \b} \epsilon^{ijklmn}}{z^2} +\frac{\e^{ijklmn} \e_{\a \g} Z^\g_\b(0) - 3 \e_{\a \b} \e^{[ijklm|p} X^{|n]}_p(0)}{z}~.\nn
\end{eqnarray}
Gluing introduces a dimension $(1,0)$ ghost system in the adjoint of $\suf(2)$ and restricting to the appropriate cohomology of the following BRST operator,
\begin{equation}
\label{eq:AS_BRST}
J_{\rm BRST}=~c^\alpha_\beta (Z_\alpha^\beta - q_\alpha \tilde q^\beta) -\frac{1}{2} (\delta_{\alpha_1}^{\alpha_6}\delta_{\alpha_3}^{\alpha_2}\delta_{\alpha_5}^{\alpha_4}-\delta_{\alpha_1}^{\alpha_4}\delta_{\alpha_3}^{\alpha_6}\delta_{\alpha_5}^{\alpha_2}) c_{\alpha_2}^{\alpha_1}b_{\alpha_4}^{\alpha_3}c_{\alpha_6}^{\alpha_5}.
\end{equation}
The cohomology can be constructed level by level using the \texttt{OPEdefs} package for \texttt{Mathematica} \cite{Thielemans:1991uw}. Up to dimension $h=\frac32$, we find the following operators,
\begin{equation}
\label{eq:AS_generators}
X^i_j~,\qquad
q_\a\tilde q^\a~,\qquad
\e_{ijklmn}\tilde q^\a Y^{lmn}_\a~,\qquad
\e^{\a\b}q_\a Y_\b^{ijk}~.
\end{equation}
Up to normalizations, these can naturally be identified with the generators of the SQCD chiral algebra,
\begin{equation}
\label{eq:argyresseibergmatch}
X_i^j \simeq J_i^j~, \qquad 
3 q_\a \tilde q^\a \simeq J~, \qquad 
\frac{1}{6} \e_{ijk l m n} \tilde q^\a Y^{l m n}_\a \simeq b_{ijk}~, \qquad
 \e^{\a \b} q_\a Y_\b^{i j k} \simeq \tilde b^{ijk}~.
\end{equation}
The equations relating chiral algebra generators in the two duality frames are the same as the ones obtained in \cite{Gaiotto:2008nz}, with the operators being viewed as generators of the Higgs branch chiral ring. In that work, establishing them at the level of the Higgs branch required a detailed understanding of the chiral ring relations on both sides. By contrast, to establish equivalence of the chiral algebras we need to check that the above operators have the same singular OPEs. Relations in the chiral ring will then show up automatically as null states.

With the \texttt{OPEdefs} package we have also computed the OPEs of the composite operators in \eqref{eq:AS_generators} and found perfect agreement with \eqref{eq:scqcdopes}. Most of the OPEs are reproduced in a fairly trivial fashion. However, the simple pole in the baryon-antibaryon OPE can only be matched by realizing that there is a null state at level two in the $(\,\widehat{\ef}_6\,)_{-3}$ algebra given by
\begin{equation}
\label{eq:t3nullstate}
Y^{ijk}_\a Y^{abc}_\b \e^{\a \b} \e_{abc l m n} + 108 \del X^{[i}_{[l} \delta^j_m \delta^{k]}_{n]} + 108 X^{[i}_{[l} X^j_m \delta^{k]}_{n]} + \frac{1}{72} Z_\a^\b Z_\b^\a \delta^{[i}_{[l} \delta^j_m \delta^{k]}_{n]}~.
\end{equation}

Thus we have shown that using our proposal for the $\goodchi[T_3]$ chiral algebra in the Argyres-Seiberg duality problem, one at least produces a self-contained $\WW$-algebra that matches between the two sides of the duality. It would be nice to prove that this $\WW$-algebra is the full chiral algebra. Indeed, if one could demonstrate this fact for the SQCD side of the duality, it seems likely that it wouldn't be too hard to prove that there can be no additional generators in the $\goodchi[T_3]$ chiral algebra beyond the affine currents.

%% file: appendices/app_A2.tex

\subsection{Reduction of \texorpdfstring{$T_3$}{T3} to free hypermultiplets}
\label{subapp:e6_to_free}

In this appendix we provide some details about the reduction of the $\goodchi[T_3]$ chiral algebra to free symplectic bosons. This corresponds to the subregular embedding $\suf(2)\hookrightarrow\suf(3)$, which is given by
\begin{equation}
\Lambda(t_0) = \frac12(T_1^{\ph{1}1} - T_3^{\ph{1}3})~,\qquad
\Lambda(t_-) = T_3^{\ph{1}1}~,\qquad
\Lambda(t_+) = T_1^{\ph{1}3}~.
\end{equation}
The grading on the Lie algebra by the Cartan element $\Lambda(t_0)$ is half-integral. In order to arrive at first-class constraints, we introduce a different Cartan element $\delta$ that gives an integral grading. More specifically, we have $\delta = \frac13 (T_1^{\ 1} + T_2^{\ 2} - 2T_3^{\ 3})$. With respect to the $\delta$-grading there are two negatively graded currents and we consequently impose the constraints $\left(J^{1}\right)_3^{\ 1} = 1$ and $\left(J^{1}\right)_3^{\ 2} = 0$. These are implemented via a BRST procedure with differential given by
\begin{equation}
d(z) = \left(\left((J^{1})_3^{\ 1} -1 \right) c_1^{\ 3}  + (J^{1})_3^{\ 2} c_2^{\ 3}\right)(z)~,
\end{equation}
where the ghost pairs $b_3^{\ 1}, c_1^{\ 3} $ and $b_3^{\ 2},c_2^{\ 3}$ have the usual singular OPEs.

Implementing the first step of the qDS procedure, one obtains the (redundant) generators of the chiral algebra at the level of vector spaces. Applying the tic-tac-toe procedure to produce genuine chiral algebra generators, we obtain the set of generators that were listed in Table \ref{tab:T3_reduced_generators}. The explicit forms of these generators are given as follows,
{\allowdisplaybreaks
\begin{align}
\label{eq:reduced_generators_def}
\JJ_{\uf(1)} &~\colonequals~  (\hat J^{1})_1^{\ 1}-2 (\hat J^{1})_2^{\ 2}+(\hat J^{1})_3^{\ 3}\nn\\
(\hat {\JJ}^{1})_1^{\ 2} &~\colonequals~ (\hat J^{1})_1^{\ 2}\nn\\
(\hat  {\JJ}^{1})_1^{\ 3} &~\colonequals~ (\hat J^{1})_1^{\ 3} -\left( -(k+1)\partial (\hat J^{1})_3^{\ 3} +(\hat J^{1})_1^{\ 1} (\hat J^{1})_3^{\ 3} -(\hat J^{1})_2^{\ 1}(\hat J^{1})_1^{\ 2}\right)\nn\\
(\hat  {\JJ}^{1})_2^{\ 3} &~\colonequals~ (\hat J^{1})_2^{\ 3} -\left( (k+2)\partial (\hat J^{1})_2^{\ 1} +(\hat J^{1})_3^{\ 3} (\hat J^{1})_2^{\ 1} -(\hat J^{1})_2^{\ 2}(\hat J^{1})_2^{\ 1}\right)\nn\\
{\WW}_{1bc} &~\colonequals~ W_{1bc} - W_{3bc} (\hat J^{1})_1^{\ 1}\nn\\
{\WW}_{2bc} &~\colonequals~ W_{2bc} - W_{3bc}(\hat J^{1})_2^{\ 1}\\
{\WW}_{3bc} &~\colonequals~ W_{3bc}\nn\\
\tilde  {\WW}^{1bc} &~\colonequals~ \tilde W^{1bc}\nn\\
\tilde  {\WW}^{2bc} &~\colonequals~ \tilde W^{2bc}\nn\\
\tilde  {\WW}^{3bc} &~\colonequals~ \tilde W^{3bc} - \left(- \tilde W^{1bc} (\hat J^{1})_1^{\ 1} - \tilde W^{2bc} (\hat J^{1})_2^{\ 1} \right)\nn\\
(\JJ^{2})_{b_1}^{\ b_2}&~\colonequals~( J^{2})_{b_1}^{\ b_2}\nn\\
(\JJ^{3})_{c_1}^{\ c_2}&~\colonequals~( J^{3})_{c_1}^{\ c_2}~\nn.
\end{align}
}

The generators $\WW_{3bc}$ and $\tilde\WW^{1bc}$ have the correct charges and mutual OPE to be identified as the expected symplectic bosons. It follows that the reduction argument will be complete if we can show that at the specific value of the level of interest $k=-3$, all the other generators listed in Eqn. \eqref{eq:reduced_generators_def} participate in a null state condition that allows them to be equated with composites of  $\WW_{3bc}$ and $\tilde\WW^{1bc}$. 

Indeed, we do find such relations to account for all additional generators. At level $h=1$, we find
\begin{align}
\JJ_{\uf(1)}  			&~=~ - {\WW}_{3bc} \tilde {\WW}^{1bc}~,\label{eq:nullJU1}\\ 
(\JJ^{2})_{b_1}^{\ b_2} &~=~ -\left({\WW}_{3b_1c} \tilde {\WW}^{1b_2c} - \frac13 \delta_{b_1}^{b_2} {\WW}_{3bc} \tilde {\WW}^{1bc} \right)~,\label{eq:nullJSU3_2}\\
(\JJ^{3})_{c_1}^{\ c_2} &~=~ -\left({\WW}_{3bc_1} \tilde {\WW}^{1bc_2} - \frac13 \delta_{c_1}^{c_2} {\WW}_{3bc} \tilde {\WW}^{1bc} \right)~,\label{eq:nullJSU3_3}\\
\tilde{\WW}_{2bc} 		&~=~ \frac{1}{2} \epsilon_{b b_1b_2} \epsilon_{c c_1c_2} \tilde {\WW}^{1b_1c_1} \tilde {\WW}^{1b_2c_2}~,\label{eq:nullWlower}\\
\tilde{\WW}^{2bc} 		&~=~ -\frac{1}{2} \epsilon^{b b_1b_2} \epsilon^{c c_1c_2}  {\WW}_{3b_1c_1}  {\WW}_{3b_2c_2}~.\label{eq:nullWupper}
\end{align}
At dimension $h=3/2$, one can find the null relations
\small
\begin{align}
(\hat{\JJ}^{1})_1^{\ 2} &~=~ \frac16\WW_{3b_1c_1}\WW_{3b_2c_2}\WW_{3b_3c_3}\epsilon^{b_1b_2b_3}\epsilon^{c_1c_2c_3}~,\\
(\hat{\JJ}^{1})_2^{\ 3} &~=~ -\frac16\tilde{\WW}^{1b_1c_1}\tilde{\WW}^{1b_2c_2}\tilde{\WW}^{1b_3c_3}\epsilon_{b_1b_2b_3}\epsilon_{c_1c_2c_3}~,\\
\WW_{1bc}				&~=~ 2\partial\WW_{3bc} + \frac{5}{12}\WW_{3b_1c_1}\WW_{3b_2c_2}\tilde{\WW}^{1b_3c_3}\epsilon^{\beta b_1b_2}\epsilon^{\gamma c_1c_2}\epsilon_{\beta b b_3}\epsilon_{\gamma c c_3} -\frac13 \WW_{3(b(c}\WW_{3b_1)c_1)}\tilde{\WW}^{1b_1c_1}~,\\
\WW^{3bc}				&~=~- \partial\tilde{\WW}^{1bc} + \frac13 \tilde{\WW}^{1b_1c_1}\tilde{\WW}^{1b_2c_2}{\WW}_{3b_3c_3}\epsilon_{\beta b_1b_2}\epsilon_{\gamma c_1c_2}\epsilon^{\beta b b_3}\epsilon^{\gamma c c_3} -\frac{2}{3}\tilde{\WW}^{1(b(c}\tilde{\WW}^{1b_1)c_1)}{\WW}_{3b_1c_1}~.
\end{align}
\normalsize
Finally, at dimension $h=2$, we find
\small
\begin{multline}
(\hat{\JJ}^{1})_1^{\ 3} = \frac{14}{9}\WW_{3bc}\partial\tilde{\WW}^{1bc}
-\frac{8}{9}\partial\WW_{3bc}\tilde{\WW}^{1bc}
+\frac{2}{9}\WW_{3(b_1(c_1}\WW_{3b_2)c_2)}\tilde{\WW}^{1(b_1(c_1}\tilde{\WW}^{1b_2)c_2)}\\
-\frac{7}{36}\WW_{3b_1c_1}\WW_{3b_2c_2}\tilde{\WW}^{1b_3c_3}\tilde{\WW}^{1b_4c_4}\epsilon^{b_1b_2b}\epsilon^{c_1c_2c}\epsilon_{b_3b_4b}\epsilon_{c_3c_4c}~.
\end{multline}
\normalsize

It is interesting to see these null relations as a consequence of the nulls in the original chiral algebra. To that effect, let us re-derive the dimension one nulls in this manner. Starting with the $\mathbf{(8,1,1)}$ null states in Table \ref{tab:relation_reductions} and specializing the indices to $(a_1,a_2)=(3,1)$, we find the null relation
\begin{equation}
0=\tfrac13 W_{3bc}\tilde W^{1bc}  + (J^{(1)})_{3}^{\ a} (J^{(1)})_{a}^{\ 1} + 3 \partial (J^{(1)})_{3}^{\ 1}  =  \tfrac13  {\WW}_{3bc} \tilde {\WW}^{1bc} +\tfrac13  \JJ_{\uf(1)} + d(\ldots)~,
\end{equation}
thus reproducing Eqn. \eqref{eq:nullJU1}. Alternatively, starting with the null states in the $\mathbf{(8,8,1)}$ and specializing the indices to $(a_1,a_2)=(3,1)$, we obtain the null relation
\begin{multline}
0=\left(W_{3b_1c}\tilde W^{1b_2c}  - \frac13 \delta_{b_1}^{b_2}W_{3bc}\tilde W^{1bc} \right) -\frac13 \beta_1 (J^{(1)})_{3}^{\ 1} (J^{(2)})_{b_1}^{\ b_2} \\
 = \left({\WW}_{3b_1c} \tilde {\WW}^{1b_2c} - \frac13  \delta_{b_1}^{b_2} {\WW}_{3bc} \tilde {\WW}^{1bc} \right)+ \left(\JJ^{(2)}\right)_{b_1}^{\ b_2}  +   d(\ldots)~,
\end{multline}
which precisely matches the null relation of Eqn. \eqref{eq:nullJSU3_2}. Similarly, one can reproduce \eqref{eq:nullJSU3_3}. It is straightforward to check that the null relations in Eqns. \eqref{eq:nullWlower}-\eqref{eq:nullWupper} can be obtained from the relations in the $\mathbf{(\bar{6},3,3)}$ and $\mathbf{(6,\bar{3},\bar{3})}$ and specializing the indices appropriately.

%% file: appendices/app_B.tex

\section{Cylinder and cap details}
\label{app:cylinders_and_caps}

This appendix describes the quantum Drinfeld-Sokolov reduction that produces the chiral algebra for cylinder and cap geometries when $\gf=\suf(3)$. We first introduce some general formulas for the Schur superconformal index associated to these geometries. These formulas prove useful for getting a basic intuition for how these chiral algebras may be described.

\input{./appendices/app_B1}
\bigskip
\input{./appendices/app_B2}
\bigskip

%% file: appendices/app_B1.tex

\subsection{Schur indices}
\label{subapp:cylinder_cap_index}

Although they are only formally defined (there is no true four-dimensional SCFT associated to the cylinder and cap geometries), the reduction rules for the Schur index allow us to define an index for these geometries that must behave appropriately under gluing. Let us determine these indices.

\paragraph{Cylinder}
Using the general results given in Eqns. \eqref{eq:SchurindexUVcurve} and \eqref{eq:psi_max}, the index of the two-punctured sphere theory can be determined immediately
\begin{align}\label{eq:indexcylinder}
\begin{split}
\II_\text{cylinder}\left(q;\ab,\mathbf{b}\right) &~=~ K_{\mathrm{max.}}(\ab;q) K_{\mathrm{max.}}(\mathbf{b};q) \sum_{\Rf} \chi_{\Rf}(\ab) \chi_{\Rf}(\mathbf{b})\\
&~=~\PE\left[ \frac{q}{1-q} \left(\chi_{\mathrm{adj}}(\ab) + \chi_{\mathrm{adj}}(\mathbf{b})\right)\right] \sum_{\Rf} \chi_{\Rf}(\ab) \chi_{\Rf}(\mathbf{b})~.
\end{split}
\end{align}
Upon using the relation $\sum_{\Rf} \chi_{\Rf}(\ab) \chi_{\Rf}(\mathbf{b}) = \delta(\ab,\mathbf{b}^{-1})$, where the delta function is defined with respect to the Haar measure, we can rewrite this index as
\begin{equation}
\II_\text{cylinder}\left(q;\ab,\mathbf{b}\right) = \PE\left[ \frac{2q}{1-q} \chi_{\mathrm{adj}}(\ab) \right] \delta(\ab,\mathbf{b}^{-1}) = I_V^{-1}(\ab;q)\ \delta(\ab,\mathbf{b}^{-1})~,
\end{equation}
where $I_V$ is the vector multiplet index \eqref{eq:vector_multiplet_index}. This makes it clear that when the gluing prescription for the index given in Eqn. \eqref{eq:indexgluing} is applied, the index $\II_{\TT}(q; \ab,\ldots)$ of a generic theory $\TT$ containing a maximal puncture with fugacities $\ab$ remains the same after gluing a cylinder to that maximal puncture
\begin{equation}
\int [d\ab] \Delta(\ab) I_V(q;\ab)\  \II_{\TT}(q; \ab,\ldots)\  \II_\text{cylinder}\left(q;\ab^{-1},\mathbf{b}\right) = \II_{\TT}(q; \mathbf{b},\ldots)~.
\end{equation}
Here $[d\ab] = \prod_{j=1}^{\text{rank} \gf}\frac{da_j}{2\pi i a_j}$ and $\Delta(\ab)$ is the Haar measure.

Returning to expression \eqref{eq:indexcylinder}, we wish to rewrite the sum over representations. Let us therefore consider the regularized sum
\begin{equation}
\label{eq:regsumreps}
\sum_{\Rf} u^{|\Rf|}\chi_{\Rf}(\ab) \chi_{\Rf}(\mathbf{b}) = \PE\left[ u\ \chi_\ff(\ab)\chi_\ff(\mathbf{b}) - u^{n} \right]\;,
\end{equation}
where $|\Rf|$ denotes the number of boxes in the Young diagram corresponding to the representation $\Rf$ of $\gf=\suf(n).$ For $\gf=\suf(2)$ we have checked this equality exactly by performing the geometric sums and for $\suf(3),$ $\suf(4)$ and $\suf(5)$ in a series expansion in $u.$ In the limit $u\rightarrow 1$ one can verify that the right hand side behaves as a $\delta$-function with respect to the Haar measure, as expected. Consequently, the cylinder index can then be rewritten in a particularly useful form,
\begin{equation}
\II_\text{cylinder}\left(q;\ab,\mathbf{b}\right)=\PE\left[ \frac{q}{1-q} \left(\chi_{\mathrm{adj}}(\ab) +\chi_{\mathrm{adj}}(\mathbf{b})\right) +   \chi_\ff(\ab)\chi_\ff(\mathbf{b}) - 1\right]~.
\end{equation}
By using $\chi_{\mathrm{adj}}(\ab) = \chi_{\ff}(\ab)\chi_{\ff}(\ab^{-1})-1 $ and the $\delta$-function constraint, one can finally rewrite the index as
\begin{align}\begin{split}
\II_\text{cylinder}\left(q;\ab,\mathbf{b}\right)&~=~\PE\left[ \frac{q}{1-q} \left(\chi_{\mathrm{adj}}(\mathbf{b}) + \left(\chi_{\ff}(\ab)\chi_{\ff}(\mathbf{b})-1 \right)\right) + \chi_\ff(\ab)\chi_\ff(\mathbf{b}) - 1\right]~,\\
&~=~ \PE\left[ \frac{q}{1-q} \chi_{\mathrm{adj}}(\mathbf{b}) + \frac{1}{1-q}\left(\chi_{\ff}(\ab)\chi_{\ff}(\mathbf{b})-1 \right)\right]~.
\end{split}\end{align}
Note that this looks like the partition function of a finitely generated chiral algebra satisfying a single relation. Namely, it appears that the chiral algebra has one set of dimension one currents transforming in the adjoint of $\suf(n)$, in addition to a bifundamental field $g_{ab}$ of dimension zero subject to a dimension zero constraint in the singlet representation. Going further, using this interpretation of the index and reintroducing the fugacity $u$ as in \eqref{eq:regsumreps}, we see that $u$ counts the power of the bifundamental generators in an operator, and the constraint should then involve $n$ bifundamental fields. A natural form for such a relation (after proper rescaling of the generators) is the following,
\begin{equation}
\label{eq:cylinderconstraint}
\frac{1}{n!}\epsilon^{a_1a_2\ldots a_n}\epsilon^{b_1b_2\ldots b_n} g_{a_1b_1}g_{a_2b_2}\ldots g_{a_nb_n} = 1.
\end{equation}
Interpreting $g_{ab}$ as a matrix, this is a unit determinant condition. This picture, guessed on the basis of the superconformal index, will be borne out in the qDS analysis below.

\paragraph{Cap} 
A similarly heuristic analysis is possible for the theory associated to a decorated cap, which is obtained by further partially closing a puncture in the cylinder theory. The index of the decorated cap theory takes the form
\begin{align}
\II_{\text{cap}({\Lambda})}\left(q;\ab,\mathbf{b}_{{\Lambda}}\right) &~=~ K_{\mathrm{max.}}(\ab;q) K_{{\Lambda}}(\mathbf{b}_{{\Lambda}},q) \sum_{\Rf} \chi_{\Rf}(\ab) \chi_{\Rf}(\mathrm{fug}_{{\Lambda}}(\mathbf{b}_{{\Lambda}};q))~,\nn\\
&~=~\PE\left[ \frac{q}{1-q} \chi_{\mathrm{adj}}(\ab) + \sum_j \frac{q^{j+1}}{1-q} \Tr_{\RR_j^{(\mathrm{adj})}}(\mathbf{b}_{{\Lambda}})\right] \sum_{\Rf} \chi_{\Rf}(\ab) \chi_{\Rf}(\mathrm{fug}_{{\Lambda}}(\mathbf{b}_{{\Lambda}};q))~,\\
&~=~I_V^{-1/2}(\ab;q)\  K_{{\Lambda}}(\mathbf{b}_{{\Lambda}},q)\  \delta(\ab^{-1},\mathrm{fug}_{{\Lambda}}(\mathbf{b}_{{\Lambda}};q))~.\nn
\end{align}
Again it is clear how gluing this index reduces the flavor symmetry of the puncture. Using \eqref{eq:indexgluing} and the general expression for a class $\SS$ index \eqref{eq:SchurindexUVcurve} for some theory $\TT$ of genus $g$ and containing $s$ punctures, of which the first is maximal with corresponding flavor fugacities $\ab$, one obtains by gluing the cap to this maximal puncture
\begin{multline}
\int [d\ab] \Delta(\ab) I_V(\ab;q)\ \II_{\text{cap}({\Lambda})}\left(q;\ab^{-1},\mathbf{b}_{{\Lambda}}\right)\  \sum_{\Rf} C_\Rf(q)^{2g-2+s}  K_{\mathrm{max.}}(\ab;q)  \chi_{\Rf}(\ab) \prod_{i=2}^s \psi_{\Rf}^{\Lambda_i}({\bf x}_{\Lambda_i} ;q) \\ = \sum_{\Rf} C_\Rf(q)^{2g-2+s}  K_{{\Lambda}}(\mathbf{b}_{{\Lambda}},q)  \chi_{\Rf}(\mathrm{fug}_{{\Lambda}}(\mathbf{b}_{{\Lambda}};q)) \prod_{i=2}^s \psi_{\Rf}^{\Lambda_i}({\bf x}_{\Lambda_i} ;q)~,
\end{multline}
where we have again used that $K_{\mathrm{max.}}(\ab;q) = I_V^{-1/2}(\ab;q).$

As in the previous paragraph we can rewrite the index in a suggestive fashion,
\begin{align}
\II_{\text{cap}({\Lambda})}\left(q;\ab,\mathbf{b}_{\Lambda}\right)
&~=~ \PE\left[ \sum_j \frac{q^{j+1}}{1-q} \Tr_{\RR_j^{(\mathrm{adj})}}(\mathbf{b}_{{\Lambda}}) + \frac{1}{1-q}\left(\chi_{\ff}(\ab)\chi_{\ff}(\mathrm{fug}_{{\Lambda}}(\mathbf{b}_{{\Lambda}};q))-1 \right)\right]~.
\end{align}
A natural interpretation of this index is as that of a chiral algebra with generators given by currents $J_{\bar\alpha}$ for $T_{\bar\alpha} \in \ker( ad_{{\Lambda}(t_+)})$ with dimensions shifted by their ${\Lambda}(t_0)$ weight. Moreover, for each $\suf(2)$ irrep in the decomposition \eqref{eq:generaldecomposition} of the fundamental representation $\ff$ there are an additional $2j+1$ generators transforming in representation $\ff\otimes\RR_j^{(\ff)}$ with dimensions $-j,-j+1,\ldots,j$. The latter generators satisfy a singlet relation of dimension zero.

%% file: appendices/app_B2.tex

\subsection{QDS argument}
\label{subapp:cylinder_cap_qDS}

Now that we have some intuition for the kinds of chiral algebras to expect, let us study the cylinder theory for $\gf=\suf(3)$ by fully closing a puncture in the $\goodchi[T_3]$ theory. Full closure is achieved via the principal embedding $\rho:\suf(2)\rightarrow \gf$, which is can be specified explicitly in components as 
\begin{equation}
\rho(t_-) = 2(T_2^{\ 1} + T_3^{\ 2})~,\qquad  \rho(t_0) = T_1^{\ 1}-T_3^{\ 3}~,\qquad \rho(t_+)=T_1^{\ 2} + T_2^{\ 3}~.
\end{equation}
The grading by $\rho(t_0)$ is integral, with the negatively graded generators being $T_3^{\ 1}$ with grade minus two and $T_2^{\ 1},T_3^{\ 2}$ with grade minus one. We should then impose the constraints
\begin{equation}
\label{eq:cylinderconstraints}
\left(J^{(1)}\right)_2^{\ 1} + \left(J^{(1)}\right)_3^{\ 2} = 1\;,\qquad \left(J^{(1)}\right)_2^{\ 1} - \left(J^{(1)}\right)_3^{\ 2} = 0\;, \qquad \left(J^{(1)}\right)_3^{\ 1}=0~.
\end{equation}
Upon introducing three $(b,c)$-ghost systems -- $(b_2^{\ 1},c_1^{\ 2})$, $(b_3^{\ 2},c_2^{\ 3})$, and $(b_3^{\ 1},c_1^{\ 3})$ -- these first-class constraints are implemented by a BRST procedure via the current
\begin{equation}
d(z) =  (J^{(1)})_2^{\ 1} c_1^{\ 2}(z) + (J^{(1)})_3^{\ 2}c_2^{\ 3}(z) + (J^{(1)})_3^{\ 1} c_1^{\ 3}(z) - \frac{1}{2} (c_1^{\ 2}+c_2^{\ 3})(z) - b_3^{\ 1} c_1^{\ 2}c_2^{\ 3}(z)~.
\end{equation}

\begin{table}[t!]
\begin{center}
\begin{tabular}{c|l}
\hline\hline
\text{dimension} & \text{generators}\\
\hline
0 & $\WW_{3bc}, \tilde{ \WW}^{1bc} $\\
1 & $\WW_{2bc}, \tilde{ \WW}^{2bc}, ( \JJ^{2})_{b_1}^{\ b_2}, (\JJ^{3})_{c_1}^{\ c_2}$\\
2 & $\hat{ \JJ}_{\text{sum}}, \WW_{1bc}, \tilde{ \WW}^{3bc}$\\
3 &$ (\hat{ \JJ}^{1})_1^{\ 3}$\\
\hline
\end{tabular}
\end{center}
\caption{\label{tab:generatorscylinder}(Redundant) generators of the cylinder theory for $\gf=\suf(3)$.}
\end{table}

This cohomological problem is partly solved by following the same approach as that advocated in Subsec. \ref{subsubsec:qDSspecseq}. The redundant generators of the reduced algebra are the tic-tac-toed versions of the currents $ (\hat J^{1})_1^{\ 3}$ and $\hat J_{\text{sum}} \equiv (\hat J^{1})_1^{\ 2} + (\hat J^{1})_2^{\ 3},$ as well as of the generators $\{(J^{2})_{b_1}^{\ b_2},\,(J^{3})_{c_1}^{\ c_2},\,W_{abc},\,\tilde W^{abc}\}$. These currents can be seen arranged according to their dimensions in Table \ref{tab:generatorscylinder}.

The explicit form of the tic-tac-toed generators of dimensions zero and one are fairly simple,
\begin{align}
\WW_{3bc}				&~\colonequals~ W_{3bc}~,\\
\tilde{\WW}^{1bc}		&~\colonequals~ \tilde W^{1bc}~,\\
\WW_{2bc}				&~\colonequals~ W_{2bc} + 2 W_{3bc}(\hat J^{(1)})_3^{\ 3}~,\\
\tilde{ \WW}^{2bc}		&~\colonequals~ \tilde W^{2bc} + 2 \tilde W^{1bc} (\hat J^{(1)})_1^{\ 1}~,\\
(\JJ^{2})_{b_1}^{\ b_2}	&~\colonequals~ (J^{2})_{b_1}^{\ b_2}~,\\
(\JJ^{3})_{c_1}^{\ c_2}	&~\colonequals~ (J^{3})_{c_1}^{\ c_2}~.
\end{align}
On the other hand, the higher dimensional generators are quite complicated,
{\small
\begin{align}\begin{split}
\hat{ \JJ}_{\text{sum}}&~\colonequals~ (\hat J^{1})_1^{\ 2} + (\hat J^{1})_2^{\ 3}  \\
& - \left(-2(2+k)\partial(\hat J^{1})_2^{\ 2} - 4(2+k)\partial(\hat J^{1})_3^{\ 3} +2(\hat J^{1})_1^{\ 1}(\hat J^{1})_2^{\ 2} + 2 (\hat J^{1})_1^{\ 1}(\hat J^{1})_3^{\ 3} + 2(\hat J^{1})_2^{\ 2}(\hat J^{1})_3^{\ 3} \right)~,
\end{split}\\
\begin{split}
\WW_{1bc}&~\colonequals~ W_{1bc} - \left(2 W_{2bc}(\hat J^{1})_1^{\ 1} -2 W_{3bc}(\hat J^{1})_2^{\ 3} \right)\\
&+ \left( -4\left((\hat J^{1})_1^{\ 1} + (\hat J^{1})_2^{\ 2}\right) W_{3bc} (\hat J^{1})_3^{\ 3} -\frac{1}{3}(-20-12k)W_{3bc}\partial(\hat J^{1})_3^{\ 3} - \frac{8}{3} \partial W_{3bc} (\hat J^{1})_3^{\ 3}\right)~,
\end{split}\\
\begin{split}
\tilde{ \WW}^{3bc}&~\colonequals~ \tilde W^{3bc} - \left( 2\tilde W^{2bc}(\hat J^{1})_3^{\ 3} + 2 \tilde W^{1bc}(\hat J^{1})_2^{\ 3} \right) +4(\hat J^{1})_3^{\ 3} \tilde W^{1bc} (\hat J^{1})_2^{\ 2}\\
& - \tilde W^{1bc}\partial\left(-\frac{4}{3}(\hat J^{1})_1^{\ 1} +(8+4k)(\hat J^{1})_3^{\ 3}  \right) - \partial \tilde W^{1bc}\left( -\frac{4}{3}(\hat J^{1})_1^{\ 1} -\frac{4}{3}(\hat J^{1})_3^{\ 3}\right)~,
\end{split}\\
\begin{split}
(\hat{ \JJ}^{1})_1^{\ 3}&~\colonequals~ (\hat J^{1})_1^{\ 3} - \left( 2(k+2)\partial(\hat J^{1})_1^{\ 2} -2 (\hat J^{1})_2^{\ 3}\left((\hat J^{1})_2^{\ 2}+(\hat J^{1})_3^{\ 3} \right) -2(\hat J^{1})_1^{\ 2}\left( (\hat J^{1})_1^{\ 1} + (\hat J^{1})_2^{\ 2} \right) \right)\\
&+ 4(4+4k+k^2)\partial^2(\hat J^{1})_1^{\ 1} - 4(2+k)(\hat J^{1})_1^{\ 1}\partial (\hat J^{1})_1^{\ 1}+ 4(2+k)(\hat J^{1})_1^{\ 1}\partial (\hat J^{1})_2^{\ 2}\\
& -4 \left( (\hat J^{1})_1^{\ 1} + (\hat J^{1})_3^{\ 3}   \right)\left((\hat J^{1})_2^{\ 2}+(\hat J^{1})_3^{\ 3} \right)\left((\hat J^{1})_1^{\ 1}+(\hat J^{1})_2^{\ 2}\right)~.
\end{split}
\end{align}}%

Our next task should be to check for redundancies by computing null relations. This analysis is substantially complicated by the presence of dimension zero fields in the cohomology. This means that we don't have an algorithm for finding such redundancies that must terminate in principle. Instead, we use the nulls of $T_3$ to predict some of the nulls in the cylinder theory.

\paragraph{Dimension zero nulls}
Starting with the $\mathbf{(8,1,1)}$ nulls and specializing the indices to $(a_1,a_2)=(3,1)$ we obtain the null relation
\begin{equation}
0=\frac{1}{3}W_{3bc}\tilde W^{1bc}  + (J^{1})_{3}^{\ a} (J^{1})_{a}^{\ 1} - 3 \partial (J^{1})_{3}^{\ 1}  = \frac{1}{4} +\frac{1}{3} {\WW}_{3bc} \tilde {\WW}^{1bc}  + d(\ldots)~.
\end{equation}
Similarly, starting with the $\mathbf{(8,8,1)}$ nulls and specializing the indices to $(a_1,a_2)=(3,1)$ we obtain the null relation
\begin{align}\begin{split}
0&~=~\left(W_{3b_1c}\tilde W^{1b_2c}  - \frac{1}{3}\delta_{b_1}^{b_2}W_{3bc}\tilde W^{1bc} \right) + (J^{(1)})_{3}^{\ 1} (J^{(2)})_{b_1}^{\ b_2}~,\\
 &~=~ \left({\WW}_{3b_1c} \tilde {\WW}^{1b_2c} - \frac{1}{3} \delta_{b_1}^{b_2} {\WW}_{3bc} \tilde {\WW}^{1bc} \right) +   d(\ldots)~.
\end{split}\end{align}
Similar nulls can be found by interchanging the second and third puncture. In summary, we have the relations
\begin{equation}
\label{WWtinverse}
{\WW}_{3b_1c} \tilde {\WW}^{1b_2c} =   -\frac{1}{4} \delta_{b_1}^{b_2}\;, \qquad  {\WW}_{3bc_1} \tilde {\WW}^{1bc_2} =  - \frac{1}{4} \delta_{c_1}^{c_2}~.
\end{equation}
This shows that, up to a rescaling, $\WW_{3bc}(z)$ and $\tilde\WW^{1bc}(z)$ can be thought of as inverses of one another.

Next, we look at the $\mathbf{(\bar{6},3,3)}$ nulls and specialize $a_1=a_2=1$, which gives us
\begin{align}\begin{split}
0&~=~2(J^{1})_{\alpha_1}^{\ 1} W_{\alpha_2bc}\epsilon^{\alpha_1\alpha_21} + \tilde W^{1b_1c_1}\tilde W^{1b_2c_2} \epsilon_{bb_1b_2}\epsilon_{cc_1c_2}~,\\ 
 &~=~\WW_{3bc} +  \tilde{\WW}^{1b_1c_1}\tilde{\WW}^{1b_2c_2} \epsilon_{bb_1b_2}\epsilon_{cc_1c_2} + d(\dots)~.
\end{split}\end{align}
Similarly from the nulls in the $\mathbf{(6,\bar{3},\bar{3})}$ we find
\begin{align}\begin{split}
0&~=~(J^{1})_{3}^{\ \ \alpha_1} \tilde W^{\alpha_2bc}\epsilon_{\alpha_1\alpha_23} +\frac{1}{2}  W_{3b_1c_1} W_{3b_2c_2} \epsilon^{bb_1b_2}\epsilon^{cc_1c_2}~,\\
 &~=~-\frac{1}{2} \tilde{\WW}^{1bc} +\frac{1}{2} \WW_{3b_1c_1} \WW_{3b_2c_2} \epsilon^{bb_1b_2}\epsilon^{cc_1c_2} + d(\ldots)~.
\end{split}\end{align}
Combining these with the previous relations, we find that
\begin{equation}
\frac{1}{3!}\tilde {\WW}^{1bc}\tilde {\WW}^{1b_1c_1}\tilde {\WW}^{1b_2c_2} \epsilon_{bb_1b_2}\epsilon_{cc_1c_2} = -\frac{1}{3!}\tilde {\WW}^{1bc}{\WW}_{3bc} = - \frac{1}{3!}{\WW}_{3bc} \tilde {\WW}^{1bc} =  \frac{1}{8}~,
\end{equation}
and 
\begin{equation}
\frac{1}{3!}{\WW}_{3bc}{\WW}_{3b_1c_1} {\WW}_{3b_2c_2} \epsilon^{bb_1b_2}\epsilon^{cc_1c_2} =  \frac{1}{3!}{\WW}_{3bc}\tilde {\WW}^{1bc} = -\frac{1}{8}~.
\end{equation}
These are conditions on the determinants of $\WW_{3bc}$ and $\tilde \WW^{1bc}$ thought of as three-by-three matrices. Note that we used the relation $\tilde {\WW}^{1bc}{\WW}_{3bc} = {\WW}_{3bc} \tilde {\WW}^{1bc}$, which is true in cohomology:
\begin{equation}
\tilde {\WW}^{1bc}{\WW}_{3bc} = {\WW}_{3bc} \tilde {\WW}^{1bc} -d(9 \partial b_3^{\ 1})~.
\end{equation}
If we now introduce rescaled operators $g_{bc} \colonequals -2 {\WW}_{3bc}$ and $\tilde g^{bc} \colonequals 2 \tilde {\WW}^{1bc}$, then $g$ and $\tilde g$ have unit determinant and are inverses of one another. Because of the determinant condition, this also means that we can rewrite $\tilde g$ in terms of positive powers of $g$, so only one needs to be considered as an honest generator of the chiral algebra.

\paragraph{Dimension one nulls}
We can continue the same analysis at dimension one. The second relation in the $\mathbf{(3,3,3)}$ representation gives us
\begin{equation}
(\JJ^{2})_{b}^{\ \beta}\WW_{3 \beta c_1} = (\JJ^{3})_{c_1}^{\ \gamma}\WW_{3 b\gamma}~.
\end{equation}
By taking the normal ordered product of both sides with $\tilde {\WW}^{1 b c_2}$ and re-ordering (ignoring BRST exact terms), we can make a sequence of replacements using the dimension zero relations of the previous paragraph and end up with the following derivation,
{\small
\begin{align}
&&\tilde {\WW}^{1 b c_2}({\JJ}^{(2)})_{b}^{\ \beta}{\WW}_{3 \beta c_1} &~=~ \tilde {\WW}^{1 b c_2}({\JJ}^{(3)})_{c_1}^{\ \gamma}{\WW}_{3 b\gamma}\nn\\
&\Longrightarrow&({\JJ}^{(2)})_{b}^{\ \beta}{\WW}_{3 \beta c_1}\tilde {\WW}^{1 b c_2} +\tfrac{8}{3}{\WW}_{3 \beta c_1}\partial\tilde {\WW}^{1 \beta c_2} &~=~ ({\JJ}^{(3)})_{c_1}^{\ \gamma}{\WW}_{3 b\gamma}\tilde {\WW}^{1 b c_2} - \tfrac{1}{3} {\WW}_{3 \beta c_1}\partial\tilde {\WW}^{1 \beta c_2} +\delta_{c_1}^{c_2} {\WW}_{3 \beta \gamma}\partial\tilde {\WW}^{1 \beta \gamma}\nn\\
&\Longrightarrow&({\JJ}^{(2)})_{b}^{\ \beta}{\WW}_{3 \beta c_1}\tilde {\WW}^{1 b c_2} &~=~  ({\JJ}^{(3})_{c_1}^{\ \gamma}{\WW}_{3 b\gamma}\tilde {\WW}^{1 b c_2} - 3 \left( {\WW}_{3 \beta c_1}\partial\tilde {\WW}^{1 \beta c_2} -\tfrac{1}{3}\delta_{c_1}^{c_2} {\WW}_{3 \beta \gamma}\partial\tilde {\WW}^{1 \beta \gamma} \right)\nn\\
&\Longrightarrow&({\JJ}^{(2)})_{b}^{\ \beta}{\WW}_{3 \beta c_1}\tilde {\WW}^{1 b c_2} &~=~ -\tfrac{1}{4} ({\JJ}^{(3)})_{c_1}^{\ c_2} - 3 \left( {\WW}_{3 \beta c_1}\partial\tilde {\WW}^{1 \beta c_2} -\tfrac{1}{3}\delta_{c_1}^{c_2} {\WW}_{3 \beta \gamma}\partial\tilde {\WW}^{1 \beta \gamma} \right)\nn\\
&\Longrightarrow&({\JJ}^{(2)})_{b}^{\ \beta}{ g}_{\beta c_1}\tilde { g}^{ b c_2} &~=~ ({\JJ}^{(3)})_{c_1}^{\ c_2} - 3 \left( {g}_{\beta c_1}\partial\tilde { g}^{ \beta c_2} -\tfrac{1}{3}\delta_{c_1}^{c_2} { g}_{\beta \gamma}\partial\tilde { g}^{\beta \gamma} \right)~.\label{eq:JintermsofJ}
\end{align}}%
At last, we see that the current $\JJ^{(3)}$ is not an independent generator. 

Other dimension one nulls can be obtained from the first equality in the $\mathbf{(3,3,3)}$. Here we find
\begin{equation}
(J^{1})_{3}^{\ \alpha}W_{\alpha bc} = (J^{2})_{b}^{\ \beta}W_{3 \beta c} ~\Longrightarrow~ \frac{1}{2}\WW_{2bc}+\frac{2}{3}\partial \WW_{3bc} = (\JJ^{2})_{b}^{\ \beta}\WW_{3 \beta c}~,
\end{equation}
which implies that the generator $\WW_{2bc}$ is not independent. Similarly, from the $\mathbf{(\bar3,\bar3,\bar3)}$ relations one finds
\begin{equation}
(J^{1})_{\alpha}^{\ 1}\tilde W^{\alpha bc} = (J^{2})_{\beta}^{\ b}\tilde W^{1 \beta c} ~\Longrightarrow~ \frac{1}{2}\tilde{\WW}^{2bc}-\frac{2}{3}\partial \tilde{\WW}^{1bc} = (\JJ^{2})_{\beta}^{\ b}\tilde {\WW}^{1 \beta c},
\end{equation}
which implies that $\tilde{\WW}^{2bc}$ is not an independent generator. 

Based on the analysis of the index in \ref{subapp:cylinder_cap_index}, we expect that all higher dimensional generators can be similarly related via null relations to composites of $\JJ^{2}$ and $g_{bc} = -2\WW_{3 b c}$. It would be interesting if this could be proven as a consequence of only the null states that are guaranteed to exist based on nulls of the unreduced theory, although such a simplification is not a necessary condition for the existence of the desired nulls.

%% file: appendices/app_C.tex

\section{Spectral sequences for double complexes}
\label{app:spectral_sequences}

In this appendix we review some of the basics of spectral sequences. Standard references are \cite{BoTu,mccleary}. One can also consult Section 5.3 of \cite{deBoer:1992sy} for a concise summary of some of the most useful statements.

One of the simplest spectral sequences makes an appearance when one considers a single cochain complex $(M^{\high{*}},d)$, where $d:M^{\high{p}}\rightarrow M^{\high{p+1}}$ is a differential of degree one satisfying $d\circ d=0$. A \emph{decreasing filtration} of $M^{\high{*}}$ is a family of subspaces $\{F^{\high{p}}M;~p\in\mathbb{Z}\}$ such that $F^{\high{p+1}}M \subseteq F^{\high{p}} M$ and $\cup_p F^{\high{p}} M = M$. We restrict our attention to \emph{bounded differential filtrations}, which satisfy two additional properties:
\begin{enumerate}
\item[$\bullet$] There exist $s,t\in\Zb$ such that $F^{\high{p}}M=M$ for $p\leqslant t$ and $F^{\high{p}}M=0$ for $p\geqslant s$. 
\item[$\bullet$] The filtration is compatible with the differential, \ie, $d(F^{\high{p}}M)\subseteq F^{\high{p}}M$. 
\end{enumerate}
We further introduce the spaces $F^{\high{p}}M^r\colonequals F^{\high{p}}M\cap M^r$. One then says that the filtration is bounded in each dimension if it is bounded for each $r$. The associated graded vector space is defined as 
\begin{equation}
E_{0}^{p,q}(M^{\high{*}},F) \colonequals F^{\high{p}}M^{\high{p+q}}/F^{\high{p+1}}M^{\high{p+q}}~.
\end{equation} 
Note that at the level of vector spaces one has $M^r \cong\oplus_{p+q=r}E_0^{p,q}$.

If $F$ is a bounded differential filtration of $(M^*,d)$, then also $(F^{\high{p}} M,d)$ is a complex. The inclusion map $F^{\high{p}}M\hookrightarrow M$ descends to a map in cohomology $H(F^{\high{p}}M,d)\rightarrow H(M,d)$ which is however not necessarily injective. We denote the image of $H(F^{\high{p}}M,d)$ under this map as $F^{\high{p}} H(M,d).$ This defines a bounded filtration on $H(M,d).$

A spectral sequence is defined as a collection of bigraded spaces $(E^{*,*}_r, d_r)$ where $r=1,2,\ldots$, the differentials $d_r$ have degrees $(r,1-r)$, and for all $p,q,r$ one has $E^{p,q}_{r+1}\cong H^{p,q}(E^{*,*}_r,d_r)$. A spectral sequence is said to converge to $N^*$ if there exists a filtration $F$ on $N^*$ such that $E_\infty^{p,q} \cong E_0^{p,q}(N^*,F)$. The main theorem of concern is then for any complex $(M,d)$ with a differential filtration $F$ bounded in each dimension, one can find a spectral sequence with $E_1^{p,q} = H^{p+q}(F^{\high{p}}M/F^{\high{p+1}}M)$ that converges to $H^*(M,d)$. In favorable situations, one may have $d_r = 0$ for $r\geqslant r_0$ in which case the spectral sequence terminates: $E^{p,q}_{r_0} = E^{p,q}_\infty$.

Let us consider the case of a double complex $(M^{*,*};d_0, d_1)$, where $M$ is bigraded and $d_0$, $d_1$ are maps of degree $(1,0)$ and $(0,1)$ respectively, satisfying $d_0\circ d_0=d_1\circ d_1=d_0\circ d_1+d_1\circ d_0$. Diagrammatically, a double complex is represented as:
\begin{equation*}
\begin{CD}
 @.                            @AAd_1A               @AAd_1A           @. \\
   @>d_0>>   M^{p,q+1}    @>d_0>>  M^{p+1,q+1}  @>d_0>> \\
 @.                            @AAd_1A               @AAd_1A           @. \\
  @>d_0>>   M^{p,q}       @>d_0>>  M^{p+1,q}   @>d_0>>  \\
 @.                           @AAd_1A               @AAd_1A            @. 
\end{CD}
\end{equation*}
The associated \emph{total complex} is defined as $\mathrm{Tot}^n M\colonequals\oplus_{p+q=n}M^{p,q}$, with total differential $d\colonequals d_0+d_1$. A double complex allows for two filtrations, namely,
\begin{equation}
F_{\rm I}^p(\mathrm{Tot}^n\ M) = \oplus_{r\geqslant p}M^{r,n-r}\;, \qquad F_{\rm II}^p(\mathrm{Tot}^n\ M) = \oplus_{r\geqslant p}M^{n-r,r}~.
\end{equation}
These filtrations are bounded in each dimension if for each $n$ only a finite number of $M^{p,q}$ with $n=p+q$ are non-zero.

Correspondingly, we can consider two spectral sequences converging to $H^*(\mathrm{Tot}\ M, d)$ with as first terms
\begin{eqnarray}
&_{\rm I} E^{p,q}_1 \cong H^{p,q}(M,d_1)\;, \qquad & _{\rm I} E^{p,q}_2 \cong H^{p,q}(H^{*,*}(M,d_1),d_0)\\
&_{\rm II} E^{p,q}_1 \cong H^{p,q}(M,d_0)\;, \qquad & _{\rm II} E^{p,q}_2 \cong H^{p,q}(H^{*,*}(M,d_0),d_1)~.
\end{eqnarray}
Note that here one can show that the first term of the spectral sequence is equal to the one mentioned in the more general case above. Higher differentials $d_{r+1}$ for $r\geqslant 1$ are defined by $d_{r+1}x=d_1 y$ where $y$ is defined by $d_0 y=d_r x$. Such a $y$ can be proven to always exist, so that the higher differentials are always well-defined.

\paragraph{Example} As a simple example of the utility of spectral sequences, let us reproduce a proof of the K\"unneth formula \cite{deBoer:1992sy}. Consider a differential graded algebra $(\Ab,d)$, \ie, a graded algebra endowed with a differential $d$ of degree one satisfying the Leibniz rule. Let it have two graded subalgebras $\Ab_1$ and $\Ab_2$ which are respected by the differential, \ie, $d\Ab_i\subseteq \Ab_i$. Let us assume the multiplication map $m:\Ab_1\otimes\Ab_2\rightarrow \Ab$ is an isomorphism of vector spaces. Then one can define the double complex $(M^{p,q}; d_0, d_1)$ by 
\begin{equation}
M^{p,q}\colonequals m(\Ab_1^p\otimes\Ab_2^q)~,\quad d_0(a_1a_2)=d(a_1)a_2~,\quad d_1(a_1a_2)=(-1)^{\mathrm{deg}(a_1)}a_1 d(a_2)~.
\end{equation} 
Assume that this double complex is bounded in each dimension; then one can make use of the spectral sequence for the double complex as described above. One finds for the first couple of levels 
\begin{equation}
E^{p,q}_1\cong m(\Ab_1^p\otimes H^q(\Ab_2,d))~,\qquad E^{p,q}_2 \cong m(H^p(\Ab_1,d)\otimes H^q(\Ab_2,d))~.
\end{equation} 
Higher differentials all manifestly vanish, so the spectral sequence terminates. At the level of vector spaces, the above-stated theorem implies that $H^*(\Ab,d) \cong m(H^*(\Ab_1,d)\otimes H^*(\Ab_2,d))$. This statement can be extended to an isomorphism of algebras because $a_1a_2$ is a representative of an element in $H^*(\Ab,d)$.